\documentclass[a4paper, 11pt, twoside, onecolumn, openright, final]{memoir}

\setbinding{0cm}
\setlrmargins{*}{*}{1.0} 
\checkandfixthelayout

\chapterstyle{verville}

\captionnamefont{\small\bfseries}
\captiontitlefont{\small}

\usepackage{graphicx}

\usepackage{amsmath}
\usepackage{amssymb}
\usepackage{mathtools}
\usepackage{dsfont}
\usepackage{dirtytalk}
\usepackage[backend=biber, style = numeric-comp, sorting = none, maxbibnames=8, maxcitenames = 8, giveninits = true]{biblatex}
\usepackage{mathrsfs}
\usepackage[mathscr]{eucal}
\usepackage[linkcolor=blue,citecolor=magenta,colorlinks=false]{hyperref}
\usepackage{physics}
\usepackage{siunitx}
\usepackage{slashed}
\usepackage[ngerman,english]{babel}

\usepackage[T1]{fontenc}

\newcommand{\p}{\partial}
\newcommand{\ii}{\mathrm{i}}
\newcommand{\ee}{\mathrm{e}}
\newcommand{\nn}{\nonumber}
\newcommand{\xperp}{\mathbf x}
\newcommand{\yperp}{\mathbf y}
\newcommand{\uperp}{\mathbf u}
\newcommand{\zperp}{\mathbf z}
\newcommand{\kperp}{\mathbf k}
\newcommand{\pperp}{\mathbf p}
\newcommand{\qperp}{\mathbf q}
\newcommand{\bperp}{\mathbf b}
\newcommand{\Bperp}{\mathbf B}
\newcommand{\sperp}{\mathbf s}
\newcommand{\vperp}{\mathbf v}

\newcommand{\eperp}{\mathbf e}
\newcommand{\zeroperp}{\mathbf 0}
\newcommand{\fermi}{\femto\meter}

\newcommand{\R}{\mathds R}
\newcommand{\elrf}{\epsilon_\mathrm{LRF}}
\DeclareMathOperator\artanh{artanh}
\DeclareMathOperator\arcosh{arcosh}

\newcommand{\trento}{T\raisebox{-.5ex}{R}ENTo}

\renewenvironment{abstract}
  {\addcontentsline{toc}{chapter}{Abstract}
  \begin{center}
   \textbf{Abstract}
   \end{center}\begin{quote}}
  {\end{quote}\cleardoublepage}
  
\newenvironment{kurzfassung}
  {\addcontentsline{toc}{chapter}{Kurzfassung}
  \begin{otherlanguage}{ngerman}
  \begin{center}
   \textbf{Kurzfassung}
   \end{center}\begin{quote}}
  {\end{quote}\end{otherlanguage}\cleardoublepage}

\newenvironment{acknowledgments}
  {\addcontentsline{toc}{chapter}{Acknowledgments}
  \begin{center}
   \textbf{Acknowledgments}
   \end{center}\begin{quote}}
  {\end{quote}\cleardoublepage}

\allowdisplaybreaks

\begin{document}

\frontmatter
\thispagestyle{empty}

	\begin{center}
	{\LARGE DISSERTATION\\[1.0cm]}
	{\LARGE\textbf{Aspects of the dilute Glasma}\\[1.0cm]}
	\end{center}
	
	\begin{center}    
		{\normalsize ausgef\"uhrt zum Zwecke der Erlangung des akademischen Grades eines Doktors der Technischen Wissenschaften unter der Leitung von}\\[0.8cm]
	\end{center}
	\begin{center}    
		{\normalsize Privatdoz.\ Dipl.-Ing.\ Dr.techn.\ Andreas Ipp\\
			Institut f\"ur Theoretische Physik,\\
			Technische Universit\"at Wien, \"Osterreich \\[0.8cm]}
	\end{center}
 \begin{center}    
		{\normalsize mitbetreut durch}\\[0.8cm]
	\end{center}
 \begin{center}    
		{\normalsize Dipl.-Ing.\ Dr.techn.\ David I. M\"uller, BSc\\
			Institut f\"ur Theoretische Physik,\\
			Technische Universit\"at Wien, \"Osterreich \\[0.8cm]}
\end{center}
	\begin{center}    
		{\normalsize eingereicht an der Technischen Universit{\"a}t Wien,\\
			Fakult{\"a}t f{\"u}r Physik} \\[0.8cm]     
	\end{center} 
	\begin{center}    
		{\normalsize von\\
			\textbf{Dipl.-Ing.\ Markus Leuthner, BSc}}\\
   Matrikelnummer 01526061\\[2.5cm]
	\end{center} 
	\begin{center} 
	\noindent\begin{tabular}{llr}
		{Wien, am \makebox[2.5cm]{\hrulefill}}&\hspace*{3cm} & \makebox[4cm]{\hrulefill}  \\[2.0cm]
	\end{tabular}\\
 
	\leftskip -1.7cm
 \begin{tabular}{p{\textwidth/4+1cm}@{}p{\textwidth/4+1cm}@{}p{\textwidth/4+1cm}@{}p{\textwidth/4+1cm}@{}}
		\makebox[\textwidth/4]{\hrulefill} & \makebox[\textwidth/4]{\hrulefill} & \makebox[\textwidth/4]{\hrulefill}
  & \makebox[\textwidth/4]{\hrulefill}\\
		Dr.\ Andreas Ipp&
  Dr.\ David I. M\"uller&
  Dr.\ Tuomas Lappi & Dr.\ Edmond Iancu\\
		(Hauptbetreuer)  &
  (Zweitbetreuer) & (Gutachter)               & (Gutachter)                 \\
	\end{tabular}
	\end{center}
 \cleardoublepage

\begin{abstract}
The Glasma is a semiclassical nonequilibrium state describing the earliest stage in relativistic heavy-ion collisions predicted by the Color Glass Condensate effective theory.
It is characterized by strong color fields, which are sourced by color currents pertaining to hard partons in the colliding nuclei.
We introduce the (3+1)D dilute Glasma framework, which incorporates the longitudinal and transverse structure of colliding particles and describes the rapidity-dependence of observables like the energy-momentum tensor. This is in stark contrast to the canonical picture of boost-invariance, where nuclei are infinitesimally thin in longitudinal direction, and the rapidity-dependence of observables is lost.
We discuss the derivation of the (3+1)D dilute Glasma field-strength tensor, which relies 
on linearizing the Yang-Mills equations in the dilute approximation, i.e.,\ assuming weak sources.
The dilute Glasma energy-momentum tensor can efficiently be evaluated numerically on a lattice.
Employing a generalized 3D McLerran-Venugopalan model, we discuss numerical results for the collisions of heavy ions at energies corresponding to experiments at RHIC and the LHC.
We discover longitudinal flow that differs significantly from Bjorken flow and argue that this is a consequence of taking into account the longitudinal extension of nuclei.
Furthermore, we find limiting fragmentation as a universal feature of the dilute Glasma analytically and numerically.
Finally, we study the applicability of the dilute Glasma to proton-proton collisions and show the necessary modifications to reproduce experimental multiplicity distributions.
\end{abstract}
\begin{kurzfassung}
    Das Glasma ist ein halbklassischer Zustand weit entfernt vom thermischen Gleichgewicht, der das fr\"uheste Stadium in Kollisionen relativistischer Atomkerne beschreibt.
    Die Existenz und Beschreibung des Glasmas folgt aus dem \glqq Color Glass Condensate\grqq, einer effektiven Theorie f\"ur Quantenchromodynamik bei hohen Energien.
    Das Glasma ist charakterisiert durch starke Farbfelder, wobei die Farbladungen der harten Partonen in den Atomkernen als Quellen fungieren.
    Im (3+1)D d\"unnen Glasma sind diese Quellen in longitudinaler Richtung ausgedehnt und weisen potenziell komplexe Substruktur sowohl in longitudinaler als auch in transversaler Richtung auf.
    Dies f\"uhrt im Glasma zu einer Rapidit\"atsabh\"angigkeit von Observablen wie etwa des Energie-Impuls-Tensors, ganz im Gegensatz zu fr\"uheren boostinvarianten Beschreibungen, wo Kerne nur infinitesimale longitudinale Ausdehnung aufweisen und die Rapidit\"atsabh\"angigkeit verloren geht.
    Die Linearisierung der Yang-Mills-Gleichungen unter der Annahme schwacher Quellen erlaubt eine analytische Beschreibung des Feldst\"arketensors f\"ur das (3+1)D d\"unne Glasma. Aus dem Feldst\"arketensor kann der Energie-Impuls-Tensor berechnet und f\"ur ein gegebenes Kernmodell numerisch ausgewertet werden. Wir analysieren numerische Resultate basierend auf einer Verallgemeinerung des Kernmodells von McLerran und Venugopalan in drei Raumdimensionen bei verschiedenen Schwerpunktsenergien, die in etwa den experimentell realisierten Energien bei RHIC und LHC entsprechen. Der Materiefluss in longitudinale Richtung unterscheidet sich dabei signifikant vom bekannten \glqq Bjorken flow\grqq, den man im ultrarelativistischen Limes erwarten w\"urde.
    Dies ist darauf zur\"uckzuf\"uhren, dass die Wechselwirkung bei longitudinal ausgedehnten Kernen nicht bei einer einzigen Zeit und an einer einzigen longitudinalen Position stattfindet.
    Wir untersuchen außerdem \glqq limiting fragmentation\grqq, ein weiteres bekanntes Ph\"anomen, dessen Auftreten im d\"unnen Glasma sowohl analytisch als auch numerisch belegbar ist.
    Zum Abschluss untersuchen wir die notwendigen Modifikationen, um Proton-Proton-Kollisionen im d\"unnen Glasma zu beschreiben, insbesondere wie die experimentell gemessene Verteilung von Teilchenzahlen reproduziert werden kann. 
\end{kurzfassung}
\begin{acknowledgments}
I thank my supervisor, Andreas Ipp, and co-supervisor, David I. M\"uller, for their guidance and support.
I also thank my colleague and friend Kayran Schmidt for years of pleasant and productive collaboration.
I further thank my international collaborators, Sören Schlichting and Pragya Singh, for their expertise and collaboration on multiple publications.
My thanks extend to all other colleagues and friends at the Institute for Theoretical Physics at TU Wien, in particular Ludwig Horer and Florian Lindenbauer, for engaging talks and discussions about physics and almost everything else.
Finally, I express my continued gratitude to my parents for their ongoing support.
\end{acknowledgments}
\tableofcontents
\cleardoublepage

\mainmatter
\chapter{Introduction}
Since the formulation of quantum chromodynamics (QCD) over 50 years ago \cite{Gross:1973id, Politzer:1973fx, Gross:1973ju, Gross:2022hyw} tremendous theoretical and experimental effort was put into improving our understanding of the strong interaction. Heavy-ion collision experiments such as those performed at the Relativistic Heavy Ion Collider (RHIC) at the Brookhaven National Laboratory and the Large Hadron Collider (LHC) at CERN probe the behavior of QCD matter under extreme conditions through the production of the quark-gluon plasma (QGP) \cite{Shuryak:2004cy, Busza:2018rrf, Gelis:2021zmx}.
Originally predicted in the 70s, the QGP was detected experimentally in 2000 \cite{Heinz:2000bk}. The QGP is a deconfined phase of QCD, where the mean free path length of quarks and gluons becomes much larger than the typical size of hadrons, which therefore break apart into a \say{soup} of quarks and gluons. However, there are still strong interactions between the constituents of the QGP, which lead to a collective behavior akin to an almost perfect fluid \cite{PHENIX:2004vcz, BRAHMS:2004adc, STAR:2005gfr, Heinz:2013th}.
The QGP is separated from the hadron gas phase of QCD via a phase transition that is predicted to be either a crossover or first-order transition depending on the values of the thermodynamic variables of the system \cite{Fukushima:2010bq}. The search for the QCD critical point, at which the first-order transition line in the phase diagram as a function of temperature and baryon chemical potential ends, remains at the forefront of theoretical and experimental research.\par
In the canonical picture of heavy-ion collisions, the QGP is described in terms of relativistic viscous hydrodynamics \cite{Schenke:2010nt, Gale:2013da, Romatschke:2017ejr}, but hydrodynamic simulations require initial conditions.
Models like the Monte Carlo Glauber Model \cite{Miller:2007ri} and {\trento} \cite{Moreland:2014oya} take a purely geometric approach, focusing on the positions of participant nucleons in the transverse plane, but ignore the initial state dynamics.
In contrast, the IP-Glasma \cite{Schenke:2012wb, Schenke:2012hg} incorporates initial state dynamics by combining the IP-Sat model \cite{Bartels:2002cj, Kowalski:2003hm} with classical Yang-Mills dynamics and provides initial conditions that successfully describe a range of observables in heavy-ion collisions. This is facilitated by the introduction of a dynamical phase at the beginning of the collision called Glasma. A similar approach is taken by the MC-KLN (Monte Carlo-Kharzeev-Levin-Nardi) model \cite{Kharzeev:2000ph, Kharzeev:2002ei} based on the KLN saturation model \cite{Kharzeev:2000ph, Kharzeev:2001gp, Kharzeev:2001yq}, which, like the IP-Glasma, contains gluons as the relevant degrees of freedom in the colliding nuclei.
However, these models do not account for the longitudinal structure of colliding nuclei in the initial condition and rely on boost-invariance to describe observables near mid-rapidity in energetic collisions. Such a description reduces the Glasma to an effectively (2+1)D system since none of the observables depend on rapidity.\par
It has been shown that the boost-invariant Glasma, upon introducing rapidity-dependent fluctuations, becomes unstable \cite{Romatschke:2005pm, Romatschke:2006nk}. The instabilities can drive the system towards equilibrium, although on a timescale that is much larger than what is observed in experiment. 
Despite subsequent efforts \cite{Bazak:2023kol} there is yet no mechanism in the Glasma that can provide fast enough hydrodynamization.
Effective kinetic theory (EKT) \cite{Arnold:2002zm} was proposed as an intermediate stage between the Glasma and the hydrodynamic phase and can describe hydrodynamization on the desired timescale \cite{Kurkela:2015qoa}.
K{\o}MP{\o}ST \cite{Kurkela:2018wud, Kurkela:2018vqr} provides a numerical implementation of EKT to bridge the gap between the highly anisotropic Glasma and the hydrodynamic phase incorporating quantum effects, which are absent from the Glasma description.
Specifically, $2\leftrightarrow 2$ elastic scatterings and $1\leftrightarrow2$ inelastic bremsstrahlung processes drive the system towards equilibrium.\par
The hydrodynamic phase is modeled by relativistic viscous hydrodynamics implemented, e.g.,\ in VISHNU \cite{Shen:2014vra} or MUSIC \cite{Schenke:2010nt}. During this phase, the system further expands in the longitudinal and transverse directions, cooling down in the process. Eventually, the QGP will transition to the confined phase and behave as a hadron gas. This transition is called particlization and numerical treatments such as iSS \cite{Shen:2014vra} or {\scshape iS3D} \cite{McNelis:2019auj} are based on a prescription by Cooper and Frye, taking into account the conservation of energy, momentum, and quantum numbers \cite{Cooper:1974mv}.
The hadron gas phase is described by models such as Ultrarelativistic Quantum Molecular Dynamics (UrQMD) \cite{Bleicher:1999xi, Bass:1998ca} or Simulating Many Accelerated Strongly-interacting Hadrons (SMASH) \cite{Petersen:2018jag}.
The hadron gas continues to expand until all interactions cease. At that point, particles move freely until they hit the detector.\par
This picture of relativistic heavy-ion collisions heavily relies on the boost-invariant approximation. Consequently, certain effects like event plane fluctuations \cite{CMS:2015xmx, ATLAS:2017rij} cannot be studied as they require good resolution of the rapidity structure of observables.
There are extensions of the (2+1)D Glasma where JIMWLK evolution \cite{Schenke:2016ksl} introduces a rapidity dependence in the Wilson lines. The JIMWLK evolution approach was also combined with modified rapidity-dependent initial conditions on the forward lightcone \cite{McDonald:2023qwc}. The \textsc{McDipper} is an eventy-by-event generator of initial conditions based on saturation physics and $\kperp_T$-factorization \cite{Garcia-Montero:2023gex}. Furthermore, there are different approaches to simulations of the (3+1)D Glasma incorporating dynamical sources \cite{Gelfand:2016yho, Ipp:2017lho, Ipp:2018hai, Ipp:2020igo, Schlichting:2020wrv, Matsuda:2023gle, Matsuda:2024moa, Matsuda:2024mmr}.
However, these approaches suffer from one of two problems: They either do not obtain the rapidity-dependence directly from the longitudinal structure of nuclei, or they require tremendous computational effort. Only if the rapidity-dependence of observables can be directly linked to nuclear models with tunable parameters can properties of nuclei be inferred from experimental data.
However, if simulations are too computationally expensive to be able to study a large number of collision events, one cannot generate the statistics required for such an analysis.
This necessitates a new framework that incorporates a tunable nuclear model while being numerically efficient at simulating events.\par
In this thesis, we study the (3+1)D dilute Glasma \cite{Ipp:2021lwz, Ipp:2022lid, Ipp:2024ykh}, a novel semianalytic approach to the description of rapidity-dependent observables in relativistic heavy-ion collisions.
We employ the dilute approximation, which linearizes the Yang-Mills equations in the limit of weak sources.
We find analytic expressions for the components of the dilute (3+1)D Glasma field strength tensor. Such analytic expressions offer new possibilities for analytic studies of observables in the (3+1)D Glasma. Additionally, the field strength tensor can be efficiently evaluated numerically for arbitrary nuclear models.
Observables in the (3+1)D Glasma can then be computed in a fraction of the time and over a larger range of spacetime rapidity than in previous simulations of the (3+1)D Glasma.
This enables the study of the longitudinal structure of observables and their dependence on the structure of colliding nuclei in unprecedented detail.\par
The structure of this thesis is as follows. We start by reviewing some background material in Chapter~\ref{ch:background}. Specifically, we discuss the Color Glass Condensate (CGC) effective theory and how the Glasma emerges as the initial state in relativistic heavy-ion collisions. We introduce rapidity and its relation to Milne coordinates and Lorentz boosts. This allows for a brief discussion of the boost-invariant Glasma. In Chapter~\ref{ch:dilute}, we describe the dilute approximation, which yields closed-form expressions for the field strength tensor of the (3+1)D Glasma. We explore limiting fragmentation as a universal feature of the dilute Glasma, and we discuss nuclear models that allow for longitudinal structure. Since the evaluation of observables in the dilute Glasma relies on numerics, we highlight some aspects of the implementation in a computer program in Chapter~\ref{ch:implementation}. Next, we discuss numerical results for collisions of Woods-Saxon-shaped nuclei in Chapter~\ref{ch:results}. We study in detail the longitudinal and transverse structure of the energy-momentum tensor and derived observables. In Chapter~\ref{ch:pp}, we consider proton-proton collisions in the dilute Glasma and discuss the modifications required to reproduce experimental multiplicity distributions. Finally, we conclude in Chapter~\ref{ch:conclusion} and give some perspectives on future research avenues in the dilute Glasma framework.
\chapter{Background}
\label{ch:background}
\section{Heavy-ion collision experiments}
Relativistic heavy-ion collision experiments are performed at the Relativistic Heavy Ion Collider (RHIC) at the
Brookhaven National Laboratory \cite{Harrison:2002es, Harrison:2003sb} and the Large Hadron Collider (LHC) at CERN \cite{Bruning:2012zz, Roland:2014jsa}. Both facilities are synchrotron accelerators featuring two circular beam pipes in which collision partners are accelerated in opposite directions before the collision. In that case, the center-of-mass energy $\sqrt{s}$ is a convenient measure of the collision energy. It is defined as the square root of the Mandelstam variable $s=(p_1^\mu+p_2^\mu)\eta_{\mu\nu}(p_1^\nu+p_2^\nu)$, where $p_{1/2}^\mu$ are the 4-momenta of the colliding particles and $\eta_{\mu\nu}$ is the Minkowski metric. For identical particles with identical speed and opposite direction $p_{1/2}^\mu = (E,\pm p^x, \pm p^y,\pm p^z)^T$ and therefore $s=(2E)^2$. In collisions of heavy-ions the center-of-mass energy per nucleon-nucleon pair $\sqrt{s_{NN}}$ is commonly used instead of $\sqrt{s}$.\par
The two detectors and corresponding experiments currently in operation at RHIC are STAR and sPHENIX (an updated version of PHENIX). Previous experiments studying heavy-ion collisions at RHIC were PHOBOS and BRAHMS. Collisions of Au nuclei at RHIC commenced in 2000, reaching the planned maximum center-of-mass energy per nucleon-nucleon pair $\sqrt{s_{NN}}=200\,\mathrm{GeV}$ in 2001. Since then, energies as low as $\sqrt{s_{NN}}=7.7\,\mathrm{Gev}$ in collider mode and even lower in fixed target mode were realized for collisions of Au nuclei in the course of the beam energy scan (BEC) program \cite{Bzdak:2019pkr}. Lower energy collisions play an important role in the search for the QCD critical point as they probe regions of the phase diagram with smaller temperature and larger baryon chemical potential compared to higher energy collisions. RHIC is also capable of colliding O, Cu, Zr, Ru, and U nuclei \cite{Arslandok:2023utm}.\par
The main experiment concerned with heavy-ion collisions at the LHC is ALICE, but data are also taken by CMS, ATLAS, and LHCb. The first Pb-Pb collisions at the LHC were recorded in 2010 with $\sqrt{s_{NN}}=2.76\,\mathrm{TeV}$, and $\sqrt{s_{NN}}=5.02\,\mathrm{TeV}$ was reached in 2015. Furthermore, in 2017 a run of Xe-Xe collisions at $\sqrt{s_{NN}}=5.44\,\mathrm{TeV}$ was performed \cite{ALICE:2022wpn}. Finally, O-O collisions are planned at LHC in the near future \cite{Arslandok:2023utm}.
\section{Color Glass Condensate}
Atomic nuclei, made up of protons and neutrons, at relativistic speeds may be described in the Color Glass Condensate effective theory.
Protons and neutrons not only consist of three valence quarks but are complicated bound states of valence quarks, sea quarks, and gluons.
Viewed in the infinite momentum frame, the valence quarks carry a significant fraction $x$ of the longitudinal proton or neutron momentum.
However, at low momentum fractions $x$, there is a large number of sea quarks and an even larger number of gluons in protons and neutrons.
The gluons become the dominant component for scattering experiments at high collision energies as the momentum fraction of scattering particles scales like $x \propto 1/\sqrt{s}$ \cite{Lappi:2007ku}. In principle, the gluons and sea quark-antiquark pairs are short-lived fluctuations. However, due to time dilation, the lifespan of these fluctuations is greatly enhanced for a relativistic particle observed in the lab frame. Thus, the sea quarks and gluons play a critical role in collision processes for relativistic protons and nuclei.\par
The Color Glass Condensate (CGC), see \cite{Iancu:2003xm, McLerran:2008es, Gelis:2010nm, Gelis:2012ri} for lecture notes and reviews, is an effective theory for high-energy QCD and applicable to the early stages of collisions of relativistic hadrons and nuclei.
The expression Color Glass Condensate comes from the state\footnote{Color Glass Condensate can, depending on the context, refer to the specific state of high energy hadrons or nuclei, or it can refer to the effective theory in which this state is described.} of hadrons or nuclei at high energies, which is described as a dense condensate of gluons. Some key features of the CGC can be understood from the name alone. \say{Color} refers to the color charges carried by quarks and gluons. A \say{glass} is a material that (supposedly) behaves like a liquid on large timescales but can be approximated as a solid if it serves as the background for processes taking part over smaller timescales.
While this picture is not applicable to glass in the sense of a commercial product \cite{ERNSBERGER19801}, it is useful to understand the CGC: There is a background of long-lived large-$x$ (i.e.,\ hard) partons acting as quasi-static sources for the small-$x$ (i.e.,\ soft) gluons.
Compared to the lifetime of the soft gluons, the background can be assumed to be static, i.e.,\ the sources appear \say{frozen} in time.
The term was also chosen as an analogy to the theory of spin glasses, where background magnetic fields are averaged over. In a similar fashion, the sources (i.e.,\ the hard partons) in CGC are stochastic quantities that need to be averaged over \cite{Iancu:2000hn}.
Finally, \say{condensate} refers to the fact that for gluons with momenta up to the saturation momentum $Q_s$, the occupation number is $\mathcal O (1/\alpha_s)$, as large as it can get.
At this occupation number, the recombination of gluons balances splitting, and the phase space density cannot grow any larger.
This behavior is called saturation.
Furthermore, $Q_s$ replaces the QCD scale $\Lambda_\mathrm{QCD}$ in the running of the coupling constant, $\alpha_s(\Lambda_\mathrm{QCD}^2)\rightarrow \alpha_s(Q_s^2)$. Since $Q_s^2\gg \Lambda^2_\mathrm{QCD}$ for small $x$ (or, equivalently, large $\sqrt{s}$), the coupling $\alpha_s$ is assumed to be small in the CGC. However, due to the large occupation numbers, which counteract the smallness of the coupling, the system cannot be treated perturbatively.
The saturation momentum is also proportional to the transverse gluon density per unit area, and its inverse marks the scale over which gluons are correlated in the transverse direction. Formally, one studies the correlator of Wilson lines in a relativistic nucleus and its falloff behavior as a function of transverse distance to define the saturation momentum \cite{Lappi:2007ku}.\par
A hadron (or a whole nucleus) moving at relativistic speeds is described in CGC as a thin sheet of color charge (the large-$x$ partons) accompanied by the produced gauge field (the small-$x$ gluons). The description of the soft gluons as a classical field is warranted due to saturation and the large gluon occupation numbers of order $\mathcal O(\alpha_s)$. The commutator $\comm{a_k}{a^\dagger_k} = 1$ of the annihilation and creation operator at a given momentum $k$ becomes negligible compared to the number operator $N_k = a^\dagger_ka_k$. The Yang-Mills field may be split into chromoelectric fields $\vec{E}$ and chromomagnetic fields $\vec{B}$. In the case of relativistic particles, one finds that at every point $\vec{E}\perp \vec{B} \perp \vec{z}$, where $\vec{z}$ is a vector in the direction of the beam axis, as will be shown explicitly in Section~\ref{sec:two_colliding_nuclei}. This is a familiar solution from electrodynamics where it can be obtained by Lorentz boosting Coulomb fields along $\vec{z}$.\par
A given observable in CGC can be calculated by first computing the observable in terms of the stochastic sources and then averaging over sources according to some probability distribution. The separation scale $\Lambda$ between hard and soft momenta and, therefore, the split of partons into sources and gauge fields is arbitrary. The validity of the CGC picture, however, requires that longitudinal momenta of the interacting soft gluons are smaller but not much smaller than $\Lambda$. Decreasing the cutoff $\Lambda$ corresponds to integrating out some of the soft gluons and including their contributions to the sources in the process. Conversely, increasing $\Lambda$ corresponds to moving contributions from the sources to the gauge field. The dependence of the sources on the cutoff scale is given in terms of the Jalilian-Marian, Iancu, McLerran, Weigert, Leonidov, Kovner (JIMWLK) equation \cite{Jalilian-Marian:1996mkd, Jalilian-Marian:1997qno, Jalilian-Marian:1997jhx, Jalilian-Marian:1998tzv, Iancu:2000hn, Iancu:2001ad, Ferreiro:2001qy}.
\section{Glasma}
The dynamic initial state right after the collision of two relativistic nuclei is called the Glasma \cite{Kovner:1995ja, Kovner:1995ts, Krasnitz:1999wc, Krasnitz:2000gz, Lappi:2006fp}. The term goes back to Larry McLerran and suggests that the system is between a glass (the CGC) and a plasma (the QGP). The Glasma is a non-equilibrium state characterized by large gluon occupation numbers. It is described as a classical gauge field with the color currents of the nuclei acting as source terms. This is analogous to the CGC. However, the properties of the color fields in the Glasma differ from those of the CGC. 
In a sheet of Color Glass Condensate, chromoelectric and chromomagnetic fields are purely transverse. After the collision of two such sheets, longitudinal chromoelectric and chromomagnetic fields form between the sheets as they move apart. The typical length of transverse correlations of these fields is $1/Q_s$. The resulting structures are called Glasma flux tubes.
The flux tubes are believed to change only very weakly as a function of rapidity and are thought to be the source of long range rapidity correlations in the spectrum of particles detected in heavy-ion collisions \cite{Dumitru:2008wn, Lappi:2009xa}. Experiments observed strong positive correlations for particle pairs separated by a small azimuthal angle but over a wide range of rapidity gaps separating the particles. This phenomenon is called the \say{near side ridge}. 
The long range rapidity correlations can be explained consistently in the flux tube picture. The azimuthal correlations, however, are not due to the flux tube structure. Instead, the azimuthal collimation of particles is explained by radial flow. Since nuclei will, in general, not collide perfectly head-on, the overlap region of two colliding nuclei in the transverse plane will typically have an oblong \say{almond} shape (see, e.g.,\ Figure~\ref{fig:elrf_slices}). This geometry leads to different pressure gradients in different transverse directions. During the hydrodynamic phase, this difference in pressure gradients is converted to momentum anisotropy, which leads to collective flow \cite{Ollitrault:1992bk}. The experimental observation of collective flow confirms the validity of the hydrodynamic description of the QGP.\par
We explore the emergence and description of the Glasma in the color glass condensate in the following sections.

\section{Two colliding nuclei}
\label{sec:two_colliding_nuclei}
According to the Color Glass Condensate effective theory, the hard partons within atomic nuclei can be modeled as classical color charges. These charges produce a color current that sources an $\mathrm{SU}(N_c)$ gauge field\footnote{The number of colors $N_c=3$ for QCD. However, simulations for $N_c=2$ are frequently carried out as they require less computational effort and are expected to mostly agree qualitatively with QCD. For the results presented later in this work, we set $N_c=3$.}, which represents the soft degrees of freedom in the nuclei. The color current and gauge field are related via the classical Yang-Mills equations.\par
\begin{figure}[t]
    \centering
    \includegraphics{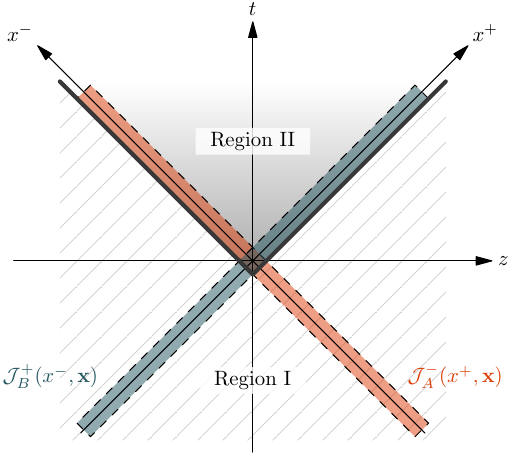}
    \caption{Minkowski diagram of a relativistic heavy-ion collision. The nuclei $A$/$B$ have finite longitudinal extent and trace out the orange/blue tracks. Their overlap is called the interaction region, which, combined with its causal future, makes up \textbf{Region II}. In \textbf{Region I}, the nuclei have not yet collided, and the single-nucleus solutions are valid. Figure adapted from \cite{Ipp:2024ykh}.}
    \label{fig:minkowski}
\end{figure}
Consider a single nucleus $A$ traveling at the speed of light in negative $z$-direction as shown in orange in Figure~\ref{fig:minkowski}. 
The source term for the classical Yang-Mills equations is given in light cone coordinates $x^\pm = (t\pm z)/\sqrt{2}$ as
\begin{align}
\label{eq:currrent_A}
    \mathcal J^\mu_A(x^+,\xperp) = \delta^{\mu}_{-} \rho_A(x^+,\xperp),
\end{align}
with the $\mathfrak{su}(N_c)$-valued color charge density $\rho_A(x^+,\xperp)$ in the lab frame.\footnote{For the limit $v\rightarrow c$ the color charge can only have infinitesimally thin support in the $x^+$-direction as the nucleus is infinitely Lorentz contracted. This is referred to as the boost-invariant limit. However, for a massive nucleus, this limit is unphysical, and therefore, it is reasonable to assume a small but finite extension in the $x^+$-direction.}
Also, $\xperp$ denotes the coordinates $x$ and $y$ perpendicular to the direction of movement (the beam axis $z$).\footnote{Note that we always set 2D vectors like $\xperp$ in boldface to set them apart from 3D vectors like $\vec{E}$, which we denote by arrows.}
As the Kronecker delta $\delta^\mu_-$ indicates, the $x^-$-component is the only nonzero vector component in the current \eqref{eq:currrent_A}.
Also, note that the current $\mathcal J^\mu_A$ does not depend on the lightcone time $x^-$. It is \say{frozen} in time as it moves along the light cone. The $x^+$-dependence then measures the longitudinal structure of the nucleus.\par
The classical Yang-Mills equations $\comm{D_\mu}{F^{\mu\nu}} =  J^\nu$ in covariant gauge $\partial_\mu  A^\mu = 0$ are (see Appendix~\ref{sec:YM_field_content} for the definition of the gauge covariant derivative and the non-Abelian field strength tensor)
\begin{align}
\label{eq:YM_eq_cov_gauge}
    \partial_\mu \partial^\mu  A^\nu - 2\ii g\comm{ A^\mu}{\partial_\mu  A^\nu} + \ii g \comm{ A_\mu}{\partial^\nu A^\mu} - g^2\comm{A_\mu}{\comm{A^\mu}{A^\nu}}=J^\nu
\end{align}
and the covariant conservation equation for the current is
\begin{align}
    \comm{D_\mu}{J^\mu} = \partial_\mu J^\mu -\ii g \comm{A_\mu}{J^\mu} = 0.
\end{align}
Consider these equations for the single nucleus $A$ with gauge field $\mathcal A^\mu_A(x^+,\xperp)$ (calligraphic symbols denote single-nucleus solutions) and its current given by \eqref{eq:currrent_A}. The conservation equation is satisfied by $\mathcal A^+_A = 0$. We can then consider \eqref{eq:YM_eq_cov_gauge} for $\nu = i$ (the two transverse directions) and we find
\begin{align}
    \partial_j \partial^j \mathcal A^i_A - 2\ii g\comm{\mathcal A^j_A}{\partial_j \mathcal A^i_A}+\ii g\comm{\mathcal A_{A,j}}{\partial^i \mathcal A^j_A} - g^2 \comm{\mathcal A_{A,j}}{\comm{\mathcal A^j_A}{\mathcal A^i_A}}
    = 0.
\end{align}
We do not allow for nontrivial solutions to this equation and demand $\mathcal A^i_A=0$.\footnote{Technically, we can get away by just demanding that $\mathcal F^{ij}_A = 0$ (in addition to $\partial_i \mathcal A^i_A=0$, which also holds due to our choice of covariant gauge). This means that $\mathcal A^i_A$ is pure gauge, but a gauge transformation that puts it to zero will keep $\mathcal A^+_A = 0$ and $\partial_- \mathcal A^-_A= 0$ unchanged, so it is compatible with covariant gauge. Therefore, we can impose $\mathcal A^i_A = 0$ as an additional gauge choice if $\mathcal F^{ij}_A = 0$.} This can be thought of as boundary conditions that forbid any color fields not initially sourced by the nuclei.
In other words, there could be a nontrivial background with which the single-nucleus solution does not interact. We choose this background to be zero. Finally, for $\nu = -$, \eqref{eq:YM_eq_cov_gauge} reads
\begin{align}
\label{eq:poisson}
    \partial_\mu\partial^\mu \mathcal A_A^-(x^+,\xperp)=-\nabla_\perp^2 \mathcal A_A^-(x^+,\xperp)=\rho_A(x^+,\xperp).
\end{align}
This result is significant. The gauge field in the presence of a single nucleus is determined by a Poisson equation in the transverse plane. Given the color charge $\rho_A$ for nucleus $A$, the corresponding gauge field can be found by solving this Poisson equation.
Since the only component of the gauge field for nucleus $A$ is $\mathcal A^-_A(x^+,\xperp)$, there are only two independent components to the corresponding field strength tensor,
\begin{align}
\mathcal F_A^{-i}(x^+,\xperp) = \partial_i \mathcal A^-_A(x^+,\xperp),
\end{align}
with $i=x,y$.
In Cartesian coordinates, this gives the contributions
\begin{align}
\label{eq:WW_F_cartesian}
    \mathcal F_A^{ti}(x^+,\xperp)=\frac{1}{\sqrt{2}}\partial_i \mathcal A^-_A(x^+,\xperp)= -\mathcal F_A^{zi}(x^+,\xperp).
\end{align}
Now it is clear that at each point, there is a chromoelectric field that points in the transverse plane (i.e.,\ it is perpendicular to the beam axis) and a chromomagnetic field that is perpendicular to the chromoelectric field and the beam axis.\par
Another nucleus that moves in positive $z$-direction can be modeled as
\begin{align}
    \mathcal J^\mu_B(x^-,\xperp)=\delta^\mu_+\rho_B(x^-,\xperp),
\end{align}
such that the overlap of the two nuclei is maximal at $t=z=0$. The Yang-Mills equations for this nucleus reduce to another Poisson equation
\begin{align}
\label{eq:poisson_B}
    -\nabla_\perp^2\mathcal A_B^+(x^-, \xperp) = \rho_B(x^-, \xperp).
\end{align}
The equations \eqref{eq:poisson} and \eqref{eq:poisson_B} are conveniently solved in momentum space, where the solution is given by
\begin{align}
    \tilde{\mathcal A}^\mp_{A/B}(x^\pm, \kperp) = \frac{1}{\kperp^2}\tilde\rho_{A/B}(x^\pm, \kperp).
\end{align}
We modify this solution by introducing an infrared (IR) cutoff $m$ and an ultraviolet (UV) cutoff $\Lambda_\mathrm{UV}$, yielding
\begin{align}
\label{eq:momentum_poisson_main}
    \tilde{\mathcal A}^\mp_{A/B}(x^\pm, \kperp) = \frac{1}{\kperp^2+m^2}\tilde\rho_{A/B}(x^\pm, \kperp)\exp\left(-\frac{\kperp^2}{2\Lambda_\mathrm{UV}^2}\right).
\end{align}
The IR cutoff $m$ regulates the divergence at $\kperp^2=0$ and serves as a screening mass in the transverse plane. It sets the size of correlated regions by suppressing any fluctuations on scales larger than $\sim 1/m$. The UV cutoff $\Lambda_\mathrm{UV}$ suppresses contributions from the largest momentum modes. Note that any numerical implementation of the Poisson equation on a discrete lattice will suppress the large momentum modes due to the finite resolution of the lattice in a way that depends on the lattice spacing. The exponential UV cutoff suppresses large momentum modes in a controllable way regardless of the lattice spacing, which makes results more comparable across different implementations. Note also that we have introduced a transverse Fourier transform defined as
\begin{align}
\label{eq:ft}
    \mathcal{A}^{\mp}_{A/B}(x^\pm, \xperp) = \intop_\kperp \tilde{ \mathcal A}^\mp_{A/B}(x^\pm, \kperp)\ee^{-\ii \kperp\cdot \xperp}
\end{align}
with $\intop_\kperp = \int \frac{\dd[2]{\kperp}}{(2\pi)^2}$.
See Appendix~\ref{ch:poisson} for a more detailed treatment of the 2D Poisson equation in the continuum and on the lattice.\par
Once we allow nuclei $A$ and $B$ to coexist and, eventually, collide, the full Yang-Mills equations
\begin{align}
\label{eq:full_YM}
    \comm{D_\mu}{F^{\mu\nu}} = J^\nu
\end{align}
and conservation equation
\begin{align}
\label{eq:full_conservation}
    \comm{D_\mu}{J^\mu} = 0
\end{align}
with
\begin{align}
    A^\mu = \mathcal A_A^\mu + \mathcal A_B^\mu + a^\mu,\label{eq:A_sum}\\
    J^\mu = \mathcal J_A^\mu + \mathcal J_B^\mu +j^\mu\label{eq:J_sum}
\end{align}
must hold. Any solution to \eqref{eq:full_YM} and \eqref{eq:full_conservation} must reduce to a superposition of the single-nucleus solutions before the interaction takes place.
The additional terms $a^\mu = a^\mu(x^+,x^-,\xperp)$ and $j^\mu=j^\mu(x^+,x^-,\xperp)$ depend on all spacetime coordinates and represent nonlinear corrections due to the non-Abelian interaction.
The term $j^\mu$ comes from rotations in color space of the currents $\mathcal J^\mu_{A/B}$ due to interactions with the color field of the other nucleus.
Therefore, $j^\mu$ is located only in the region where either $\mathcal J^\mu_A$ or $\mathcal J^\mu_B$ is nonzero, and the interaction has already taken place, which is around the boundary of the forward lightcone.
In contrast, $a^\mu$ is nonzero everywhere in the causal future of the interaction region.
Specifically, well inside the forward lightcone it is the only contribution to the total color field. In other words, it is the gauge field left behind in the wake of the colliding nuclei, and $a^\mu$ is, therefore, our Glasma field.\par
Any solution to the collision problem will depend on the sources $\rho_{A/B}$. However, the color charge corresponding to the hard partons within a nucleus in CGC is a stochastic quantity. The probability functional $W[\rho]: \rho \rightarrow \R$ assigns a probability to every color charge configuration $\rho: \R^3\rightarrow\mathfrak{su}(N_c)$ such that
\begin{align}
\label{eq:W_formal_integral}
    \int \mathrm D\rho\,W[\rho] = 1.
\end{align}
This integral, which formally runs over all possible charge configurations $\rho$, has a clear interpretation for discrete color charge densities.
If $\rho$ is represented on an $N_\perp \times N_\perp \times N_l$ lattice, then \eqref{eq:W_formal_integral} becomes an $((N_c^2-1)\times N_\perp\times N_\perp \times N_l)$-dimensional integral, one for every color component in a basis expansion of $\rho$ at each lattice site. Note that in a collision of two nuclei, the collision partners may be described by two different models $W_{A/B}[\rho_{A/B}]$.
Whenever we refer to a specific nuclear model in the context of CGC, we mean a specific choice of this probability density functional.\par
Observables like $n$-gluon inclusive spectra or the energy-momentum tensor and its higher order correlations, generically denoted here by $\mathcal O[\rho_A, \rho_B]$ can only be a functional of the color charges that act as sources for the involved nuclei.
The CGC expectation value for these observables is then given as
\begin{align}
\label{eq:CGC_observable}
    \langle \mathcal O[\rho_A, \rho_B] \rangle = \int \mathrm D\rho_A \int \mathrm D\rho_B\, \mathcal O[\rho_A, \rho_B]\,W_A[\rho_A]W_B[\rho_B].
\end{align}
The direct evaluation of \eqref{eq:CGC_observable} is not feasible in numerical computations. Instead, there are two more practical approaches. Simple observables can often be expressed in terms of a few $n$-point functions
\begin{align}
    \langle \rho(x_1)\rho(x_2)\dotsc \rho(x_n) \rangle = \int \mathrm D\rho\, \rho(x_1)\rho(x_2)\dotsc\rho(x_n)W[\rho]
\end{align}
for each nucleus. If the corresponding $n$-point functions are known, the computation of the observables is straightforward.
Alternatively, if one can find a process by which to randomly sample charge configurations that obey the desired distribution, it is possible to sample a large number of such configurations and calculate the observable for each configuration. In the context of heavy-ion or proton collisions, this is akin to simulating individual collision events.
Although a large number of events is needed to obtain the expectation value for those observables that exhibit large event-by-event fluctuations, one gains access to the fluctuations in addition to the mean values.
To simplify things further, we choose $W[\rho]$ to be a Gaussian\footnote{See Appendix~\ref{ch:gaussian} for a more detailed treatment of Gaussian probability functionals.}, which is fully determined by its 1- and 2-point functions.
All the higher $n$-point functions can then be described in terms of 1- and 2-point functions analogous to Wick's theorem.
We will, therefore, only ever consider 1- and 2-point functions of color charges when specifying a nuclear model. The first such model we will discuss is the classic McLerran-Venugopalan model in Section~\ref{sec:boost_invariance}.\par
Tying into the previous discussion about the CGC, the functional $W[\rho] = W_\Lambda[\rho]$ is strictly speaking only valid for a single value of the cutoff $\Lambda$ separating hard and soft momenta. We make no explicit mention of $\Lambda$ here and assume the nuclear models discussed in the following to be equally valid over a large range of collision energies.\par
There is no known analytic solution to the full collision problem presented in this section, i.e.,\ there is no closed-form expression for $a^\mu$ in terms of $\mathcal J^\mu_{A/B}$.
To discuss the conditions under which a solution can be found, we first introduce the important concept of rapidity and how it relates to Milne coordinates and Lorentz boosts in Sections \ref{sec:rapidity} and \ref{sec:milne}.
We then briefly touch on the boost-invariant limit in Section \ref{sec:boost_invariance}.
However, the main focus of this thesis is the treatment of the (3+1)D Glasma, which breaks boost-invariance.
Such a description becomes feasible through the dilute approximation, which we begin to discuss in Chapter~\ref{ch:dilute}.

\section{Rapidity}
\label{sec:rapidity}
Rapidity is a kinematic variable that parametrizes the longitudinal distribution of particles produced in heavy-ion collisions. There are two slightly different notions of rapidity, which have to be distinguished carefully. The simple term \say{rapidity} usually refers to
\begin{align}
    y = \frac{1}{2}\ln\left(\frac{E+p_z}{E-p_z}\right) = \frac{1}{2}\ln\left(\frac{1+v_z}{1-v_z}\right)=\artanh v_z.
\end{align}
Rapidity is purely determined by a particle's longitudinal velocity $v_z$ and does not depend on the transverse kinematics. Two useful relations for the energy $E$ and longitudinal momentum $p_z$ are
\begin{align}
E = m_\perp \cosh y,\hspace{2cm} p_z = m_\perp \sinh y,
\end{align}
where $m_\perp = \sqrt{m^2 + \pperp^2}=\sqrt{m^2 + p_x^2+p_y^2}$ is called the transverse mass.

On the other hand, the pseudorapidity
\begin{align}
    \eta = \frac{1}{2}\ln\left(\frac{|p|+p_z}{|p|-p_z}\right) = \artanh \left(\frac{p_z}{|p|}\right),
\end{align}
where $|p| = \sqrt{p_z^2+\pperp^2}$,
only depends on the ratio between longitudinal and transverse momentum.
It is fully determined by the angle between the beam axis and the particle's trajectory and is a purely geometric quantity in that sense. Also,
\begin{align}
    |p| = |\pperp|\cosh\eta,\hspace{2cm}p_z = |\pperp|\sinh\eta.
\end{align}
Note that for a highly relativistic particle ($m \ll |p|$), the two notions of rapidity coincide.
A general relation is given by
\begin{align}
    y=\frac{1}{2}\ln\left(\frac{\sqrt{\cosh^2\eta+m^2/\pperp^2}+\sinh\eta}{\sqrt{\cosh^2\eta+m^2/\pperp^2}-\sinh\eta}\right).
\end{align}
The Jacobian
\begin{align}
\label{eq:rapidity_jacobian}
    \dv{y}{\eta} = \frac{\cosh\eta}{\sqrt{\cosh^2\eta+m^2/\pperp^2}}
\end{align}
can then be used to relate distributions in rapidity to distributions in pseudorapidity. Strictly speaking, \eqref{eq:rapidity_jacobian} is only valid for a single particle species with mass $m$ and only at fixed transverse momentum $|\pperp|$. In practice, effective mass and transverse momentum parameters can be used to approximate the relation between rapidity and pseudorapidity for all charged particles in a collision \cite{Albacete:2012xq}.
\section{Milne coordinates and boosts}
\label{sec:milne}
The lab frame in heavy-ion collisions is aptly described in Minkowski coordinates. By convention, the coordinate $z$ describes the direction of the beam axis (the longitudinal direction). The transverse coordinates $x$ and $y$, collectively denoted by $\xperp$, describe the two directions perpendicular to the beam. The fields of the nuclei before the collision are conveniently described in light cone coordinates, where the Minkowski coordinates $t$ and $z$ are replaced with
\begin{align}
    x^\pm = \frac{t\pm z}{\sqrt{2}}.
\end{align}
However, in the literature on heavy-ion collisions, observables are often given in terms of Milne coordinates
\begin{alignat}{2}
    \tau &= \sqrt{2x^+x^-}=\sqrt{t^2-z^2},\\
    \eta_s &=  \frac{1}{2}\ln\left(\frac{x^+}{x^-}\right) = \artanh\left(\frac{z}{t}\right).
\end{alignat}
Here, $\tau$ is called proper time, and $\eta_s$ is the spacetime rapidity coordinate\footnote{To distinguish the spacetime rapidity $\eta_s$ from the pseudorapidity $\eta$, we use the subscript $s$. However, we do not extend that notation to tensor components, where a simple $\eta$ refers to a spacetime rapidity component, as should be clear from the context.}. Neither of these names are by accident. Consider a particle starting at the coordinate origin and traveling in the $z$-direction with constant velocity $v_z = z/t$. We can then identify
\begin{align}
    \eta_s = \frac{1}{2}\ln\left(\frac{1+v_z}{1-v_z}\right)=y.
\end{align}
In words, $\eta_s$ labels where a particle emitted at the origin and traveling with rapidity $y$ can be found at any point in time.
The complementary coordinate $\tau$ measures the proper time that has elapsed for such a particle if its transverse velocity is zero.\par
The Milne frame is a curvilinear coordinate system with line element
\begin{align}
    \dd s^2 = \dd \tau^2 - \tau^2\dd \eta_s^2 - \dd x^2 - \dd y^2.
\end{align}
It depends on a specific choice of origin and, therefore, singles out a particular spacetime point.\footnote{The usual treatment of Minkowski space (in contrast to the treatment as an affine space) also requires a choice of origin. However, in Minkowski coordinates, the components of a tensor do not depend on the choice of origin, whereas in Milne coordinates, a change of origin mixes the $\tau$- and $\eta_s$-components.}
For a collision of nuclei with infinitesimal longitudinal extent, the origin can be naturally put at the interaction point in the $t$-$z$-plane.
The significance of the Milne frame also becomes clear in such a boost-invariant setup.
For infinitesimally thin nuclei, all observables of the produced medium are independent of spacetime rapidity (see also Section~\ref{sec:boost_invariance}).
To highlight the connection between the spacetime rapidity $\eta_s$ and Lorentz boosts, we study how the coordinate vector in Milne coordinates transforms under a Lorentz boost\footnote{The Lorentz boosts in this chapter are to be understood as passive coordinate transformations. The physical system of two colliding nuclei and the medium created in the collision is always the same, and all of the discussion here concerns different descriptions of that system.} with parameter $\xi$ in longitudinal direction:
\begin{align}
    \begin{pmatrix}
        t'\\
        x'\\
        y'\\
        z'
    \end{pmatrix}
    =
    \begin{pmatrix}
        \cosh \xi & 0 & 0 & -\sinh \xi\\
        0 & 1 & 0 & 0 \\
        0 & 0 & 1 & 0 \\
        -\sinh \xi & 0 & 0 & \cosh \xi
    \end{pmatrix}
    \begin{pmatrix}
        t\\
        x\\
        y\\
        z
    \end{pmatrix}
    =
    \begin{pmatrix}
        t\cosh\xi-z\sinh\xi\\
        x\\
        y\\
        -t\sinh\xi+z\cosh\xi
    \end{pmatrix}.
\end{align}
Then,
\begin{align}
    \tau' &= \sqrt{t'^2-z'^2} = \sqrt{t^2-z^2} = \tau,\\
    \eta_s' &= \artanh\left(\frac{z'}{t'}\right) = \artanh\left(\frac{z}{t}\right) - \xi = \eta_s - \xi,
\end{align}
so a longitudinal boost amounts to nothing more than a shift in spacetime rapidity. A setup where both nuclei move with the speed of light and are infinitely Lorentz contracted is not affected by a finite boost. However, in the description of that system, a boost still shifts the spacetime rapidity. This leads to a contradiction unless none of the observables depend on spacetime rapidity.\par
If we ignore the transverse kinematics for a moment (a reasonable approximation for the center of the transverse plane), the Milne coordinates perfectly capture the symmetries of a system expanding with relativistic velocity 
\begin{align}
\label{eq:bjorken_v_mink}
    \begin{pmatrix}
        u^t\\
        u^z
    \end{pmatrix}
    =
    \frac{1}{\sqrt{t^2-z^2}}
    \begin{pmatrix}
        t\\
        z
    \end{pmatrix}.
\end{align}
Such a system is referred to as Bjorken flow \cite{Bjorken:1982qr}. The corresponding velocity in Milne coordinates takes the simple form
\begin{align}
    \begin{pmatrix}
        u^\tau\\
        u^\eta
    \end{pmatrix}
    =
    \begin{pmatrix}
        1\\
        0
    \end{pmatrix}.
\end{align}
In general, the transformation from Minkowski coordinates to Milne coordinates is
\begin{align}
\label{eq:Mink_to_Milne}
    \begin{pmatrix}
        \pdv{\tau}{t} & \pdv{\tau}{z}\\
        \pdv{\eta}{t} & \pdv{\eta}{z}
    \end{pmatrix}
    =
    \begin{pmatrix}
        \cosh\eta_s & -\sinh\eta_s\\
        -\frac{\sinh \eta_s}{\tau} & \frac{\cosh\eta_s}{\tau}
    \end{pmatrix}
\end{align}
for contravariant tensor components and the inverse of \eqref{eq:Mink_to_Milne} for covariant components. This looks suspiciously like a Lorentz boost. Indeed, the transformation \eqref{eq:Mink_to_Milne} is nothing but a Lorentz boost acting on a contravariant component along with a rescaling of all $\eta_s$-components by a factor of $1/\tau$, i.e.,
\begin{align}
    \begin{pmatrix}
        \cosh\eta_s & -\sinh\eta_s\\
        -\frac{\sinh \eta_s}{\tau} & \frac{\cosh\eta_s}{\tau}
    \end{pmatrix}
    =
    \begin{pmatrix}
        1 & 0\\
        0 & \frac{1}{\tau}
    \end{pmatrix}
    \begin{pmatrix}
        \cosh\eta_s & -\sinh\eta_s\\
        -\sinh \eta_s & \cosh\eta_s
    \end{pmatrix}.
\end{align}
Now, the second factor corresponds to a Lorentz boost with the velocity \eqref{eq:bjorken_v_mink}. Putting all of this together yields the following insight: Tensor components in the Milne frame are what an observer moving with Bjorken flow would observe in their rest frame in Minkowski coordinates (properly taking into account the additional $1/\tau$ factor). Since in idealized conditions (boost-invariance, no transverse flow), the medium produced in a heavy-ion collision undergoes Bjorken flow, Milne components can be seen as an approximation to quantities observed in the local rest frame of the medium. Most importantly, the energy-momentum tensor component $T^{\tau\tau}$ can be interpreted as an approximation of the local rest frame energy density. In a more realistic model of a heavy-ion collision, the validity of this approximation is undermined by the breaking of boost-invariance. The medium is no longer required to evolve with $u^\tau = 1$ and $u^\eta = 0$. It is produced in an extended region within the $t$-$z$-plane, a feature that is ill-described by the Milne frame, which relies on a specific choice of coordinate origin. Also, the picture of Bjorken flow does not account for transverse kinematics.

\section{The boost-invariant Glasma}
\label{sec:boost_invariance}
The relativistic nuclei introduced in Section \ref{sec:two_colliding_nuclei} move along lightlike trajectories and, due to infinite time dilation, appear frozen in time for an observer in the lab frame. The longitudinal extent of the nuclei, however, was not constrained. This will be crucial when considering the (3+1)D dilute Glasma. In this section, however, we will briefly explore the consequences of assuming an infinite Lorentz contraction factor for both colliding nuclei.\par
We modify \eqref{eq:currrent_A} by imposing a delta profile in longitudinal direction on the nucleus,
\begin{align}
    \label{eq:JA_boost_inv}
    \mathcal J^\mu_A(x^+,\xperp)=\delta^\mu_-\delta(x^+)\bar\rho_A (\xperp).
\end{align}
Here, $\bar\rho_A(\xperp)$ denotes a purely transverse (i.e.,\ 2D) color charge density.
This additional assumption about the sources is called the boost-invariant approximation or boost-invariant limit.
One can also understand this limit in terms of the Heisenberg uncertainty principle: We are interested in the large $x$ degrees of freedom in a particle moving with large $p^-$. These contributions are then highly localized in $x^+$.\par
Recall that in covariant gauge, the Yang-Mills equations simplify to a single transverse Poisson equation
\begin{align}
-\nabla_\perp^2\mathcal A^-_A(x^+,\xperp) = \delta(x^+)\bar\rho_A(\xperp).
\end{align}
To describe the collision of nuclei in the boost-invariant limit, it is convenient to perform a gauge transformation to lightcone gauge, where ${\mathcal A}^-_{A,\mathrm{LC}} = 0$.
From the general form
\begin{align}
    \mathcal A^\mu_A(x) \rightarrow {\mathcal A}^\mu_{A,\mathrm{LC}}(x) = U_A(x)\left(\mathcal A_A^\mu (x) - \frac{1}{\ii g}\partial^\mu\right)U_A^\dagger(x)
\end{align}
of a gauge transformation (see Appendix \ref{ch:YM} for more details), we see that this demand is fulfilled if
\begin{align}
    \partial^- U_A^\dagger(x)= \ii g \mathcal A^-_A(x^+,\xperp)U_A^\dagger(x).
\end{align}
The solution to this equation is given by the Wilson line
\begin{align}
    U_A^\dagger(x^+,\xperp) = \mathcal P \exp\left(\ii g \int_{-\infty}^{x^+} \dd{z^+} \mathcal A_A^-(z^+,\xperp)\right),
\end{align}
where $\mathcal P$ denotes path ordering. In the new gauge ${\mathcal A}_{A,\mathrm{LC}}^+=0$, since $\mathcal A_A^+ = 0$ and $\partial^+ U_A^\dagger(x^+,\xperp)=0$. All information about the gauge field is now in the transverse components
\begin{align}
   {\mathcal A}_{A,\mathrm{LC}}^i(x^+,\xperp) = \frac{\ii}{g} U_A(x^+,\xperp)\partial^iU_A^\dagger(x^+,\xperp).
\end{align}
Note that since $\mathcal A^-_A(x^+,\xperp)$ has only delta-like longitudinal support, $U^\dagger_A(x^+,\xperp)$ as a function of $x^+$ is exceptionally simple. It is 1 for $x^+<0$ and some function of $\xperp$ for $x^+>0$ with a discontinuity at $x^+=0$. We can, therefore, define
\begin{align}
U^\dagger_A(\xperp)\coloneqq\lim_{x^+\rightarrow \infty}U^\dagger_A(x^+,\xperp)&=\mathcal P \exp\left( \ii g \int_{-\infty}^{\infty}\dd{z^+}\mathcal A^-_A(z^+,\xperp) \right)\nn\\
    &=\exp\left(\ii g \mathcal A^-_A(\xperp)\right),
    \label{eq:U_dagger_2d}
\end{align}
where
\begin{align}
    -\nabla_\perp^2\mathcal A^-_A(\xperp) = \bar\rho_A(\xperp).
\end{align}
We may then write
\begin{align}
\label{eq:AiA_lc_gauge}
    {\mathcal A}_{A,\mathrm{LC}}^i(x^+,\xperp)=\theta(x^+)\alpha^i_A(\xperp)
\end{align}
with
\begin{align}
    \alpha_A^i(\xperp) = \frac{\ii}{g}U_A(\xperp)\partial^iU^\dagger_A(\xperp).
\end{align}
Note that the path ordering operator becomes ambiguous in the high energy limit of delta-like support. The Wilson line \eqref{eq:U_dagger_2d} is therefore not the correct high energy limit of a nucleus with longitudinal extent that is infinitely Lorentz contracted and should be viewed as a purely 2D object \cite{Fukushima:2007ki}.\par
In lightcone gauge, the only nonzero components of the gauge field ${\mathcal A}^\mu_{A,\mathrm{LC}}(x)$ are the transverse components ${\mathcal A}^i_{A,\mathrm{LC}}(x^+,\xperp)$. These components are, in general, nonzero for $x^+>0$ as opposed to the solutions in covariant gauge, which were localized at $x^+=0$. The lightcone gauge solutions, however, are pure gauge for $x^+>0$, and the only physical contribution comes from the discontinuity at $x^+=0$. Note that the gauge transformation also affects the source term \eqref{eq:JA_boost_inv}, which transforms as
\begin{align}
    \mathcal J^\mu_A(x^+,\xperp) \rightarrow {\mathcal J}^\mu_{A,\mathrm{LC}}(x^+,\xperp) = U_A(x^+,\xperp)\mathcal J^\mu_A(x^+,\xperp) U^\dagger_A(x^+,\xperp)
\end{align}
just like the field strength tensor.\par
The transformation to light cone gauge can be analogously applied to nucleus $B$ with current
\begin{align}
\label{eq:JB_boost_invariant}
    \mathcal J_B^\mu(x^-,\xperp) = \delta^\mu_+ \delta(x^-)\bar \rho_B(\xperp),
\end{align}
where the gauge condition is ${\mathcal A}_{B,\mathrm{LC}}^+=0$. The only nontrivial components of the gauge field in lightcone gauge are then
\begin{align}
\label{eq:AiB_lc_gauge}
    {\mathcal A}_{B,\mathrm{LC}}^i(x^-,\xperp)=\theta(x^-)\alpha^i_B(\xperp)
\end{align}
with
\begin{align}
    \alpha_B^i(\xperp) = \frac{\ii}{g}U_B(\xperp)\partial^iU^\dagger_B(\xperp)
\end{align}
and
\begin{align}
    U^\dagger_B(\xperp)= \exp\left( \ii g \mathcal A^+_B(\xperp) \right).
\end{align}
To understand the term boost-invariance, we apply a Lorentz boost with rapidity $\xi$ in $z$-direction to the sources $\mathcal J^\mu_A$ and $\mathcal J^\mu_B$. This will also boost the gauge field produced by these currents, and therefore, the Glasma and all of its observables will be shifted by $\xi$ in spacetime rapidity. In lightcone coordinates the corresponding boost matrix $\Lambda^\mu{}_\nu$ has components
\begin{align}
    \Lambda^+{}_+(\xi) &= \ee^{-\xi}\\
    \Lambda^-{}_-(\xi) &= \ee^{+\xi}\\
    \Lambda^+{}_-(\xi) &= \Lambda^-{}_+(\xi) = 0.
\end{align}
Therefore, under the boost $\mathcal J^\mu_{A/B}(x)\rightarrow \Lambda^\mu{}_\nu \mathcal J^\nu_{A/B}(\Lambda^{-1}x)$
\begin{align}
    \mathcal J_A^-(x^+,\xperp) &\rightarrow \ee^{+\xi} \mathcal J_A^-(\ee^{+\xi}x^+,\xperp),\\
    \mathcal J_B^+(x^-, \xperp)&\rightarrow \ee^{-\xi} \mathcal J_B^+(\ee^{-\xi}x^-,\xperp).
\end{align}
Since $\delta(\ee^{\pm\xi} x^\pm) = \ee^{\mp \xi}\delta(x^\pm)$ this boost does not change the currents and the solution to the collision problem cannot change either. This contradicts the shift in $\eta_s$ under the boost unless none of the observables of the Glasma depend on spacetime rapidity. A direct consequence of the boost-invariant limit is that observables cannot depend on spacetime rapidity, and all rapidity profiles of observables are completely flat. In heavy-ion collision experiments, this is a good approximation for the region around mid-rapidity for very energetic collisions \cite{Iancu:2003xm}.\par
We now consider the collision of two nuclei in the boost-invariant limit.
The Fock-Schwinger gauge condition
\begin{align}
    x^+ A^-(x) + x^-A^+(x) = 0
\end{align}
conveniently combines the lightcone gauges of the single nuclei along the lightcone boundary where the corresponding nucleus is located.
Expressing the gauge field in Milne coordinates, an equivalent condition is
\begin{align}
    A^\tau(x) = 0.
\end{align}
We will employ this gauge condition to treat the collision problem.
Assuming that the nuclei move without recoil, the spacetime structure of the currents cannot change in the presence of the other nucleus, and both currents are conserved individually. However, both nuclei could suffer rotations in color space when passing through the gauge field of the other nucleus, as is clear from the conservation equations
\begin{align}
    \comm{D_\mu}{\mathcal J_A^\mu(x)}&=\partial_- \mathcal J_A^-(x^+,\xperp)-\ii g \comm{A^+(x)}{\mathcal J^-_A(x^+,\xperp)}=0,\\
    \comm{D_\mu}{\mathcal J_B^\mu(x)}&=\partial_+ \mathcal J_B^+(x^-,\xperp)-\ii g \comm{A^-(x)}{\mathcal J^+_B(x^-,\xperp)}=0.
\end{align}
A color rotation can be avoided if $A^+=0$ where $\mathcal J^-_A$ is localized and $A^-=0$ where $\mathcal J^+_B$ is localized. It is often argued that this is trivially realized through the Fock-Schwinger gauge condition. However, the Fock-Schwinger gauge condition does not constrain the fields at $x^+=x^-=0$. Since the nuclei only have delta-like support, this is the only point where the nuclei come into contact and interact. A more careful treatment, see, e.g.,\ \cite{Blaizot:2008yb}, properly defines the infinitesimal width of colliding nuclei through a limit and confirms that the currents of the nuclei do not rotate in color space.
\begin{figure}
    \centering
    \includegraphics{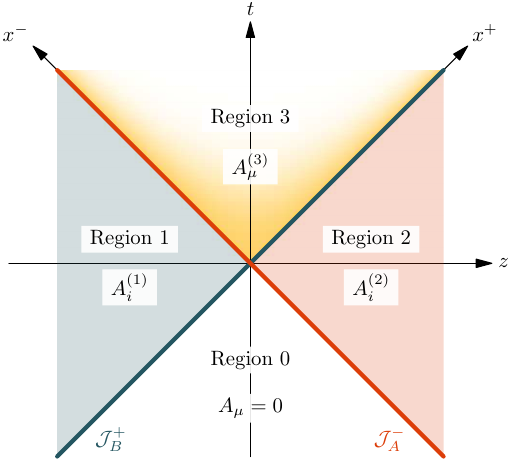}
    \caption{Minkowski diagram of two colliding nuclei in the boost-invariant limit. The tracks of the nuclei $A$ (orange) and $B$ (blue) are constrained to the boundaries of the lightcones, which divide the spacetime into four numbered regions with different solutions for the gauge field $A_\mu(x)$.}
    \label{fig:boost-invariant_minkowski}
\end{figure}
Now, since we assume that the sources do not suffer any recoil and do not color-rotate, the total gauge current is given by the sum of the single-nucleus currents
\begin{align}
    J^\mu(x) = \mathcal J^\mu_A (x^+,\xperp) + \mathcal J^\mu_B (x^-, \xperp).
\end{align}
We show the spacetime picture of a collision of two relativistic nuclei in the boost-invariant limit in the $t$-$z$-plane in Figure~\ref{fig:boost-invariant_minkowski}. The tracks of the nuclei, which coincide with the boundaries of the lightcones, divide spacetime into four distinct regions with distinct solutions for the gauge field $A_\mu(x)$. Region 0 is causally disconnected from any of the sources. Just like for the single-nucleus solutions \eqref{eq:AiA_lc_gauge} and \eqref{eq:AiB_lc_gauge} in lightcone gauge, we may put $A_\mu(x)=0$ in Region 0. Region 1 is causally disconnected from nucleus $A$, so the single-nucleus solution \eqref{eq:AiB_lc_gauge} for nucleus $B$ is valid, i.e.,\ $A_i^{(1)}(x)=\alpha_i^B(\xperp)$. Analogously, $A_i^{(2)}(x)=\alpha_i^A(\xperp)$. As argued in Section~\ref{sec:two_colliding_nuclei}, in Region 3, the full Yang-Mills equations need to be solved to obtain a solution for $A_\mu^{(3)}(x)$. We express the gauge field in all of spacetime using Heaviside functions
\begin{align}
    A^\tau(x) 
    &= 0,\\
    A^\eta(x) &= \theta(x^+)\theta(x^-)\alpha^\eta(\tau, \xperp),\\
    A^i(x) &= \theta(x^+)\theta(-x^-) \alpha^i_A(\xperp) + \theta(-x^+)\theta(x^-)\alpha^i_B(\xperp) + \theta(x^+)\theta(x^-) \alpha^i(\tau, \xperp),
\end{align}
where we used all known information about the gauge field but had to introduce two unknown functions $\alpha^\eta(\tau, \xperp)$ and $\alpha^i(\tau, \xperp)$. Inserting this gauge field into the Yang-Mills equations yields explicit expressions for the fields on the boundary of the forward lightcone \cite{Kovner:1995ja}
\begin{align}
    \alpha^\eta(\tau \rightarrow 0, \xperp) 
    &= \frac{\ii g}{2}\comm{\alpha_A^i(\xperp)}{\alpha_B^i(\xperp)},\\
    \alpha^i(\tau\rightarrow 0, \xperp) &= \alpha^i_A(\xperp) + \alpha^i_B(\xperp),
\end{align}
and the derivatives
\begin{align}
    \partial_\tau \alpha^\eta(\tau\rightarrow 0, \xperp) &=0,\\
    \partial_\tau \alpha^i(\tau \rightarrow 0, \xperp) &= 0.
\end{align}
Thus, at the boundary of the forward lightcone, the gauge fields and their conjugate momenta are completely defined. These data can be used as initial conditions in numerical simulations of the boost-invariant Glasma in the forward lightcone \cite{Krasnitz:1998ns, Lappi:2003bi}. Analytical computations have been performed as an expansion in powers of $\tau$ \cite{Fries:2006pv, Fujii:2008km, Chen:2015wia, Carrington:2020ssh}. Another approach is to linearize the Yang-Mills equations in the forward lightcone by assuming weak sources and neglecting higher-order terms in the sources. This allows for an analytic solution of the Yang-Mills equations, which can be matched to the initial conditions on the forward lightcone \cite{Kovner:1995ts}. This is conceptually similar to the approach taken in this thesis, although we will go beyond boost-invariant initial conditions.\par
So far, we have not made any remark about the structure of the colliding nuclei other than that the color charge corresponding to the hard degrees of freedom needs to be supplied via some model. One such model was devised by McLerran and Venugopalan \cite{McLerran:1993ka, McLerran:1993ni} and is correspondingly called the McLerran-Venugopalan model or MV model for short. In its simplest iteration, the MV model states that color charges should be drawn from a Gaussian probability functional (see Appendix~\ref{ch:gaussian}) fully defined by the 1- and 2-point functions
\begin{align}
\label{eq:MV_2d_simple_1}
    \langle \bar \rho^a_{A/B}(\xperp) \rangle &=0,\\
    \langle \bar\rho^a_{A/B}(\xperp) \bar\rho^b_{A/B} (\yperp)\rangle &= g^2\mu^2 \delta^{ab}\delta^{(2)}(\xperp - \yperp)
    \label{eq:MV_2d_simple_2}
\end{align}
of the 2D color charge density components $\bar\rho^a(\xperp)$.
The Latin indices refer to coefficients with respect to some basis of $\mathfrak{su}(N_c)$. The Yang-Mills coupling $g$ takes a fixed value. The phenomenological MV parameter $\mu$ has the dimension of an energy and corresponds to the average color charge of the hard partons per transverse area and per color \cite{McLerran:1993ni, Iancu:2003xm}. The expression $g^2\mu$ is proportional to the saturation momentum $Q_s$ with an $\mathcal O (1)$ proportionality constant \cite{Iancu:2003xm, Lappi:2007ku}.
Nuclei in the MV model have infinite transverse extent and are, in practice, modeled using periodic boundary conditions in the transverse plane. The model is, therefore, only valid in the center region of the fireball and cannot describe anisotropies in the transverse plane. Indeed, the delta function in \eqref{eq:MV_2d_simple_2} ensures that the model is invariant under transverse rotations and translations. In the longitudinal direction, the nuclei in the MV model can be imagined as infinitesimally thin sheets. It was later realized that the longitudinal delta functions in \eqref{eq:JA_boost_inv} and \eqref{eq:JB_boost_invariant} can be problematic \cite{Jalilian-Marian:1996mkd}. This is mitigated by the generalized MV model
\begin{align}
\label{eq:MV_2d_generalized_1}
        \langle \rho^a_{A/B}(x^\pm, \xperp) \rangle &=0,\\
    \langle \rho^a_{A/B}(x^\pm, \xperp) \rho^b_{A/B} (y^\pm, \yperp)\rangle &= g^2\mu^2(x^\pm) \delta^{ab}\delta(x^\pm - y^\pm)\delta^{(2)}(\xperp - \yperp),
    \label{eq:MV_2d_generalized_2}
\end{align}
which allows for a finite longitudinal extent and nontrivial longitudinal structure of nuclei by promoting the MV parameter to a function $\mu(x^\pm)$. We will further generalize this model in Chapter~\ref{ch:dilute} to treat collisions of nuclei with longitudinal correlations away from the boost-invariant limit.
In the limit of infinite collision energies, the generalized 
MV model \eqref{eq:MV_2d_generalized_1}--\eqref{eq:MV_2d_generalized_2} does not reduce to \eqref{eq:MV_2d_simple_1}--\eqref{eq:MV_2d_simple_2} \cite{Fukushima:2007ki}. In the picture of infinitesimal sheets of color charge, the proper high energy limit of \eqref{eq:MV_2d_generalized_1}--\eqref{eq:MV_2d_generalized_2} can be seen as an infinite number of uncorrelated color sheets collapsing on each other, described by
\begin{align}
\label{eq:MV_2d_inf_sheets_1}
    \langle \bar \rho^{n,a}_{A/B}(\xperp) \rangle &=0,\\
    \langle \bar\rho^{n,a}_{A/B}(\xperp) \bar\rho^{m,b}_{A/B} (\yperp)\rangle &= g^2\mu^2\frac{1}{N_s}\delta^{nm}\delta^{ab}\delta^{(2)}(\xperp - \yperp).
    \label{eq:MV_2d_inf_sheets_2}
\end{align}
Here, the indices $n$ and $m$ label the $N_s$ different sheets. This formulation of the high energy (i.e.,\ boost-invariant) limit is suitable for numerical simulations, where $N_s\rightarrow \infty$ is replaced by $N_s \gg 1$.
There is a separate Poisson equation for each sheet,
\begin{align}
    -\nabla_\perp^2 \mathcal A^{n,\mp}_{A/B}(\xperp) = \bar \rho^n_{A/B}(\xperp),
\end{align}
and the lightlike Wilson lines are given by
\begin{align}
    U^\dagger_{A/B}(x^\pm, \xperp)=\prod_{n=1}^{N_s}\exp\left(ig\mathcal A^{n,\mp}_{A/B}(\xperp)\right).
\end{align}
\chapter{The dilute Glasma}
\label{ch:dilute}
In this chapter, we introduce the dilute Glasma, which is a solution to the collision problem outlined in Section~\ref{sec:two_colliding_nuclei}. In particular, the dilute Glasma makes no a priori assumption about the longitudinal or transverse structure of the colliding nuclei and does not rely on boost invariance.
Instead, the dilute approximation is employed to solve the Yang-Mills equations in the forward lightcone. The term \say{dilute} refers to the assumption that the sources pertaining to the hard partons within the nuclei are weak, and terms of higher order in the sources can be neglected. This linearizes the Yang-Mills equations and allows us to derive analytic expressions for the Glasma field and the corresponding field strength tensor. 
\section{The dilute approximation}
\label{sec:dilute_approximation}
We will now return to the general collision problem away from the boost-invariant limit. A solution to \eqref{eq:full_YM} and \eqref{eq:full_conservation} can be found by employing the dilute approximation \cite{Ipp:2021lwz, Ipp:2022lid, Ipp:2024ykh}, which starts by expanding the equations \eqref{eq:full_YM} and \eqref{eq:full_conservation} in powers of $\mathcal J_A$ and $\mathcal J_B$. More formally,
\begin{align}
    a^\mu = \sum_{i,k}a^{\mu, i,k},\label{eq:a_sum}\\
    j^\mu = \sum_{i,k}j^{\mu, i,k},
\end{align}
where the sum runs over all ordered pairs of positive integers. The terms $a^{i,k}$ and $j^{i,k}$ contain all contributions of order $(\mathcal J_A)^i(\mathcal J_B)^k$. Assuming that the sources $\mathcal J_{A/B}$ are weak, we may drop higher-order contributions in $a^\mu$ and $j^\mu$ and only consider the leading order terms $a^{\mu,1,1}$ and $j^{\mu,1,1}$. Then, the conservation equation \eqref{eq:full_conservation} becomes
\begin{align}
    \comm{D_\mu}{J^\mu} &= \partial_\mu J^\mu - \ii g \comm{A_\mu}{J^\mu}\nn\\
    &= \comm{\mathcal D_{A,\mu}}{\mathcal J^\mu_A} + \comm{\mathcal D_{B,\mu}}{\mathcal J^\mu_B}\nn\\
    &\quad+ \partial_\mu j^{\mu,1,1}-\ii g\comm{\mathcal A_{A,\mu}}{\mathcal J_B^\mu}-\ii g\comm{\mathcal A_{B,\mu}}{\mathcal J_A^\mu}=0.
\end{align}
The first two terms are just the conservation equations for the single-nucleus solutions and are, therefore, 0. Thus,
\begin{align}
    \label{eq:j11}
    \partial_\mu j^{\mu,1,1}=\ii g\comm{\mathcal A_{A,\mu}}{\mathcal J_B^\mu}+\ii g\comm{\mathcal A_{B,\mu}}{\mathcal J_A^\mu}.
\end{align}
Similarly, for the Yang-Mills equations \eqref{eq:full_YM},
\begin{align}
    \comm{D_\mu}{F^{\mu\nu}}&=\comm{\mathcal D_{A,\mu}}{\mathcal F_A^{\mu\nu}}+\comm{\mathcal D_{B,\mu}}{\mathcal F_B^{\mu\nu}}\nn\\
    &\quad+\partial_\mu (\partial^\mu a^{\nu,1,1}-\partial^\nu a^{\mu,1,1}-\ii g  \comm{\mathcal A_A^\mu}{\mathcal A_B^\nu}-\ii g  \comm{\mathcal A_B^\mu}{\mathcal A_A^\nu})\nn\\
    &\quad -\ii g \comm{\mathcal A_{A,\mu}}{\partial^\mu \mathcal A_{B}^\nu - \partial^\nu \mathcal A_{B}^\mu}-\ii g \comm{\mathcal A_{B,\mu}}{\partial^\mu \mathcal A_{A}^\nu - \partial^\nu \mathcal A_{A}^\mu}\nn\\
    &=  \mathcal J_A^\nu + \mathcal J_B^\nu +j^{\nu,1,1}.
\end{align}
Considering that the Yang-Mills equations hold for the single-nucleus solutions and the covariant gauge condition $\partial_\mu A^\mu = 0$, this can be simplified to
\begin{align}
\label{eq:a11}
    \partial_\mu \partial^\mu a^{\nu,1,1}&=\ii g  \comm{\mathcal A_{A,\mu}}{2\partial^\mu\mathcal A_B^\nu- \partial^\nu \mathcal A_{B}^\mu}+\ii g  \comm{\mathcal A_{B,\mu}}{2\partial^\mu\mathcal A_A^\nu - \partial^\nu \mathcal A_{A}^\mu} +j^{1,1,\nu}.
\end{align}
To avoid clutter, we will write $j^\mu$ and $a^\mu$ instead of $j^{\mu,1,1}$ and $a^{\mu,1,1}$ in the following, but we still mean the leading order terms and never consider any terms of higher order than $\mathcal J_A\mathcal J_B$.\par
Equation \eqref{eq:j11} is a single equation that cannot fix all components of $j^\mu$.
We, therefore, make the additional assumption that the nuclei continue past the collision without recoil.
After the collision, where the currents of the two nuclei can be meaningfully separated again, they cannot have obtained additional vector components, and their localization in spacetime has to be the same as if the collision had never occurred. Therefore, there can only be a $j^-$-component, localized in the same region as $\mathcal J_A^-$, and a $j^+$-component, localized in the same region as $\mathcal J_B^+$. The spatial localization allows for two equations to be extracted from \eqref{eq:j11}, which are
\begin{align}
\partial_+j^+(x)&=\ii g \comm{\mathcal A_{A}^-(x^+,\xperp)}{\mathcal J^+_B(x^-,\xperp)},\\
    \partial_-j^-(x)&=\ii g \comm{\mathcal A_{B}^+(x^-,\xperp)}{\mathcal J^-_A(x^+,\xperp)}\label{eq:YM_j-}.
\end{align}
Note that the symbol $x$ means all coordinates $(x^+,x^-,\xperp)$ collectively.
These equations can be integrated to obtain
\begin{align}
j^+(x) &= \ii g\int_{-\infty}^{x^+}\dd{w^+}\comm{\mathcal A^-_A(w^+,\xperp)}{\mathcal J^+_B(x^-,\xperp)},\\
    j^-(x) 
    &= \ii g\int_{-\infty}^{x^-}\dd{w^-}\comm{\mathcal A^+_B(w^-,\xperp)}{\mathcal J^-_A(x^+,\xperp)},
\end{align}
Then, \eqref{eq:a11} is a wave equation for each component $a^\nu$ with source terms that depend only on the single-nucleus solutions,
\begin{align}
\label{eq:YM_ap}
    \partial_\mu\partial^\mu a^+(x) &= \ii g  \comm{\mathcal A_{A}^-(x^+,\xperp)}{\partial_-\mathcal A_B^+(x^-,\xperp)} +j^+(x),\\
    \partial_\mu\partial^\mu a^-(x)&= \ii g  \comm{\mathcal A_{B}^+(x^-,\xperp)}{\partial_+\mathcal A_A^-(x^+,\xperp)} +j^-(x),\\
    \partial_\mu\partial^\mu a^i(x)&= \ii g  \comm{\mathcal A_{A}^-(x^+,\xperp)}{ \partial_i \mathcal A_{B}^+(x^-,\xperp)}+\ii g  \comm{\mathcal A_{B}^+(x^-,\xperp)}{ \partial_i \mathcal A_{A}^-(x^+,\xperp)}.\label{eq:YM_ai}
\end{align}
Following \cite{Ipp:2021lwz}, we express these equations as
\begin{align}
\label{eq:a_wave_eq}
    \partial_\mu \partial^\mu a^\nu(x) = S^\nu(x).
\end{align}
The components of the inhomogeneity vector $S^\nu(x)$ are
\begin{align}
\label{eq:Sp}
    S^+(x) &= \ii g \comm{\mathcal A_{A}^-(x^+,\xperp)}{\partial_-\mathcal A_B^+(x^-,\xperp)}\nn\\
    &\quad+\ii g\int_{-\infty}^{x^+}\dd{w^+}\comm{\mathcal A^-_A(w^+,\xperp)}{\mathcal J_B^+(x^-,\xperp)},\\
    S^-(x) &= \ii g\comm{\mathcal A_{B}^+(x^-,\xperp)}{\partial_+\mathcal A_A^-(x^+,\xperp)} \nn\\
    &\quad+\ii g\int_{-\infty}^{x^-}\dd{w^-}\comm{\mathcal A^+_B(w^-,\xperp)}{\mathcal J^-_A(x^+,\xperp)},\\
    S^i(x)&=\ii g \comm{\mathcal A_{A}^-(x^+,\xperp)}{ \partial_i \mathcal A_{B}^+(x^-,\xperp)}+  \ii g\comm{\mathcal A_{B}^+(x^-,\xperp)}{ \partial_i \mathcal A_{A}^-(x^+,\xperp)}.
    \label{eq:Si}
\end{align}
We solve the wave equation \eqref{eq:a_wave_eq} using the method of Green's function,
\begin{align}
\label{eq:a_G}
    a^\mu(x) = \intop_y G_\mathrm{ret}(x-y)S^\mu(y)
\end{align}
with the retarded propagator
\begin{align}
\label{eq:Gret}
    G_\mathrm{ret}(z) =-\frac{1}{2\pi}\Theta(z^0)\delta(z^\mu z_\mu),
\end{align}
which ensures causality. The symbol $\intop_y$ is a shorthand for $\int \dd[4]{y}$.
We rewrite the expressions \eqref{eq:Sp}--\eqref{eq:Si} using Fourier transforms in the transverse directions. We recall that
\begin{align}
    \mathcal{A}^{\mp}_{A/B}(x^\pm, \xperp) = \intop_\pperp \tilde{ \mathcal A}^\mp_{A/B}(x^\pm, \pperp)\ee^{-\ii \pperp\cdot \xperp}
\end{align}
with $\intop_\pperp = \int \frac{\dd[2]{\pperp}}{(2\pi)^2}$. Also, $\pperp \cdot \xperp = p_i x_i = p^i x^i$.
Note that from the Poisson equations \eqref{eq:poisson} and \eqref{eq:poisson_B} in momentum space $\tilde{\mathcal J}_{A/B}^\mp(x^\pm, \pperp) = \pperp^2 \tilde{\mathcal A}_{A/B}^\mp(x^\pm, \pperp)$.\par
We write the integral \eqref{eq:a_G} for the component $a^i$, inserting \eqref{eq:Si}, \eqref{eq:Gret} and using \eqref{eq:ft},
\begin{align}
    a^i(x) &= \frac{g}{2\pi} \intop_y \intop_\pperp \intop_\qperp \Theta(x^0-y^0)\delta((x^\mu-y^\mu)(x_\mu-y_\mu))(p^i-q^i)\nn\\
    &\hspace{3cm}\times
    \comm{\tilde{\mathcal A}_{A}^-(y^+,\pperp)}{ \tilde{\mathcal A}_{B}^+(y^-,\qperp)}  \ee^{-\ii(\pperp+\qperp)\cdot\yperp}.
\end{align}
We introduce new coordinates $v^\mu = x^\mu - y^\mu$,
\begin{align}
    a^i(x) &= \frac{g}{2\pi} \intop_{v^+}\intop_{v^-}\int_0^\infty \dd{|\vperp|}|\vperp|\int_0^{2\pi} \dd{\theta_\vperp} \intop_\pperp \intop_\qperp \Theta(v^0)\delta(2v^+v^--\vperp^2)(p^i-q^i)\nonumber\\
    &\hspace{1cm}\times
    \comm{\tilde{\mathcal A}_{A}^-(x^+-v^+,\pperp)}{ \tilde{\mathcal A}_{B}^+(x^--v^-,\qperp)}  \ee^{-\ii(\pperp+\qperp)\cdot\xperp}\ee^{\ii|\pperp+\qperp||\vperp|\cos\theta_\vperp},
\end{align}
where we have used polar coordinates $(|\vperp|, \theta_\vperp)$ for the $\vperp$-plane. Without loss of generality, the angle $\theta_\vperp$ was oriented such that it is zero if $\vperp$ is parallel to $\pperp + \qperp$. Both, $|\vperp|$ and $\theta_\vperp$ can then be integrated out, yielding
\begin{align}
    a^i(x) &= \frac{g}{2} \intop_{v^+}\intop_{v^-} \intop_\pperp \intop_\qperp \Theta(v^0)\Theta(v^+v^-)(p^i-q^i)\nonumber\\
    &\hspace{0.9cm}\times
    \comm{\tilde{\mathcal A}_{A}^-(x^+-v^+,\pperp)}{ \tilde{\mathcal A}_{B}^+(x^--v^-,\qperp)}  \ee^{-\ii(\pperp+\qperp)\cdot\xperp}J_0(\sqrt{2v^+v^-}|\pperp + \qperp|)
\end{align}
with $J_0$ the $0$-th Bessel function of the first kind.
From the two Heavyside functions, it is clear that the $v^+$- and $v^-$-integrals only span the forward lightcone. We turn our attention to $a^+$, which, using \eqref{eq:Sp}, \eqref{eq:a_G}, \eqref{eq:Gret}, and \eqref{eq:ft}, we write as
\begin{align}
    a^+(x)&=-\frac{\ii g}{2\pi} \intop_y\intop_\pperp\intop_\qperp\Theta(x^0-y^0)\delta((x^\mu-y^\mu)(x_\mu-y_\mu))\ee^{-\ii (\pperp+\qperp)\cdot\yperp}\nn\\
    &\hspace{-0.2cm}\times\left(\comm{\tilde{\mathcal A}_{A}^-(y^+,\pperp)}{\partial_-^{(y)}\tilde{\mathcal A}_B^+(y^-,\qperp)}+\int_{-\infty}^{y^+}\dd{w^+}\comm{\tilde{\mathcal A}^-_A(w^+,\pperp)}{\tilde{\mathcal J}_B^+(y^-,\qperp)}\right).
\end{align}
We perform the same change of coordinates $v^\mu = x^\mu - y^\mu$ as before,
\begin{align}
    a^+(x)&=-\frac{\ii g}{2\pi} \intop_{v^+}\intop_{v^-}\int_0^\infty \dd{|\vperp|}|\vperp|\int_0^{2\pi} \dd{\theta_\vperp}\intop_\pperp\intop_\qperp\Theta(v^0)\delta(2v^+v^--\vperp^2)\nn\\
    &\quad\times\ee^{-\ii(\pperp+\qperp)\cdot\xperp}\ee^{\ii|\pperp+\qperp||\vperp|\cos\theta_\vperp}\left(-\comm{\tilde{\mathcal A}_{A}^-(x^+-v^+,\pperp)}{\partial_-^{(v)}\tilde{\mathcal A}_B^+(x^--v^-,\qperp)}\vphantom{\int_{-\infty}^{x^+-v^+}}\right.\nn\\
    &\hspace{3.5cm}\left.+\int_{-\infty}^{x^+-v^+}\dd{w^+}\comm{\tilde{\mathcal A}^-_A(w^+,\pperp)}{\tilde{\mathcal J}_B^+(x^--v^-,\qperp)}\right),
\end{align}
employing polar coordinates for the $\vperp$-plane. Carrying out the $|\vperp|$- and $\theta_\vperp$-integrals yields
\begin{align}
\label{eq:ap_int}
    a^+(x)&=-\frac{\ii g}{2} \intop_{v^+}\intop_{v^-}\intop_\pperp\intop_\qperp\Theta(v^0)\Theta(v^+v^-)\ee^{-\ii(\pperp+\qperp)\cdot\xperp}J_0(\sqrt{2v^+v^-}|\pperp+\qperp|)\nn\\
    &\hspace{0.4cm}
    \times\left(-\comm{\tilde{\mathcal A}_{A}^-(x^+-v^+,\pperp)}{\partial_-^{(v)}\tilde{\mathcal A}_B^+(x^--v^-,\qperp)}\vphantom{\int_{-\infty}^{x^+-v^+}}\right.\nn\\
    &\hspace{2cm}\left.+\int_{-\infty}^{x^+-v^+}\dd{w^+}\comm{\tilde{\mathcal A}^-_A(w^+,\pperp)}{\tilde{\mathcal J}_B^+(x^--v^-,\qperp)}\right)\nn\\
    &\eqqcolon \frac{\ii g}{2}\left(I_1(x)-I_2(x).\right)
\end{align}
We have split up the two terms contributing to $a^+(x)$ and treat them separately in the following.
Again, the Heaviside functions ensure that the $v^+$- and $v^-$-integrals only run over the forward lightcone, restricting $v^+$ and $v^-$ to nonnegative numbers. We implicitly assume these boundaries for the integrals in the following, dropping the Heaviside functions. The first term can then be simplified by partial integration in $v^-$. The resulting boundary terms do not contribute as we assume the gauge fields of the nuclei to be nonzero only in a track of finite width. Therefore, both $v^-\rightarrow \infty$ and $v^-=0$ lead to evaluating $\tilde{\mathcal A}_B^+(x^--v^-,\qperp)$ outside this track as long as $x^-$ is sufficiently far inside the future lightcone in the $t$-$z$-plane such that it lies outside of the track of nucleus $B$. Then,
\begin{align}
    I_1(x)&= \intop_{v^+}\intop_{v^-}\intop_\pperp\intop_\qperp\ee^{-\ii(\pperp+\qperp)\cdot\xperp}J_0(\sqrt{2v^+v^-}|\pperp+\qperp|)\nn\\
    &\hspace{3cm}\times\comm{\tilde{\mathcal A}_{A}^-(x^+-v^+,\pperp)}{\partial_-^{(v)}\tilde{\mathcal A}_B^+(x^--v^-,\qperp)}\vphantom{\int_{-\infty}^{x^+-v^+}}\nn\\
    &=\intop_{v^+}\intop_{v^-}\intop_\pperp\intop_\qperp\ee^{-\ii(\pperp+\qperp)\cdot\xperp}\frac{|\pperp+\qperp|v^+}{\sqrt{2v^+v^-}}J_1(\sqrt{2v^+v^-}|\pperp+\qperp|)\nn\\
    &\hspace{3cm}\times\comm{\tilde{\mathcal A}_{A}^-(x^+-v^+,\pperp)}{\tilde{\mathcal A}_B^+(x^--v^-,\qperp)}\vphantom{\int_{-\infty}^{x^+-v^+}}.
\end{align}
The second term in \eqref{eq:ap_int} is simplified by changing the order of the $v^+$- and $w^+$-integrals
\begin{align}
\int_{0}^{\infty}\dd{v^+}\int_{-\infty}^{x^+-v^+}\dd{w^+} \rightarrow \int_{-\infty}^{x^+}\dd{w^+}\int_{0}^{x^+-w^+}\dd{v^+}
\end{align}
which leaves the integration domain unchanged. Employing
\begin{align}
    &\hphantom{=}\,\int_0^{x^+-w^+}\dd{v^+} J_0(\sqrt{2v^+v^-}|\pperp+\qperp|) \nn\\
    &= \frac{\sqrt{2(x^+-w^+)v^-}}{v^-|\pperp+\qperp|}J_1(\sqrt{2(x^+-w^+)v^-}|\pperp+\qperp|)
\end{align}
yields
\begin{align}
    I_2(x)&= \intop_{v^+}\intop_{v^-}\intop_\pperp\intop_\qperp\ee^{-\ii(\pperp+\qperp)\cdot\xperp}J_0(\sqrt{2v^+v^-}|\pperp+\qperp|)\nn\\
    &\hspace{3cm}\times\int_{-\infty}^{x^+-v^+}\dd{w^+}\comm{\tilde{\mathcal A}^-_A(w^+,\pperp)}{\tilde{\mathcal J}_B^+(x^--v^-,\qperp)}\nn\\
    &=\int_{-\infty}^{x^+}\dd{w^+}\intop_{v^-}\intop_\pperp\intop_\qperp\ee^{-\ii(\pperp+\qperp)\cdot\xperp}
    \frac{\sqrt{2(x^+-w^+)v^-}}{v^-|\pperp+\qperp|}\nn\\
    &\hspace{1cm}\times J_1(\sqrt{2(x^+-w^+)v^-}|\pperp+\qperp|)
    \comm{\tilde{\mathcal A}^-_A(w^+,\pperp)}{\tilde{\mathcal J}_B^+(x^--v^-,\qperp)}\nn\\
    &=\intop_{v^+}\intop_{v^-}\intop_\pperp\intop_\qperp\ee^{-\ii(\pperp+\qperp)\cdot\xperp}
    \frac{\sqrt{2v^+v^-}}{v^-|\pperp+\qperp|}\nn\\
    &\hspace{1cm}\times J_1(\sqrt{2v^+v^-}|\pperp+\qperp|)
    \comm{\tilde{\mathcal A}^-_A(w^+,\pperp)}{\tilde{\mathcal J}_B^+(x^--v^-,\qperp)}.
\end{align}
The redefinition $w^+\rightarrow x^+-v^+$ restores the familiar integration over nonnegative $v^+$ in the last line. Combining $I_1(x)$ and $I_2(x)$ finally yields
\begin{align}
    a^+(x)&=\frac{\ii g}{2}\intop_{v^+}\intop_{v^-}\intop_\pperp\intop_\qperp\ee^{-\ii(\pperp+\qperp)\cdot\xperp}
    \frac{ v^+(|\pperp+\qperp|^2-2\qperp^2)}{\tau'|\pperp+\qperp|}J_1(\tau'|\pperp+\qperp|)\nn\\
    &\hspace{3cm}\times
    \comm{\tilde{\mathcal A}_{A}^-(x^+-v^+,\pperp)}{\tilde{\mathcal A}_B^+(x^--v^-,\qperp)}\vphantom{\int_{-\infty}^{x^+-v^+}},
\end{align}
where we have used the suggestive notation $\tau'=\sqrt{2v^+v^-}$. Although the variables $v^+$ and $v^-$ are just integration variables, they parametrize a lightcone, for which the use of \say{Milne-like} coordinates $\tau'$ and $\eta'=\ln \left(v^+/v^-\right)/2$ is convenient. The component $a^-(x)$ can be found in an analogous fashion or by a simple symmetry argument. We employ the latter and obtain $a^-(x)$ from $a^+(x)$ by exchanging the gauge fields in the commutator (yielding a minus sign) and exchanging $v^+ \leftrightarrow v^-$ and $\pperp \leftrightarrow \qperp$.

\section{Dilute Glasma field strength tensor}
\label{sec:dilute_glasma_f}
The solutions to \eqref{eq:YM_ap}--\eqref{eq:YM_ai} are
\begin{align}
    \label{eq:ap} a^+(x)&=\frac{\ii g}{2}\intop_{v^+}\intop_{v^-}\intop_\pperp\intop_\qperp\ee^{-\ii(\pperp+\qperp)\cdot\xperp}
    \frac{ v^+(|\pperp+\qperp|^2-2\qperp^2)}{\tau'|\pperp+\qperp|}J_1(\tau'|\pperp+\qperp|)\nn\\
    &\hspace{3cm}\times
    \comm{\tilde{\mathcal A}_{A}^-(x^+-v^+,\pperp)}{\tilde{\mathcal A}_B^+(x^--v^-,\qperp)}\vphantom{\int_{-\infty}^{x^+-v^+}},\\
    \label{eq:am} a^-(x) &= \frac{\ii g}{2}\intop_{v^+}\intop_{v^-} \intop_{\pperp}\intop_\qperp
    \ee^{-\ii(\pperp+ \qperp)\cdot \xperp}
    \frac{v^-(2 \pperp^2-|\pperp + \qperp|^2)}{\tau' |\pperp+\qperp|}J_1(  \tau'|\pperp + \qperp |)\nn\\
    &\hspace{3cm}\times
    \comm{\tilde {\mathcal A}^{-}_A( x^+ - v^+,\pperp)}{ \tilde  {\mathcal A}^{+}_B(x^- - v^-, \qperp)},\\
    \label{eq:ai} a^i(x) &= \frac{g}{2} \intop_{v^+}\intop_{v^-} \intop_{\pperp}\intop_\qperp \ee^{-\ii(\pperp+ \qperp)\cdot \xperp}(p^i- q^i)J_0( \tau'|\pperp + \qperp | )\nn\\
    &\hspace{3cm}\times  \comm{\tilde {\mathcal A}^{-}_A( x^+\! - v^+\!,\pperp)}{ \tilde {\mathcal A}^{+}_B(x^-\! - v^-\!, \qperp)}.
\end{align}
We reiterate that the solutions \eqref{eq:ap}--\eqref{eq:ai} for the dilute Glasma field are strictly speaking only valid inside the future lightcone, away from the tracks of the nuclei. If the color charges of the nuclei do not have compact support, the tracks, as shown in Figure~\ref{fig:minkowski}, strictly speaking, have infinite extent. However, we assume that the fields associated with the individual nuclei fall off after some distance and that compact support is a reasonable approximation.\par
To make progress toward gauge-invariant observables, an intermediate step is computing the field strength tensor of the dilute Glasma. Note that the commutator term usually present in a non-Abelian field strength tensor is, for the dilute Glasma field, of higher order in the sources, which was already neglected in the gauge field. We, therefore, define the dilute Glasma field strength tensor
\begin{align}
    f^{\mu\nu}(x) = \partial^\mu a^\nu(x) - \partial^\nu a^\mu(x)
\end{align}
as the Abelian part of the Yang-Mills field strength tensor.
We first consider $\partial^+ a^-(x)$, the first term contributing to $f^{+-}(x)$, and compute
\begin{align}
    \partial^+a^-(x) &=  \frac{\ii g}{2}\intop_{v^+}\intop_{v^-} \intop_{\pperp}\intop_\qperp
    \ee^{-\ii(\pperp+ \qperp)\cdot \xperp}
    \frac{v^-(2 \pperp^2-|\pperp + \qperp|^2)}{\tau' |\pperp+\qperp|}
    J_1(  \tau'|\pperp + \qperp |)\nn\\
    &\hspace{3cm}\times
    \comm{\tilde {\mathcal A}^{-}_A( x^+ - v^+,\pperp)}{ \partial_-^{(x)}\tilde  {\mathcal A}^{+}_B(x^- - v^-, \qperp)}
\end{align}
We rewrite $\partial_-^{(x)}$ as $-\partial_-^{(v)}$ and integrate by parts to obtain
\begin{align}
\label{eq:fpm_cont_1}
    \partial^+ a^-(x)&=\frac{\ii g}{4} \intop_{v^+}\intop_{v^-} \intop_{\pperp}\intop_\qperp\ee^{-\ii(\pperp+ \qperp)\cdot \xperp} (2 \pperp^2-|\pperp + \qperp|^2) J_0( \tau'|\pperp + \qperp | ) 
    \nn \\
    &\hspace{3cm}\times
    \comm{\tilde{\mathcal A}^-_A( x^+ - v^+,\pperp)}{\tilde {\mathcal A}^+_B(x^- - v^-, \qperp)}.
\end{align}
Again, the boundary terms do not contribute as they lead to the gauge field being evaluated outside of the track of the nucleus. Analogously, we obtain
\begin{align}
\label{eq:fpm_cont_2}
    -\partial^- a^+(x)&= \frac{\ii g}{4}\intop_{v^+}\intop_{v^-} \intop_{\pperp}\intop_\qperp \ee^{-\ii(\pperp+ \qperp)\cdot \xperp}  (  2 \qperp^2 - |\pperp + \qperp|^2)J_0( \tau'|\pperp + \qperp | ) 
    \nn\\
    &\hspace{3cm}\times
    \comm{\tilde{\mathcal A} ^-_A( x^+\! - v^+,\pperp)}{ \tilde{\mathcal A}^+_B(x^-\! - v^-, \qperp)}.
\end{align}
Combining \eqref{eq:fpm_cont_1} and \eqref{eq:fpm_cont_2} yields
\begin{align}
    f^{+-}(x)&= -\ii g \intop_{v^+}\intop_{v^-} \intop_{\pperp}\intop_\qperp
    \ee^{-\ii(\pperp+ \qperp)\cdot \xperp}
    \pperp \cdot \qperp
    J_0( \tau'|\pperp + \qperp | )\nn\\
    &\hspace{3cm}\times
 \comm{ \tilde{\mathcal A}^-_A( x^+ - v^+,\pperp)}{\tilde {\mathcal A}^+_B(x^- - v^-, \qperp)}.
\end{align}
Now we turn to $f^{+i}(x)$ and through rewriting $\partial_-^{(x)}\rightarrow -\partial_-^{(v)}$ and partial integration we obtain
\begin{align}
\label{eq:fpi_cont_1}
    \partial^+ a^i(x)&=-\frac{g}{2} \intop_{v^+}\intop_{v^-} \intop_{\pperp}\intop_\qperp
    \ee^{-\ii(\pperp+ \qperp)\cdot \xperp}
    (p^i- q^i)\frac{v^+|\pperp+ \qperp|}{\tau'}J_1( \tau'|\pperp + \qperp | )\nn\\
    &\hspace{3cm}\times
    \comm{\tilde {\mathcal A}^-_A( x^+\! - v^+,\pperp)}{\tilde{\mathcal A} ^+_B(x^-\! - v^-, \qperp)} .
\end{align}
Also, $\partial^i a^+ = -\partial_i a^+$ is straightforwardly evaluated to
\begin{align}
\label{eq:fpi_cont_2}
    \partial^i a^+(x) &= -\frac{g}{2}  \intop_{v^+}\intop_{v^-} \intop_{\pperp}\intop_\qperp
    \ee^{-\ii(\pperp+ \qperp)\cdot \xperp}
    (p^i +q^i)
    \frac{v^+(  |\pperp + \qperp|^2 - 2 \qperp^2 )}
    {\tau'|\pperp + \qperp | }
    J_1(\tau' |\pperp + \qperp | )
    \nn\\
    &\hspace{3cm}\times
    \comm{\tilde{\mathcal A}^-_A( x^+\!  - v^+, \pperp)}{ \tilde{\mathcal A}^+_B(x^-\! - v^- , \qperp)}.
\end{align}
Putting \eqref{eq:fpi_cont_1} and \eqref{eq:fpi_cont_2} together yields
\begin{align}
    f^{+i}(x) &=  g \intop_{v^+}\intop_{v^-} \intop_{\pperp}\intop_\qperp
    \ee^{-\ii(\pperp+ \qperp)\cdot \xperp}
    \frac{v^+\left(q^i(\pperp^2+2\pperp\cdot\qperp)-p^i\qperp^2\right)}{\tau'|\pperp + \qperp | }J_1(\tau' |\pperp + \qperp | )
    \nn\\
    &\hspace{3cm}\times
    \comm{\tilde {\mathcal A}^-_A( x^+ - v^+, \pperp)}{\tilde{\mathcal A} ^+_B(x^-\! - v^- , \qperp)}
    \label{eq:fpi}
\end{align}
Analogously,
\begin{align}
    f^{-i}(x) &=  - g\intop_{v^+}\intop_{v^-} \intop_{\pperp}\intop_\qperp
    \ee^{-\ii(\pperp+ \qperp)\cdot \xperp}
    \frac{v^-\left(p^i(\qperp^2+2\pperp\cdot\qperp)-q^i\pperp^2 \right)}{\tau'|\pperp + \qperp | }J_1( \tau'|\pperp + \qperp | )
    \nn\\
    &\hspace{3cm}\times
    \comm{ \tilde{\mathcal A}^-_A( x^+-  v^+, \pperp)}{\tilde{\mathcal A}^+_B(x^-\! - v^- , \qperp)}. 
\end{align}
For the components $f^{ij}$ we find by straightforward differentiation
\begin{align}
    f^{ij}(x) &= \ii g \intop_{v^+}\intop_{v^-} \intop_{\pperp}\intop_\qperp
    \ee^{-\ii(\pperp+ \qperp)\cdot \xperp}(q^i p^j - q^j p^i)J_0( \tau'|\pperp + \qperp | ) \nn\\
    &\hspace{3cm}\times\comm{\tilde{\mathcal A} ^-_A( x^+ - v^+,\pperp)}{
    \tilde{\mathcal A} ^+_B(x^- - v^-, \qperp)}.
\end{align}
All components of the dilute Glasma field strength tensor have thus been expressed as six-dimensional integrals. However, there are further simplifications, which reduce the number of integrals from six to three. We start by introducing
\begin{align}
\label{eq:beta_tilde_A}
    \tilde \beta^{i}_{A}(x^+-v^+,\pperp) &= \ii p^i\tilde {\mathcal A}^-_A(x^+-v^+,\pperp),\\
    \tilde \beta^{i}_{B}(x^--v^-,\qperp) &= \ii q^i\tilde{\mathcal A} ^{+}_{B}(x^--v^-,\qperp),
    \label{eq:beta_tilde_B}
\end{align}
which are just the Fourier transforms of
\begin{align}
\beta^{i}_A(x^+-v^+,\uperp) &= \partial^i_{(u)} \mathcal A_A^-(x^+-v^+,\uperp),\\
\beta^{i}_B(x^--v^-,\sperp) &= \partial^i_{(s)}\mathcal A^+_B(x^--v^-,\sperp).
\end{align}
Note that these are just the single nucleus field strength tensors prior to the collision.
Using the expressions \eqref{eq:beta_tilde_A}--\eqref{eq:beta_tilde_B} we write
\begin{align}
    f^{+-}(x) &= \ii g \intop_{v^+}\intop_{v^-} \intop_\pperp\intop_\qperp
    \ee^{-\ii (\pperp+\qperp) \cdot \xperp}
     J_0(\tau' |\pperp + \qperp | )\nn\\
     &\hspace{3cm}\times
     \comm{\tilde \beta^{i}_A(x^+ - v^+,\pperp)}{\tilde \beta^{i}_B(x^- - v^-, \qperp)}.
\end{align}
We transform back to coordinate space
\begin{align}
\label{eq:fpm_before}
        f^{+-}(x) &= \ii g  \intop_{v^+}\intop_{v^-} \intop_\pperp\intop_\qperp \intop_\uperp\intop_\sperp
        \ee^{-\ii (\pperp+\qperp) \cdot \xperp}
        \ee^{\ii (\pperp \cdot \uperp + \qperp \cdot \sperp)}
        J_0( \tau'|\pperp + \qperp | ) \nn\\
        &\hspace{3cm}\times\comm{\beta^{i}_A(x^+ - v^+, \uperp)}{\beta^{i}_B(x^- - v^-, \sperp)}.
\end{align}
The exponent in the second exponential can be rewritten as
\begin{align}
    \pperp \cdot \uperp + \qperp \cdot \sperp = \frac{1}{2}(\pperp + \qperp) \cdot (\uperp + \sperp) + \frac{1}{2} (\pperp - \qperp) \cdot (\uperp - \sperp),
\end{align}
which allows us to introduce the coordinates $\kperp = \pperp + \qperp$ and $\Delta \kperp = \frac{1}{2}(\pperp - \qperp)$. It is then straightforward to integrate out $\Delta \kperp$, yielding a delta function that puts $\sperp = \uperp$. We integrate out $\sperp$ too and obtain
\begin{align}
        f^{+-}(x) &= \ii g \intop_{\kperp} \intop_{v^+}\intop_{v^-} \intop_{\uperp}
        e^{-\ii \kperp \cdot (\xperp - \uperp)}
        J_0(\tau'|\kperp| ) \comm{\beta^{i}_A(x^+ - v^+, \uperp)}{\beta^{i}_B(x^- - v^-, \uperp)}.
\end{align}
Writing $\kperp \cdot (\xperp - \uperp) = k |\xperp-\uperp| \cos \theta$ with $k\coloneqq |\kperp|$, we may integrate out $\theta$, yielding another Bessel function
\begin{align}
\label{eq:fpm_after}
        f^{+-}(x) &= \ii g\int_{0}^{\infty} \frac{\dd{k} k}{2\pi} \intop_{v^+}\intop_{v^-} \intop_{\uperp}   J_0( \tau'k) J_0(k|\xperp-\uperp|)\nn\\
        &\hspace{3cm}\times\comm{ \beta^{i}_A(x^+ - v^+, \uperp)}{\beta^{i}_B(x^- - v^-, \uperp)}.
\end{align}
We can now make use of the closure relation
\begin{align}
\label{eq:bessel_closure}
\int_0^{\infty} \dd{k}kJ_\nu(ka)J_\nu(kb)=\frac{\delta(a-b)}{a},
\end{align}
which holds for all values of $\nu$ (no sum implied). Applying \eqref{eq:bessel_closure}, we obtain
\begin{align}
        f^{+-}(x) &=  \frac{\ii g}{2\pi} \intop_{v^+}\intop_{v^-} \intop_{\uperp}    \frac{\delta(\tau' - |\xperp-\uperp|)}{\tau'}
        \comm{\beta^{i}_A(x^+ - v^+, \uperp)}{\beta^{i}_B(x^- - v^-, \uperp)}. 
    \label{eq:fpm_intermediate}
\end{align}
We make the change of integration variables
\begin{align}
    \int_0^{\infty} \dd{v^+}\int_0^{\infty}\dd{v^-} = \int_{-\infty}^{\infty}\dd{\eta'}\int_0^{\infty}\dd{\tau'}\tau'
\end{align}
where $\eta' = \ln(v^+/v^-)/2$ and $\tau' = \sqrt{2v^+v^-}$ as introduced before. Now the delta function in Eq.~\eqref{eq:fpm_intermediate} can be removed by integrating over $\tau'$, yielding
\begin{align}
        f^{+-}(x) &=  \frac{\ii g}{2\pi} \int_{-\infty}^{\infty} \dd{\eta'} \intop_{\uperp}  \comm{\beta^{i}_A(x^+\! - \frac{|\xperp- \uperp|}{\sqrt{2}} \ee^{+\eta'}, \uperp)}{\beta^{i}_B(x^-\! - \frac{|\xperp-\uperp|}{\sqrt{2}} \ee^{-\eta'}, \uperp)}.
\end{align}
Finally, it is convenient to make a shift in the integration variable $\uperp= \xperp - \vperp$, which gives
\begin{align}
        f^{+-}(x)&=   \frac{\ii g}{2\pi}\int_{-\infty}^{\infty} \dd{\eta'} \intop_{\vperp}  \comm{\beta^{i}_A(x^+\! - \frac{|\vperp|}{\sqrt{2}} \ee^{+\eta'}, \xperp - \vperp)}{\beta^{i}_B(x^-\! - \frac{|\vperp|}{\sqrt{2}} \ee^{-\eta'}, \xperp - \vperp)}.
\end{align}
To simplify $f^{+i}(x)$, we rewrite the terms in Eq.~\eqref{eq:fpi} as
\begin{align}
    &\comm{\tilde {\mathcal A}^-_A(x^+-v^+,\pperp)}{ \tilde {\mathcal A}^+_B(x^--v^-,\qperp)} p^i \qperp^2\nn\\
    &\hspace{3cm}= 
    \ii \comm{\tilde \beta^{i}_A(x^+-v^+,\pperp)}{ \widetilde{ \partial^j\beta^{j}_B}(x^--v^-,\qperp)}, \\
    &-\comm{\tilde {\mathcal A}^-_A(x^+-v^+, \pperp)}{ \tilde {\mathcal A}^+_B(x^--v^-, \qperp)} q^i \pperp^2\nn\\
    &\hspace{3cm}=
    - \ii \comm{\widetilde{\partial^j \beta^{j}_A}(x^+-v^+,\pperp)}{\tilde \beta^{i}_B(x^--v^-,\qperp)}, \\    
    &-2\comm{\tilde {\mathcal A}^-_A(x^+-v^+,\pperp)}{ \tilde {\mathcal A}^+_B(x^--v^-,\qperp)}  q^i \pperp \cdot \qperp\nn\\
    &\hspace{3cm}=
    - 2 \ii \comm{\tilde \beta^{j}_A(x^+-v^+,\pperp)}{ \widetilde{\partial^i \beta^{j}_B}(x^--v^-,\qperp)},
\end{align}
where $\widetilde{(\dots)}$ denotes a Fourier transformation in the transverse plane.
Defining
\begin{align}
    \tilde G^{i}(x^+\! - v^+, x^-\!- v^-, \pperp, \qperp) &\coloneqq  \comm{\tilde\beta^{i}_A(x^+-v^+,\pperp)}{ \widetilde{\partial^j \beta^{j}_B}(x^--v^-,\qperp)}\nn\\
    &\quad- \comm{\widetilde{\partial^j\beta^{j}_A }(x^+-v^+,\pperp)}{\tilde \beta^{i}_B(x^--v^-,\qperp)}\nn\\&\quad - 2 \comm{\tilde \beta^{j}_A(x^+-v^+,\pperp)}{ \widetilde{\partial^i \beta^{j}_B}(x^--v^-,\qperp)},
\end{align}
we can write $f^{+i}(x)$ as
\begin{align}
    f^{+i}(x)&= -\ii g  \intop_{\pperp} \intop_\qperp \intop_{v^+}\intop_{v^-}
    \ee^{-\ii(\pperp + \qperp)\cdot \xperp}
    \frac{1}{|\pperp+\qperp|} \frac{v^+}{\tau'} J_1(\tau'|\pperp+\qperp| )\nn\\
    &\hspace{3cm}\times\tilde G^{i}(x^+- v^+, x^-- v^-, \pperp, \qperp) ,
\end{align}
Performing a Fourier transformation in the coordinates $\pperp$ and $\qperp$ yields
\begin{align}
    f^{+i}(x) &= -\ii g \intop_{\pperp}\intop_\qperp\intop_{v^+}\intop_{v^-} \intop_{\uperp}\intop_{\sperp}
    \ee^{\ii(\pperp \cdot \uperp + \qperp \cdot \sperp)} \ee^{-\ii(\pperp + \qperp)\cdot \xperp}
    \frac{1}{|\pperp+\qperp|} \frac{v^+}{\tau'} J_1( \tau'|\pperp+\qperp|) \nn\\
    &\hspace{4cm}\times G^{i}(x^+\!- v^+, x^-\!- v^-, \uperp, \sperp)
\end{align}
with
\begin{align}
\label{eq:G_pos_space}
    G^{i}(x^+\!- v^+, x^-\!- v^-, \uperp, \sperp) &=  \comm{\beta^{i}_A(x^+-v^+,\uperp)}{  \partial^j\beta^{j}_B(x^--v^-,\sperp)}\nn\\
    &\quad-\comm{\partial^j \beta^{j}_A(x^+-v^+,\uperp)}{  \beta^{i}_B(x^--v^-,\sperp)}\nn\\&\quad - 2  \comm{\beta^{j}_A(x^+-v^+,\uperp)}{ \partial^i \beta^{j}_B(x^--v^-,\sperp)}.
\end{align}
We now repeat the steps that took us from \eqref{eq:fpm_before} to \eqref{eq:fpm_after}. We write $\kperp = \pperp + \qperp$ and $\Delta \kperp = \frac{1}{2}(\pperp - \qperp)$, integrate out $\Delta \kperp$ and $\sperp$, then introduce polar coordinates and integrate out the angle $\theta$ between $\kperp$ and $\xperp - \uperp$ yielding another $J_0$,
\begin{align}
    f^{+i}(x) &= - \frac{\ii g}{2\pi} \int_0^{\infty} \dd{k}\intop_{v^+}\intop_{v^-} \intop_{\uperp} \frac{v^+}{\tau'}J_0(k|\xperp-\uperp|) J_1( \tau'k)\nn\\
    &\hspace{4cm}\times G^{i}(x^+- v^+, x^-- v^-, \uperp, \uperp).
\end{align}
Note that $G^{i}(x^+- v^+, x^-- v^-, \uperp, \uperp)$ is a total derivative, i.e.,
\begin{align}
    G^{i}(x^+- v^+, x^-- v^-, \uperp, \uperp) &= \partial^j\left( \comm{\beta^{i}_A(x^+-v^+,\uperp)}{\beta^{j}_B(x^--v^-,\uperp)}\right.\nn\\
    &\quad-\comm{\beta^{j}_A(x^+-v^+,\uperp)}{\beta^{i}_B(x^--v^-,\uperp)}\nn\\
    &\quad\left.-\delta^{ij}\comm{\vphantom{\beta^j_A}\beta^{k}_A(x^+-v^+,\uperp)}{\beta^{k}_B(x^--v^-,\uperp)}\right)\nn\\
    &\eqqcolon \partial^jG^{ij}(x^+-v^+,x^--v^-,\uperp).
\end{align}
Under the reasonable assumption\footnote{We assume here that the color fields of the nuclei, or, more precisely, their derivatives, go to zero as we approach infinity in transverse space. Even for nuclear models without compact support, this is usually fulfilled. However, this assumption is not compatible with periodic boundary conditions in the transverse plane.} that $G^{ij}\rightarrow 0$ as $|\uperp|\rightarrow \infty$ we can  partially integrate in $\uperp$, which yields
\begin{align}
f^{+i}(x) &= -\frac{\ii g}{2\pi}  \int_0^{\infty} \dd{k}\intop_{v^+}\intop_{v^-} \intop_{\uperp}  \frac{v^+}{\tau'}k\,w^jJ_1(k|\xperp-\uperp|) J_1( \tau'k)\nn\\
&\hspace{4cm}\times G^{ij}(x^+- v^+, x^-- v^-, \uperp).
\end{align}
The symbol $w^j$ denotes the unit vector
\begin{align}
    w^j = \partial^j_{(u)} |\xperp- \uperp| =  \frac{x^j- u^j}{|\xperp-\uperp|}.
\end{align}
Using again the closure relation \eqref{eq:bessel_closure} for the Bessel functions gives
\begin{align}
\label{eq:fpi_intermediate}
f^{+i}(x) &= -\frac{\ii g}{2\pi} \intop_{v^+}\intop_{v^-} \intop_{\uperp}   \frac{v^+}{\tau'}w^j\frac{\delta(\tau'-|\xperp-\uperp|)}{\tau'}G^{ij}(x^+- v^+, x^-- v^-, \uperp).
\end{align}
We perform the familiar coordinate transformation $(v^+,v^-)\rightarrow (\tau',\eta')$ and integrate out the delta function. After performing the coordinate shift $\uperp = \xperp - \vperp$, the result is
\begin{align}
    f^{+i}(x)
    = -\ii \frac{g}{2\pi}\int_{-\infty}^{\infty}\dd{\eta'} \intop_{\vperp} \frac{\ee^{+\eta'}}{\sqrt{2}}w^j G^{ij}(x^+- \frac{|\vperp|}{\sqrt{2}}\ee^{+\eta'}, x^-- \frac{|\vperp|}{\sqrt{2}}\ee^{-\eta'}, \xperp-\vperp).
\end{align}
This concludes the simplification of $f^{+i}(x)$. The component $f^{-i}(x)$ can be worked out completely analogously. Finally, $f^{ij}(x)$ is similar to $f^{+-}(x)$. To write all the field strength tensor components in a concise way, we introduce the symbols
\begin{align}
\label{eq:V}
    V(x,\eta',\vperp) &\coloneqq -\ii\comm{\beta^i_A(x^+-\frac{|\vperp|}{\sqrt{2}}\ee^{+\eta'},\xperp-\vperp)}{\beta^i_B(x^-- \frac{|\vperp|}{\sqrt{2}}\ee^{-\eta'}, \xperp-\vperp)},\\
\label{eq:Vij}
    V^{ij}(x,\eta', \vperp)&\coloneqq -\ii\left( \comm{\beta^i_A(x^+-\frac{|\vperp|}{\sqrt{2}}\ee^{+\eta'},\xperp-\vperp)}{\beta^j_B(x^-- \frac{|\vperp|}{\sqrt{2}}\ee^{-\eta'}, \xperp-\vperp)}\right.\nn\\
    &\hspace{1cm}\left.-\comm{\beta^j_A(x^+-\frac{|\vperp|}{\sqrt{2}}\ee^{+\eta'},\xperp-\vperp)}{\beta^i_B(x^-- \frac{|\vperp|}{\sqrt{2}}\ee^{-\eta'}, \xperp-\vperp)}\right).
\end{align}
We can then write all independent components of the dilute Glasma field strength tensor in lightcone coordinates as
\begin{align}
    f^{+-}(x) &= -\frac{g}{2\pi}\int_{-\infty}^\infty \dd{\eta'}\intop_\vperp V(x,\eta',\vperp),\label{eq:fpm}\\
    f^{+i}(x) &= \frac{g}{2\pi} \int_{-\infty}^\infty\dd{\eta'} \intop_\vperp \left(V^{ij}(x,\eta',\vperp)-\delta^{ij}V(x,\eta',\vperp)\right)w^j\frac{\ee^{+\eta'}}{\sqrt{2}},\\
    f^{-i}(x) &= \frac{g}{2\pi} \int_{-\infty}^\infty \dd{\eta'}\intop_\vperp \left(V^{ij}(x,\eta',\vperp)+\delta^{ij}V(x,\eta',\vperp)\right)w^j\frac{\ee^{-\eta'}}{\sqrt{2}},\\
    f^{ij}(x) &= -\frac{g}{2\pi}\int_{-\infty}^\infty \dd{\eta'}\intop_\vperp V^{ij}(x,\eta',\vperp).\label{eq:fij}
\end{align}
These expressions were first presented in \cite{Ipp:2022lid}. They match the previously found expressions for the field strength tensor of the (2+1)D dilute Glasma \cite{Guerrero-Rodriguez:2021ask} in the boost-invariant limit. The corresponding calculation is carried out in Appendix \ref{ch:boost-invariant_limit}. The structure of the integrals \eqref{eq:fpm}--\eqref{eq:fij} reveals a lot about the essence of the dilute approximation, as will be discussed in the following.\par
\begin{figure}
    \centering
    \includegraphics{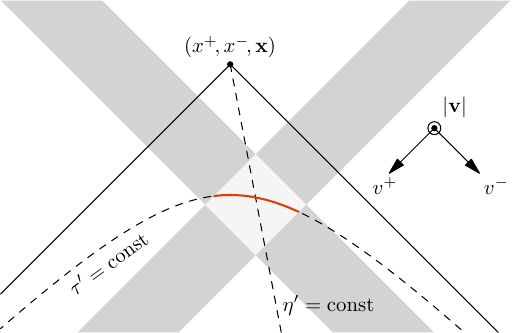}
    \caption{Integration path for the field strength tensor $f^{\mu\nu}$ in the $v^+$-$v^-$-plane. The point $(x^+,x^-,\xperp)$ marks where the field strength tensor is evaluated in the lab frame and marks the origin of the $v^+$-$v^-$ coordinate system, which spans the past lightcone attached at that point. Using $\tau'$ and $\eta'$ instead of $v^+$ and $v^-$, the integral runs only over $\eta'$ along the dashed hyperbola. The hyperbola is highlighted inside the overlap region (light gray) of the tracks of both nuclei (dark gray), where the integrand is nonzero. Original figure from \cite{Ipp:2024ykh}.}
    \label{fig:backwards_milne_integral}
\end{figure}
\begin{figure}
    \centering
    \includegraphics[width=0.7\textwidth]{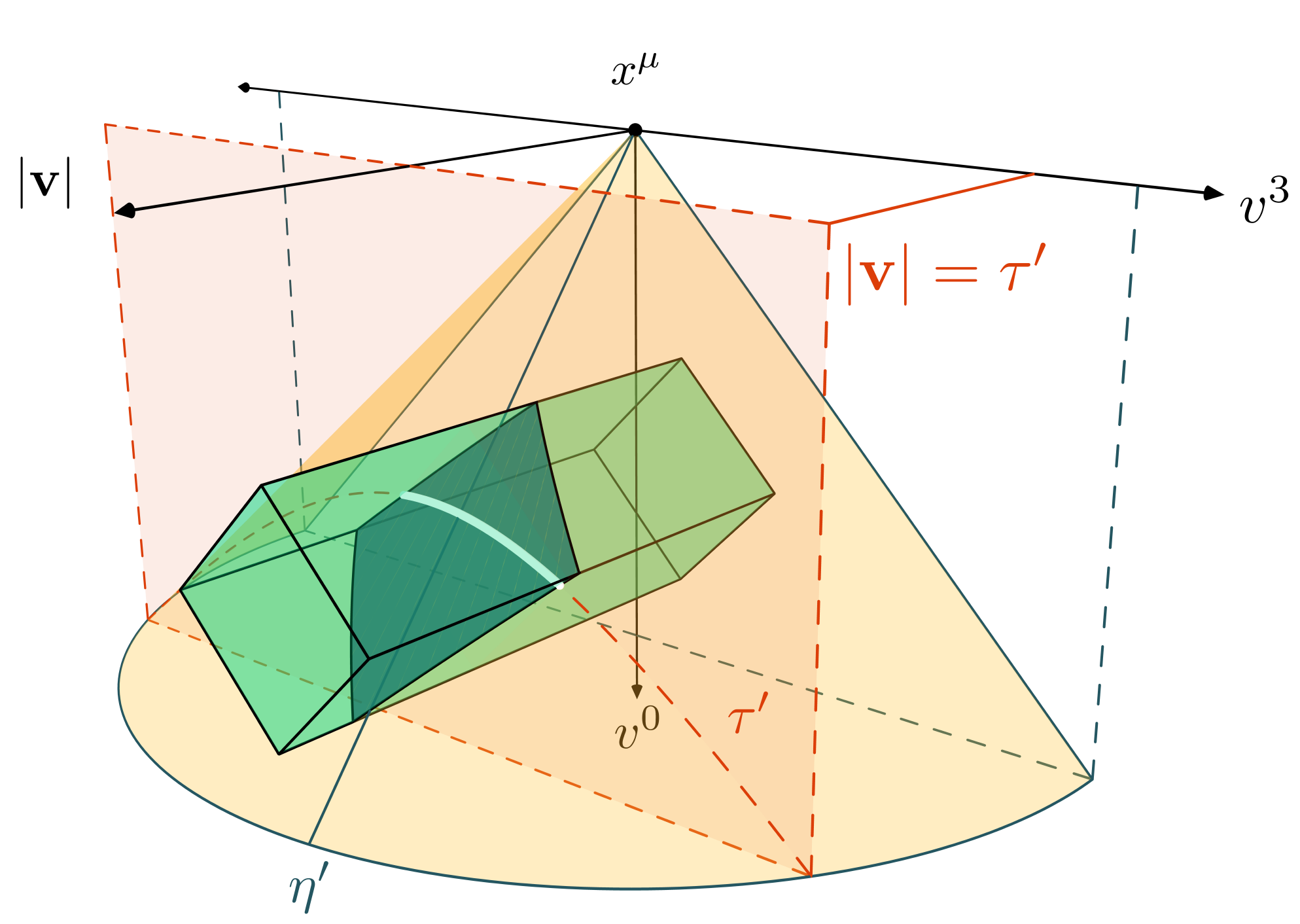}
    \caption{Three-dimensional rendering of the integration hypersurface in the field strength tensor. The light green cuboid is constructed by taking the diamond-shaped interaction region in the $v^+$-$v^-$-plane (the overlap of the tracks of the nuclei) and extruding it in the $|\vperp|$ direction. We assume finite support for nuclei in both longitudinal and transverse directions, but due to Lorentz contraction, the nuclei are much larger in the transverse direction. The past lightcone (yellow) attaches at $x^\mu$ and intersects the extruded collision region diamond in the dark green patch. Disregarding an additional angular coordinate that we are not able to show in this rendering, the dark green patch is the area that needs to be integrated over. Fixing $|\vperp|$ cuts through this surface as indicated by the orange plane. At fixed $|\vperp|$, the integration path is the white hyperbola, the same hyperbola that is shown in Figure~\ref{fig:backwards_milne_integral}. Its shape as a conic section appears naturally in this three-dimensional sketch.}
    \label{fig:3d_integrand}
\end{figure}

First, note that the integrals in \eqref{eq:fpm}--\eqref{eq:fij} can be seen as running over a three-dimensional hypersurface in the four-dimensional space parametrized by
\begin{align}
    v^\mu = (v^+,v^-,\vperp) = (\frac{\tau'}{\sqrt{2}}\ee^{+\eta'}, \frac{\tau'}{\sqrt{2}}\ee^{-\eta'}, \vperp).
\end{align}
This is generic Minkowski space with Milne coordinates for the longitudinal direction. The hypersurface that the integral runs over is defined by $\tau'=|\vperp|$. This identification becomes apparent in the delta functions of \eqref{eq:fpm_intermediate} and \eqref{eq:fpi_intermediate}, which are integrated out in the final results \eqref{eq:fpm}--\eqref{eq:fij}. With respect to the four-dimensional space spanned by $v^\mu$, this identification selects lightlike vectors, i.e.,\ the integration hypersurface is a lightcone.
In Figure~\ref{fig:backwards_milne_integral}, we show a projection of this lightcone on the $v^+$-$v^-$-plane and the corresponding integration path. A more complete picture of the integration surface is shown in Figure~\ref{fig:3d_integrand}, where the modulus of the transverse direction $|\vperp|$ is drawn as a third coordinate. The fourth (angular) coordinate is missing from the three-dimensional sketch in Figure~\ref{fig:3d_integrand}, but the overall structure of the integral becomes clear.\par
Note that the four-vector $v^\mu=x^\mu - u^\mu$ measures the displacement between the point $x^\mu$, at which the field strength tensor is evaluated, and the point $u^\mu$, at which the nuclear fields are evaluated. Taking the viewpoint of a particle picture, this means that gluons are produced in the interaction region, from which they propagate on lightlike paths until they are \say{measured} by evaluating $f^{\mu\nu}$.
There are no further interactions among produced gluons, a direct consequence of the dilute approximation and the Abelian dilute Glasma field strength tensor.\par
The formulae \eqref{eq:fpm}--\eqref{eq:fij} are a key result of \cite{Ipp:2024ykh} and a centerpiece of this thesis. They provide a starting point to efficiently compute observables in the dilute Glasma, as will be explained in the following sections.\par

\section{Landau matching and local rest frame energy density}
A rigorous approach to finding the local rest frame for the energy-momentum tensor of a flowing medium is solving the Landau condition
\begin{align}
    T^\mu{}_\nu u^\nu = \elrf u^\mu
\end{align}
with $u^\mu$ the only timelike eigenvector of $T^\mu{}_\nu$. The corresponding eigenvalue $\elrf$ is the local rest frame energy density. 
To study the conditions under which such a solution exists, we discuss a two-dimensional system, which only features time and a single (longitudinal) spatial direction. A general energy-momentum tensor for this system can be written in Minkowski coordinates as
\begin{align}
    T^{\mu\nu} = \begin{pmatrix}
        e & s\\
        s & p
    \end{pmatrix}.
\end{align}
This tensor is only physical if $T^{00}=e \geq 0$ and if that condition cannot be violated by a Lorentz boost, which can be phrased as
\begin{align}
\label{eq:energy_positivity_condition}
    e \cosh^2 \xi - 2s\cosh\xi\sinh\xi+p\sinh^2\xi \geq 0
\end{align}
for any real $\xi$. The inequality \eqref{eq:energy_positivity_condition} is also referred to as the weak energy condition \cite{rezzolla2013relativistic}. Let us consider the limits $\xi\rightarrow \pm \infty$, from which we obtain
\begin{align}
\label{eq:nec_energy_positivity}
    e + p &\geq 2|s|
\end{align}
as a necessary condition for a two-dimensional energy-momentum tensor to be physical.
Note that as long as this inequality is fulfilled, $s$ and $p$ are allowed to be negative. Another reasonable condition is that the energy flux and momentum density are timelike or lightlike, i.e.,
\begin{align}
    \label{eq:flux_inequality_abstract}
    (T^{00})^2 \geq (T^{01})^2
\end{align}
in any frame of reference. This is the dominant energy condition \cite{rezzolla2013relativistic}. Using the boosted components
\begin{align}
    T^{00} = e \cosh^2\xi - 2s\cosh\xi\sinh\xi+p\sinh^2\xi, \\
    T^{01} = s (\cosh^2\xi+\sinh^2\xi) - (e+p)\cosh\xi\sinh\xi,
\end{align}
we consider $\xi=0$, which yields
\begin{align}
\label{eq:e_s_inequality}
    e \geq |s|,
\end{align}
and the limit $\xi\rightarrow \pm \infty$, which yields 
\begin{align}
(e-p)(e+p-2s)\geq 0,\\
(e-p)(e+p+2s)\geq 0.
\end{align}
Invoking \eqref{eq:nec_energy_positivity}, we see that the second factor never becomes negative and therefore
\begin{align}
\label{eq:e_p_inequality}
    e \geq p.
\end{align}
The inequalities \eqref{eq:e_s_inequality} and \eqref{eq:e_p_inequality} are further necessary conditions for a physically reasonable energy-momentum tensor.\par
Finding the rest frame of our two-dimensional energy-momentum tensor means finding the frame where the off-diagonal components vanish. In that frame, there is no momentum and no energy flux in the spatial direction. Thus, the medium is considered to be at rest. Diagonalizing the energy-momentum tensor implies solving the eigenvalue problem
\begin{align}
    T^\mu{}_\nu u^\nu = \lambda u^\mu.
\end{align}
The eigenvalues are
\begin{align}
\label{eq:ev_pos}
    \lambda_1 &= \frac{e-p + \sqrt{(e+p)^2 - (2s)^2}}{2},\\
    \lambda_2 &= \frac{e-p - \sqrt{(e+p)^2 - (2s)^2}}{2}.
\end{align}
Conveniently, the inequality \eqref{eq:nec_energy_positivity} ensures that these eigenvalues are real. Furthermore, for nonnegative $e$, the inequality \eqref{eq:e_p_inequality} ensures that at least $\lambda_1$ is nonnegative.
Only if both inequalities saturate, $\lambda_1$ (and $\lambda_2$ too) will be zero.
This corresponds to an EMT with $e= p = |s|$. 
In that case, there is no meaningful notion of a rest frame, just like for a massless particle. In fact, this is true for the energy-momentum tensor of the single-nucleus solution \eqref{eq:WW_F_cartesian}. The corresponding nucleus was assumed to move with the speed of light, so there can be no rest frame.\par
The corresponding (non-normalized) eigenvectors are
\begin{align}
    u_1 = \begin{pmatrix} \frac{e+p+\sqrt{(e+p)^2-(2s)^2}}{2s}\\1 \end{pmatrix},\qquad u_2 =  \begin{pmatrix} \frac{e+p-\sqrt{(e+p)^2-(2s)^2}}{2s}\\1 \end{pmatrix}.
\end{align}
The Minkowski norm of $u_1^\mu = u_1$ is
\begin{align}
    u_1^\mu u^1_\mu = \left[\frac{e+p}{2|s|}+ \sqrt{\left(\frac{e+p}{2|s|}\right)^2-1}\right]^2-1 \geq 0
\end{align}
and it is zero only if \eqref{eq:nec_energy_positivity} saturates. Analogously, $u_2^\mu u_\mu^2 \leq 0$.
Therefore, we have arrived at the following statement:
For any physical two-dimensional energy-momentum tensor that does not saturate \eqref{eq:nec_energy_positivity}, there exists exactly one timelike eigenvector, and it has a positive corresponding eigenvalue.
This implies the existence of a local rest frame. In other words, apart from the edge case where \eqref{eq:nec_energy_positivity} is saturated, the conditions that ensure the physicality of the energy-momentum tensor are precisely the same conditions that ensure the existence of a timelike eigenvector with positive eigenvalue and thus the existence of a local rest frame. The relativistic velocity of the rest frame can be obtained by normalizing $u_1^\mu$, such that $u_1^\mu u^1_\mu = 1$ and picking the overall sign such that $u^\mu$ is future-pointing.\\
The local rest frame must be related to the lab frame by a Lorentz boost. We denote the energy-momentum tensor in a boosted frame as
\begin{align}
    \tilde T^{\mu}{}_\nu = \Lambda^\mu{}_\rho \Lambda_{\nu}{}^{\sigma} T^{\rho}{}_{\sigma},
\end{align}
where in matrix notation
\begin{align}
    \Lambda^\mu{}_\rho = \begin{pmatrix}
        \cosh\xi & -\sinh \xi \\
        -\sinh\xi & \cosh \xi
    \end{pmatrix}.
\end{align}
The components of $\tilde T^\mu{}_\nu$ are then
\begin{align}
\tilde T^0{}_0 &= e\cosh^2\xi-2s\sinh\xi\cosh\xi +p\sinh^2\xi,\\
\tilde T^0{}_1&=-s\cosh 2\xi+(e+p)\cosh\xi\sinh\xi,\\
\tilde T^1{}_0&=
        s\cosh 2\xi-(e+p)\cosh\xi\sinh\xi,\\
       \tilde T^1{}_1&=-p\cosh^2\xi +2s\sinh \xi\cosh\xi- e\sinh^2\xi.
\end{align}
Demanding this matrix to be diagonal, i.e.,\ $\tilde T^0{}_1 = \tilde T^1{}_0=0$, yields
\begin{align}
    \xi = \frac{1}{4}\ln\left(\frac{e+p+2s}{e+p-2s}\right)
\end{align}
and puts the energy-momentum tensor in the form
\begin{align}
    \tilde T^\mu{}_\nu = \begin{pmatrix}
       \frac{e-p+\sqrt{(e+p)^2-(2s)^2}}{2} & 0\\
        0 & \frac{e-p-\sqrt{(e+p)^2-(2s)^2}}{2}
    \end{pmatrix}.
\end{align}
Note that these are exactly the eigenvalues from the original eigenvalue problem. The boost velocity is given by
\begin{align}
    \tanh\xi = v = \frac{\sqrt{e+p+2s} - \sqrt{e+p-2s}}{\sqrt{e+p+2s}+\sqrt{e+p-2s}}.
\end{align}
A brief calculation shows that this velocity is completely equivalent to 
\begin{align}
    v = \frac{u^1}{u^0} = \frac{2s}{e+p+\sqrt{(e+p)^2-(2s)^2}}.
\end{align}
Therefore, the velocity that was obtained as the timelike eigenvector from the eigenvalue problem is exactly what is required to boost into the local rest frame, invoking a Lorentz transformation. This closes the circle. The picture of an eigenvalue problem for the energy-momentum tensor is computationally convenient for finding the local rest frame. On a conceptual level, it may be preferable to think about a local Lorentz boost applied to the energy-momentum tensor, which yields the components in the local rest frame. Note that the local rest frame energy density is a Lorenz scalar since, under a Lorentz transformation, the eigenvalue problem
\begin{align}
    (T^\mu{}_\nu -\delta^\mu_\nu \lambda)u^\nu = 0 \rightarrow \Lambda^\mu{}_\rho(T^\rho{}_\nu - \delta^\rho_\nu \lambda)u^\nu = 0
\end{align}
has the same solution.\par
Most of the previous discussion was carried out for a two-dimensional system. In general, a four-dimensional energy-momentum tensor can have additional components, and the eigenvalue problem becomes much more complicated. 
Numerically, the full energy-momentum tensor as a $4\times 4$ matrix can still be diagonalized, and we take the corresponding positive eigenvalue as the local rest frame energy density $\elrf$.

\section{Limiting fragmentation}
\label{sec:limiting_fragmentation}
The term limiting fragmentation describes the phenomenon that rapidity profiles of a given experimental observable $\mathcal O$ are independent of the collision energy for large rapidities if a shift in rapidity by the corresponding beam rapidity is applied.
Mathematically,
\begin{align}
\label{eq:lf_beam_rapidity}
    \left.\mathcal O_{\sqrt{s_1}}(\eta_s-Y_\mathrm{beam}(\sqrt{s_1}))\right|_{\eta_s \gg 0} = \left.\mathcal O_{\sqrt{s_2}}(\eta_s-Y_\mathrm{beam}(\sqrt{s_2}))\right|_{\eta_s \gg 0},
\end{align}
where $Y_\mathrm{beam}(\sqrt{s})$ is the beam rapidity at the collision energy $\sqrt{s}$.
A first theoretical description of limiting fragmentation was given in \cite{Benecke:1969sh}, and a first experimental account was given for charged particle distributions in proton-antiproton collisions \cite{UA5:1986yef}.
Since then, limiting fragmentation has been studied experimentally for different collision systems, e.g.,\ for p-A collisions \cite{Elias:1979cp}, d-Au collisions \cite{PHOBOS:2004fzb} and Au-Au collisions at RHIC \cite{BRAHMS:2001llo, Back:2002wb}.
In the dilute Glasma, limiting fragmentation holds for all observables that are local in spacetime rapidity. Specifically, limiting fragmentation holds for observables local and nonlocal in the transverse plane. We show a proof of this statement in the following.\par
Note that the expressions \eqref{eq:fpm}--\eqref{eq:fij} for the dilute Glasma field strength tensor can be constructed from terms of the form
\begin{align}
&\varphi_0^{ij}(\tau, \eta_s, \xperp) \nn\\
&=\int \dd{\eta'} \intop_\vperp
\left[{\beta^i_A(\frac{\tau}{\sqrt{2}}\ee^{+\eta_s}-\frac{|\vperp|}{\sqrt{2}}\ee^{+\eta'},\xperp-\vperp)},{\beta^j_B(\frac{\tau}{\sqrt{2}}\ee^{-\eta_s}- \frac{|\vperp|}{\sqrt{2}}\ee^{-\eta'}, \xperp-\vperp)}\right]
\end{align}
and
\begin{align}
&\varphi^{ijk}_\pm(\tau, \eta_s, \xperp) \nn\\
&= \int \dd{\eta'} \intop_\vperp
\comm{\beta^i_A(\frac{\tau}{\sqrt{2}}\ee^{+\eta_s}-\frac{|\vperp|}{\sqrt{2}}\ee^{+\eta'},\xperp-\vperp)}{\beta^j_B(\frac{\tau}{\sqrt{2}}\ee^{-\eta_s}- \frac{|\vperp|}{\sqrt{2}}\ee^{-\eta'}, \xperp-\vperp)}\nn\\
&\hspace{3cm}\times w^k\frac{\ee^{\pm \eta'}}{\sqrt{2}},
\end{align}
where we make the Milne coordinates in the spacetime argument explicit. We rewrite the longitudinal argument
\begin{align}
    \frac{\tau}{\sqrt{2}}\ee^{\pm\eta_s}-\frac{|\vperp|}{\sqrt{2}}\ee^{\pm\eta'} = \ee^{\pm \eta_s}(\frac{\tau}{\sqrt{2}}-\frac{|\vperp|}{\sqrt{2}}\ee^{\pm\eta'\mp \eta_s} )
\end{align}
and perform a shift $\eta' \rightarrow \eta'+\eta_s$ in the integration variable, yielding
\begin{align}
    &\varphi_0^{ij}(\tau, \eta_s, \xperp) \nn\\
    &=\int \dd{\eta'} \intop_\vperp
\comm{\beta^i_A(\frac{\ee^{+\eta_s}}{\sqrt{2}}
(\tau-|\vperp|\ee^{+\eta'}),\xperp-\vperp)}
{\beta^j_B(\frac{\ee^{-\eta_s}}{\sqrt{2}}(\tau- |\vperp|\ee^{-\eta'}), \xperp-\vperp)},\\
&\varphi^{ijk}_\pm(\tau, \eta_s, \xperp) \nn\\
&= \int\dd{\eta'} \intop_\vperp
\comm{\beta^i_A(\frac{\ee^{+\eta_s}}{\sqrt{2}}(\tau-|\vperp|\ee^{+\eta'}),\xperp-\vperp)}
{\beta^j_B(\frac{\ee^{-\eta_s}}{\sqrt{2}}(\tau- |\vperp|\ee^{-\eta'}), \xperp-\vperp)}\nn\\
&\hspace{3cm}\times w^k\frac{\ee^{\pm \eta'\pm\eta_s}}{\sqrt{2}},
\end{align}
In the limit $\eta_s \gg 0$ the field $\beta^i_A$ is only nonzero for $\eta'\approx \ln (\tau/|\vperp|)$. We may therefore evaluate $\beta^j_B$ and the factor $\ee^{\pm\eta'\pm\eta_s}$ at this value. We obtain
\begin{align}
        &\varphi_0^{ij}(\tau, \eta_s, \xperp) =  \frac{\sqrt{2}}{\tau}\ee^{-\eta_s}\intop_\vperp
\comm{B^i_A(\xperp-\vperp)}
{\beta^j_B(\frac{\ee^{-\eta_s}}{\sqrt{2}}(\tau- |\vperp|^2/\tau), \xperp-\vperp)},\\
&\varphi^{ijk}_\pm(\tau, \eta_s, \xperp)\nn\\
&= \frac{1}{\tau^{1 \mp 1}}\ee^{(\pm 1-1)\eta_s}\intop_\vperp
\comm{B^i_A(\xperp-\vperp)}
{\beta^j_B(\frac{\ee^{-\eta_s}}{\sqrt{2}}(\tau- |\vperp|^2/\tau), \xperp-\vperp)}w^k{|\vperp|}^{\mp 1},
\end{align}
where we used
\begin{align}
    &\hphantom{=}\,\int_{-\infty}^\infty \dd{\eta'}\beta^i_A(\frac{\ee^{+\eta_s}}{\sqrt{2}}
(\tau-|\vperp|\ee^{+\eta'}),\xperp-\vperp)\nn\\
&= \int_{0}^\infty \dd{y^+}\frac{1}{y^+}\beta^i_A(x^+ - y^+,\xperp-\vperp)\nn\\
&= \int_{-\infty}^{x^+} \dd{y^+}\frac{1}{x^+-y^+}\beta^i_A(y^+,\xperp-\vperp)\nn\\
&\approx \frac{1}{x^+}\int_{-\infty}^{\infty} \dd{y^+}\beta^i_A(y^+,\xperp-\vperp)\nn\\
   &= \frac{1}{x^+}B^i_A(\xperp - \vperp)\nn\\
   &=\frac{\sqrt{2}}{\tau}\ee^{-\eta_s}B^i_A(\xperp - \vperp)
\end{align}
In the second line, we introduced a new integration variable $y^+ = |\vperp|\ee^{+\eta'}/\sqrt{2}$. In the third line, we changed $y^+\rightarrow x^+-y^+$.
Note that there is no problem from the pole at $x^+=y^+$ since we assume that we always evaluate the field strength tensor outside the tracks of the nuclei.
Therefore, $x^+ > y^+$ for all nonzero contributions to the integral.
Furthermore, we understand the limiting fragmentation limit $\eta_s \gg 0$ to mean that $x^+$ is much larger than the longitudinal extent of the nucleus, and we used this in the fourth line. Finally, we introduced the variable $B^i_A(\xperp-\vperp)$ as the longitudinal integral of $\beta^i_A(y^+,\xperp-\vperp)$.\par
An increase in collision energy is realized by boosting both nuclei in their respective direction of movement, yielding
\begin{align}
B^i_A(\xperp - \vperp) &\rightarrow B^i_A(\xperp - \vperp),\\
\ee^{-\eta_s}\beta^j_B(\frac{\ee^{-\eta_s}}{\sqrt{2}}(\tau- |\vperp|^2/\tau), \xperp-\vperp) &\rightarrow \ee^{-\eta_s}\ee^{\zeta}\beta^j_B(\ee^{\zeta}\frac{\ee^{-\eta_s}}{\sqrt{2}}(\tau- |\vperp|^2/\tau), \xperp-\vperp).
\end{align}
Importantly, the boost does not affect $B^i_A(\xperp - \vperp)$.
The beam rapidity of both nuclei increases by $\zeta$ in the process.
The components of the field strength tensor
\begin{align}
    f^{+-} &= \frac{\ii g}{2\pi} \varphi^{ii}_0,\\
    f^{\pm i} &= -\frac{\ii g}{2\pi}(\epsilon^{ij}\epsilon^{kl}\mp \delta^{ij}\delta^{kl})\varphi^{klj}_\pm,\\
    f^{ij} &= \frac{\ii g}{2\pi}\epsilon^{ij}\epsilon^{kl}\varphi_0^{kl}.
\end{align}
change under such a transformation as
\begin{align}
    f^{+-}(\tau, \eta_s, \xperp) &\rightarrow f^{+-}(\tau, \eta_s - \zeta, \xperp),\\
    f^{\pm i}(\tau, \eta_s, \xperp)&\rightarrow \ee^{\pm \zeta}f^{\pm i}(\tau, \eta_s - \zeta, \xperp),\\
    f^{ij}(\tau, \eta_s, \xperp) &\rightarrow f^{ij}(\tau, \eta_s - \zeta, \xperp),
\end{align}
which is exactly the behavior under an active Lorentz boost with parameter $\zeta$ in $z$-direction,
\begin{align}
    f^{\mu\nu}(x) \rightarrow \Lambda^\mu{}_\rho(\zeta) \Lambda^\nu{}_\sigma(\zeta) f^{\rho\sigma}(\Lambda^{-1}(\zeta)x).
\end{align}
Since $\zeta$ is the difference in beam rapidity between the less energetic and more energetic system, this reproduces \eqref{eq:lf_beam_rapidity} except for the outer Lorentz transformation. If we consider observables where the outer transformations becomes 1, we exactly recover the original definition of limiting fragmentation given at the start of this section.

\section{Nuclear models}
\label{sec:nuclear_models}
\subsection{Gauss and Woods-Saxon model}
In all the models considered in this work, the charge density components of nuclei are taken to be neutral on average, i.e.,
\begin{align}
    \langle \rho^a(x^\pm, \xperp) \rangle = 0.
\end{align}
For the 2-point function, often referred to as color charge correlator, we make an Ansatz akin to a generalized McLerran-Venugopalan (MV) model  \cite{McLerran:1993ka, McLerran:1993ni}
\begin{align}
    \label{eq:correlator_general}
    \langle \rho^a(x^\pm,\xperp), \rho^b(y^\pm,\yperp)\rangle =g^2\mu^2\delta^{ab} T(\frac{x^\pm+y^\pm}{2},\frac{\xperp+\yperp}{2})U_\xi(x^\pm-y^\pm)\delta^{(2)}(\xperp-\yperp).
\end{align}
Note that because the dilute approximation only admits terms linear in \eqref{eq:correlator_general}, the MV parameter $\mu$ enters observables in the dilute Glasma through an overall prefactor. The $\delta^{ab}$ indicates that different color components of $\rho$ are completely uncorrelated. The function $T$ can be interpreted as an enveloping profile. We demand the normalization
\begin{align}
\label{eq:T_norm}
    \int \dd{x^\pm} T(x^\pm, \zeroperp) = 1,
\end{align}
which is needed to reproduce the 2D MV model in a certain limit, as will be shown below.
Finally, the function $U_\xi$ governs the longitudinal correlations of color charge within our nuclei. 
We take it to be a Gaussian of width $\xi$
\begin{align}
\label{eq:U_xi}
    U_\xi(x^\pm-y^\pm) =  \frac{1}{\sqrt{2\pi}\xi}\exp\left(-\frac{(x^\pm-y^\pm)^2}{2\xi^2}\right)
\end{align}
and we call $\xi$ the longitudinal correlation length. There are several reasonable choices for the enveloping profile $T$. We demand the nucleus to be spherically symmetric in its rest frame.\footnote{Spherical symmetry is a reasonable assumption for large nuclei, particularly for the Pb and Au nuclei considered in this section. However, deviations from spherical nuclei have been studied in the context of heavy-ion collisions \cite{Schenke:2014tga, Schenke:2020mbo, Magdy:2022cvt, Fortier:2024yxs}.} Then $T(x^\pm, \xperp)$ can only be a function of $2\gamma^2(x^\pm )^2+\xperp^2$. We choose $T$ to be a Gaussian of width $R$ in the rest frame, which fixes
\begin{align}
    T(\frac{x^\pm+y^\pm}{2}, \frac{\xperp+\yperp}{2}) = \frac{\gamma}{\sqrt{\pi}R}\exp\left(-\frac{2\gamma^2(x^\pm+y^\pm )^2+(\xperp+\yperp)^2}{8R^2}\right).
\end{align}
We can insert these expressions for $U_\xi$ and $T$ into the correlator \eqref{eq:correlator_general}, yielding
\begin{align}
    &\langle \rho^a(x^\pm,\xperp), \rho^b(y^\pm,\yperp)\rangle\nn\\ &\hspace{0.9cm}=\frac{g^2\mu^2\gamma}{\sqrt{2}\pi R\xi}\delta^{ab} \ee^{-\frac{(x^\pm + y^\pm)^2}{8R_l^2}}\ee^{-
    \vphantom{\frac{(x^\pm + y^\pm)^2}{8R_l^2}}
    \frac{(\xperp+\yperp)^2}{8R^2}} \ee^{-
    \vphantom{\frac{(x^\pm + y^\pm)^2}{8R_l^2}}
    \frac{(x^\pm-y^\pm)^2}{2\xi^2}}\delta^{(2)}(\xperp-\yperp),
\end{align}
where $R_l = R/(\sqrt{2}\gamma)$ is the Lorentz-contracted longitudinal radius. We may rewrite this correlator as
\begin{align}
    &\langle \rho^a(x^\pm,\xperp), \rho^b(y^\pm,\yperp)\rangle \nn\\
    &\hspace{0.9cm}=\frac{g^2\mu^2\gamma}{\sqrt{2}\pi R\xi}\delta^{ab} \ee^{-\frac{(x^\pm)^2}{4 R_l^2}}\ee^{-\frac{(y^\pm)^2}{4 R_l^2}}\ee^{\frac{(x^\pm-y^\pm)^2}{8 R_l^2}}\ee^{-
    \vphantom{\frac{(x^\pm)^2}{4 R_l^2}}
    \frac{(x^\pm-y^\pm)^2}{2 \xi^2}}\ee^{-
    \vphantom{\frac{(x^\pm)^2}{4 R_l^2}}
    \frac{\xperp^2}{4R^2}}\ee^{-
    \vphantom{\frac{(x^\pm)^2}{4 R_l^2}}\frac{\yperp^2}{4R^2}}\delta^{(2)}(\xperp-\yperp).
\end{align}
Identifying $T$ and $U_\xi$ yields
\begin{align}
\label{eq:correlator_factorized}
    &\langle \rho^a(x^\pm,\xperp), \rho^b(y^\pm,\yperp)\rangle\nn\\
    &\hspace{0.9cm}=g^2\mu^2\delta^{ab} \sqrt{T(x^\pm, \xperp)}\sqrt{T(y^\pm, \yperp)}U_\xi^\mathrm{mod}(x^\pm - y^\pm)\delta^{(2)}(\xperp-\yperp)
\end{align}
with
\begin{align}
\label{eq:U_mod}
    U_\xi^\mathrm{mod}(x^\pm - y^\pm)= \ee^{\frac{(x^\pm-y^\pm)^2}{8 R_l^2}}U_\xi(x^\pm -y^\pm).
\end{align}
The correlator \eqref{eq:correlator_factorized} is only equivalent to \eqref{eq:correlator_general} for Gaussian profiles $T$ and $U_\xi$. We will continue to use the Gaussian \eqref{eq:U_xi} for $U_\xi$, but we take \eqref{eq:correlator_factorized} as a starting point and allow for a different profile $T$. This is also what was done in \cite{Ipp:2024ykh}, where a Woods-Saxon profile
\begin{align}
\label{eq:long_normalized_WS}
    T(x^\pm, \xperp) = \frac{\gamma}{\sqrt{2}d\ln\left(1+\ee^{R/d}\right)}\frac{1}{1+\exp(\frac{\sqrt{2(\gamma \, x^\pm)^2+\xperp^2}-R}{d})},
\end{align}
parametrized by the Woods-Saxon radius $R$ and skin depth $d$, was used.\footnote{In the notation of \cite{Ipp:2024ykh}, the function $U_\xi$ refers to what is called $U_\xi^\mathrm{mod}$ in this work. Apart from this purely notational difference, the correlator used in \cite{Ipp:2024ykh} is equivalent to \eqref{eq:correlator_factorized} with \eqref{eq:long_normalized_WS} inserted for the enveloping profile.} The Woods-Saxon distribution \cite{Woods:1954zz} is a standard measure for the distribution of nucleons in large nuclei and is therefore used here to govern the overall shape of the nucleus.
Note, however, that the nuclei, which follow \eqref{eq:correlator_factorized} with \eqref{eq:long_normalized_WS} do not feature any substructure that mirrors the individual nucleons, i.e.,\ the nuclei are not \say{lumpy}.
Instead, the nontrivial substructure in longitudinal direction is realized through the function $U^\mathrm{mod}_\xi$ and governed by the correlation length parameter $\xi$. The scale of correlations in the transverse direction is set by the infrared regulator employed when solving the transverse Poisson equation.
\begin{figure}
    \centering
    \includegraphics{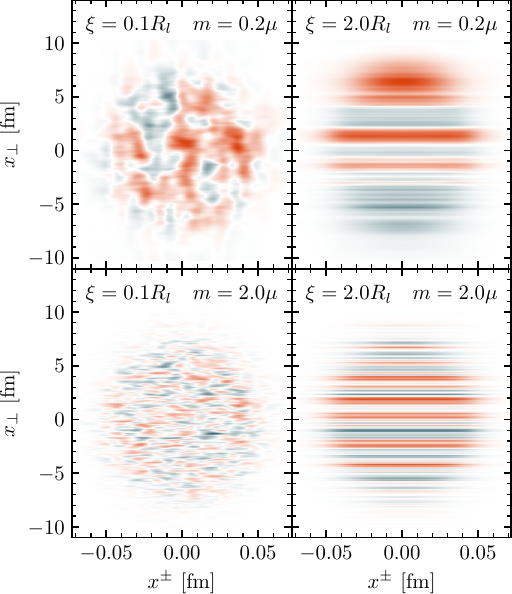}
    \caption{Color field component $\mathcal A^{\mp, a}_{A/B}$ of a nucleus sampled from the Woods-Saxon model \eqref{eq:correlator_factorized} for different values of the longitudinal correlation length parameter $\xi$ and the IR regulator $m$. A central slice through one transverse direction is shown. Hence, the color field depends on the longitudinal coordinate $x^\pm$ and the remaining transverse coordinate $x_\perp \in (x,y)$. Positive values are colored orange, and negative values are colored blue.}
    \label{fig:nuclear_model}
\end{figure}
This is illustrated in Figure~\ref{fig:nuclear_model}, where the color field of a random nucleus sampled from the model \eqref{eq:correlator_factorized} with \eqref{eq:U_mod} and \eqref{eq:long_normalized_WS} is shown. In the longitudinal direction $x^\pm$, the correlation length parameter $\xi$ determines the scale of fluctuations. Specifically, the limit $\xi = 2R_l$ describes completely coherent nuclei in longitudinal direction. In the transverse direction $x_\perp$, the IR cutoff $m$ governs the size of fluctuations.\par
It is worthwhile to study the limiting cases $\xi \rightarrow 0$ (no longitudinal correlations) and $\xi \rightarrow 2 R_l$ (longitudinally coherent nuclei) analytically. In the first case the function $U^\mathrm{mod}_\xi(x^\pm - y^\pm)$ becomes $\delta^{(2)}(x^\pm-y^\pm)$, which puts the correlator \eqref{eq:correlator_factorized} in the form
\begin{align}
\label{eq:3d_mv}
    &\langle \rho^a(x^\pm,\xperp) \rho^b(y^\pm,\yperp)\rangle=g^2\mu^2\delta^{ab}T(x^\pm, \xperp)\delta(x^\pm-y^\pm)\delta^{(2)}(\xperp-\yperp).
\end{align}
This is the traditional MV model supplemented with nontrivial longitudinal extent \cite{Iancu:2002aq, Fukushima:2007ki}.
One may define the transverse charge density
\begin{align}
    \rho_\perp^a(\xperp) = \int_{-\infty}^\infty \dd{x^{\pm}}\, \rho^a(x^\pm,\xperp),
\end{align}
which allows to integrate out the longitudinal dependence in \eqref{eq:3d_mv} and obtain
\begin{align}
    \langle \rho_\perp^a(\xperp) \rho_\perp^b(\yperp)\rangle = g^2\mu^2\delta^{ab}\delta^{(2)}(\xperp\!-\! \yperp) \int_{-\infty}^{+\infty}\dd{x^\pm}\,T(x^\pm, \xperp).
\end{align}
Due to the normalization \eqref{eq:T_norm}, at the center of the nucleus, we exactly recover the original two-dimensional MV model \cite{McLerran:1993ka, McLerran:1993ni}.\par
In the limit $\xi \rightarrow 2 R_l$, the function $U_\xi$ takes the constant value $1/\sqrt{8 \pi R_l^2}$. Consequently, the charge densities do not fluctuate in the longitudinal direction. Instead, they can be sampled from a two-dimensional MV model and modulated with $\sqrt{T}$ afterward.
\subsubsection{Sampling charge densities}
To show how we can sample color charge distributions whose statistics are captured by the correlator \eqref{eq:correlator_factorized}, we start by generating some Gaussian noise $\chi^a(x^\pm, \xperp)$ fulfilling
\begin{align}
    \langle \chi^a(x^\pm, \xperp)\rangle &= 0,\\
    \langle \chi^a(x^\pm, \xperp)\chi^b(y^\pm, \yperp)\rangle &= \delta^{ab}\delta(x^\pm - y^\pm) \delta^{(2)}(\xperp - \yperp).
\end{align}
We Fourier transform this noise in the longitudinal direction to obtain
\begin{align}
    \tilde \chi^a(k^\pm, \xperp) = \int \dd{x^\pm} \ee^{\ii k^\pm x^\pm} \chi^a(x^\pm, \xperp)
\end{align}
with the correlator
\begin{align}
    \langle \tilde \chi^a(k^\pm, \xperp) \tilde \chi^b(p^\pm, \yperp) \rangle &= \delta^{ab}\int \dd{x^\pm}\int\dd{y^\pm} \ee^{\ii k^\pm x^\pm}\ee^{\ii p^\pm y^\pm}\delta(x^\pm - y^\pm) \delta^{(2)}(\xperp - \yperp)\nn\\
    &=\delta^{ab}\int \dd{x^\pm} \ee^{\ii(k^\pm +p^\pm)x^\pm} \delta^{(2)}(\xperp - \yperp)\nn\\
    &=2\pi\delta^{ab}  \delta(k^\pm + p^\pm) \delta^{(2)}(\xperp - \yperp).
\end{align}
We introduce a new field
\begin{align}
    \tilde \zeta^a (k^\pm, \xperp) = \sqrt{\tilde U_\xi^\mathrm{mod}(k^\pm)}\tilde \chi^a(k^\pm, \xperp).
\end{align}
Because $\tilde U^\mathrm{mod}_\xi(k^\pm) = \tilde U^\mathrm{mod}_\xi(-k^\pm)$, the correlator for this field becomes
\begin{align}
    \langle \tilde \zeta^a(k^\pm, \xperp) \tilde \zeta^b(p^\pm, \yperp) \rangle = 2\pi \delta^{ab}\tilde U^\mathrm{mod}_\xi(k^\pm)\delta(k^\pm + p^\pm) \delta^{(2)}(\xperp - \yperp).
\end{align}
We can now transform $\tilde \zeta^a(k^\pm, \xperp)$ back to position space,
\begin{align}
    \zeta^a(x^\pm, \xperp) = \int \frac{\dd{k^\pm}}{2\pi} \ee^{-\ii k^\pm x^\pm} \tilde \zeta^a(k^\pm, \xperp),
\end{align}
which gives a correlator
\begin{align}
    &\langle \zeta^a(x^\pm, \xperp)\zeta^b(y^\pm, \yperp)\rangle\nn\\
    &\hspace{1cm}= \delta^{ab}\int \frac{\dd{k^\pm}}{2\pi}\int\dd{p^\pm} \ee^{-\ii k^\pm x^\pm}  \ee^{-\ii p^\pm y^\pm}\tilde U^\mathrm{mod}_\xi(k^\pm)\delta(k^\pm + p^\pm) \delta^{(2)}(\xperp - \yperp) \nn\\
    &\hspace{1cm}= \delta^{ab}\int \frac{\dd{k^\pm}}{2\pi}\ee^{-\ii k^\pm (x^\pm-y^\pm)} \tilde U^\mathrm{mod}_\xi(k^\pm)\delta^{(2)}(\xperp - \yperp) \nn\\
    &\hspace{1cm}= \delta^{ab}U^\mathrm{mod}_\xi(x^\pm - y^\pm)\delta^{(2)}(\xperp-\yperp).
\end{align}
Finally, we define
\begin{align}
    \rho^a(x^\pm,\xperp) = g\mu\sqrt{T(x^\pm, \xperp)}\zeta^a(x^\pm, \xperp),
\end{align}
which reproduces the correlator \eqref{eq:correlator_factorized}.\par
From the previous discussion, it is not entirely obvious that the samples of color charges obtained through the procedure outlined above come from a Gaussian probability functional. An infinite number of probability functionals produce the same 1- and 2-point functions but there is a unique functional that is also Gaussian. Note that the field $\chi^a(x^\pm, \xperp)$ introduced above is sampled as Gaussian noise and can be seen as being drawn from the Gaussian probability functional
\begin{align}
W[\chi]\propto \exp\left(-\frac{1}{2}\sum_a \intop_x \left(\chi^a(x^\pm, \xperp)\right)^2\right),
\end{align}
where $\intop_x$ stands for $\int \dd{x^\pm}\int\dd[2]{\xperp}$.
Then, $\rho$ is obtained from $\chi$ through a complicated, but linear procedure
\begin{align}
    \rho^a(x^\pm, \xperp) = \intop_y M( x^\pm, \xperp, y^\pm, \yperp)\chi^a(y^\pm, \yperp)
\end{align}
that can also be inverted
\begin{align}
    \chi^a(x^\pm, \xperp) = \intop_y M^{-1}( x^\pm, \xperp, y^\pm, \yperp)\rho^a(y^\pm, \yperp).
\end{align}
The inverse $M^{-1}( x^\pm, \xperp, y^\pm, \yperp)$ is defined via the relation
\begin{align}
    \intop_y M^{-1}( x^\pm, \xperp, y^\pm, \yperp) M( y^\pm, \yperp, z^\pm, \zperp)=\delta^{(3)}(x^\pm-z^\pm, \xperp-\zperp).
\end{align} 
Then,
\begin{align}
W[\rho]\propto \ee^{-\frac{1}{2}\sum_a \intop_x\intop_y\intop_z M^{-1}( x^\pm, \xperp, y^\pm, \yperp)M^{-1}( x^\pm, \xperp, z^\pm, \zperp)\rho^a(y^\pm, \yperp)
 \rho^a(z^\pm, \zperp)}
\end{align}
is the functional from which we have effectively drawn $\rho$. It is clearly Gaussian as well.
\subsection{Gauge invariance}
Under a gauge transformation, the charge density transforms as
\begin{align}
    \rho^a(x)\rightarrow \tilde \rho^a(x) = U^{ab}(x) \rho^b(x)
\end{align}
with the orthogonal $(N_c^2-1)\times (N_c^2 -1)$ matrix $U^{ab}$ (see Appendix~\ref{ch:YM} for more details). From this, it is immediately obvious that
\begin{align}
    \langle \tilde \rho^a(x) \rangle = U^{ab}(x) \langle \rho^b(x) \rangle = 0.
\end{align}
However,
\begin{align}
\langle \tilde \rho^a(x) \tilde \rho^b(y) \rangle = U^{ac}(x)U^{bd}(y) \langle \rho^c(x)\rho^d(y)\rangle. 
\end{align}
Only if $\langle \rho^c(x) \rho^d(y) \rangle = \delta^{cd} \delta^{(3)}(x-y)\lambda(x)$, with some generic function $\lambda(x)$, does the 2-point function become gauge invariant, because in that case
\begin{align}
    \langle \tilde \rho^a(x) \tilde \rho^b(y) \rangle &= U^{ac}(x)U^{bd}(y) \delta^{cd} \delta^{(3)}(x-y) \lambda(x)\nn\\
    &=U^{ac}(x)U^{bc}(x)  \delta^{(3)}(x-y) \lambda(x)\nn\\
    &= U^{ac}(x)(U^{-1})^{cb}(x)  \delta^{(3)}(x-y) \lambda(x)\nn\\
    &= \delta^{ab} \delta^{(3)}(x-y) \lambda(x).
\end{align}
Note that our model, while diagonal in color space, is not of this specific form, as it does not feature delta-shaped correlations in the spacetime coordinates. This makes it impossible to contract the gauge transformation matrices, which are, in general, evaluated at different points. One can insert a Wilson line into the two-point function to mitigate this, essentially transporting the $\rho(y)$ from $y$ to $x$. In covariant gauge, due to the vanishing transverse part of the gauge field, the transverse part of this Wilson line is just a factor of 1 and, therefore, vanishes. The longitudinal part of the Wilson line, however, does not vanish. The inclusion of the Wilson line in the correlator means that the correlator is not just a 2-point-function anymore, as the Wilson line introduces an additional dependence on $\rho$. In conclusion, there is no way to define a gauge invariant 2-point-function with nontrivial (i.e.,\ non delta-like) longitudinal structure. However, in the dilute limit to leading order, all Wilson lines are just 1, and it can be concluded that our correlator is valid and gauge invariant in that limit.
\chapter{Implementation details}
\label{ch:implementation}
The dilute Glasma is a semi-analytic technique for computing observables in the earliest stage of relativistic heavy-ion and hadron collisions.
The formulae \eqref{eq:fpm}--\eqref{eq:fij} provide analytic closed-form expressions for the components of the field-strength tensor, from which all observables in a classical field theory can, in principle, be calculated.
However, in computing observables, we rely on numerics. If we are only interested in expectation values of simple observables, we may express these observables in terms of the color charge correlators of the two colliding particles.
This approach was chosen in \cite{Ipp:2021lwz}, but it results in higher dimensional integrals that must be solved numerically.
The results in this thesis were obtained using a different approach made feasible by the reformulation of the field strength tensor components \eqref{eq:fpm}--\eqref{eq:fij} as three-dimensional integrals.
By sampling individual nuclei or hadrons from the Gaussian probability distribution fixed by the corresponding color charge correlator, we may compute the field strength tensor on an event-by-event basis by solving the integrals \eqref{eq:fpm}--\eqref{eq:fij} using Monte Carlo integration.
Computing two statistically independent estimates for $f^{\mu\nu}$ gives access to an unbiased estimate for the energy-momentum tensor $T^{\mu\nu}$.
Using only a single estimate for $f^{\mu\nu}$ to compute $T^{\mu\nu}$ would lead to a biased estimator, as is highlighted in Appendix \ref{ch:mc}, which also serves as a brief and general introduction to Monte Carlo integration techniques. More details on the Monte Carlo integrals that appear in the dilute Glasma can be found in Section~\ref{sec:mc_procedure}.
We repeat the process of sampling unbiased estimates of $T^{\mu\nu}$ for several independent events and, therefore, gain access to the expectation value and event-by-event fluctuations of the energy-momentum tensor.\par
While computations in the dilute Glasma framework are less expensive than traditional simulations of the (3+1)D Glasma using discrete time stepping, they still require substantial numerical effort and are only made feasible by efficient parallelization. The corresponding computer code was developed in Python and makes use of \texttt{Numba} \cite{10.1145/2833157.2833162} and \texttt{CUDA} \cite{10.1145/1365490.1365500} for increased performance, especially when running on Nvidia GPUs. In particular, the evaluation of the field strength tensor is parallelized over spacetime points on the GPU.
We stress that this procedure is several orders of magnitude faster than comparable simulations of the (3+1)D Glasma, where an initial condition is evolved over sufficiently small time steps.
Additionally, results in the dilute Glasma can be obtained at an arbitrary time (for example, the matching time to hydrodynamics) without the need to compute all intermediate time steps, which further reduces computation time.
\section{Discretization of continuous quantities}
\label{sec:discretization}
Note that the numerical implementation of the dilute Glasma does not require a single large spacetime lattice that encompasses both nuclei and the Glasma. However, certain spacetime domains require discretization.\par
\begin{figure}
    \centering
    \includegraphics[scale=0.9]{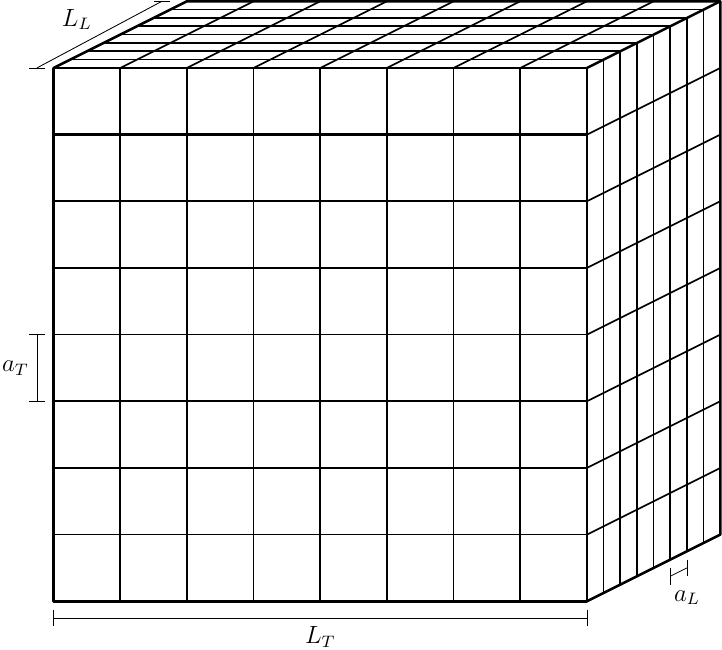}
    \caption{Discrete lattice on which the color fields of the nuclei are represented. The total longitudinal (transverse) extent of the lattice is $L_L$ ($L_T)$. The distance between two lattice sites in longitudinal (transverse) direction is $a_L$ ($a_T$). The dimensions in $x$- and $y$-direction are always the same. The color fields are defined on the cells between grid lines.}
    \label{fig:grid}
\end{figure}
First of all, the color charges pertaining to the hard degrees of freedom in the colliding particles have to be sampled on discrete three-dimensional lattices, as shown in Figure~\ref{fig:grid}. We always treat the two transverse directions identically, and therefore, the lattice is determined by four parameters: the longitudinal and transverse extent, $L_L$ and $L_T$, and the longitudinal and transverse lattice spacing, $a_L$ and $a_T$.
The color charges, which are sampled in the cells located between the grid lines in 
Figure~\ref{fig:grid} act as sources in the transverse Poisson equation, which is solved numerically on the same lattice.
Details of this procedure can be found in Appendix~\ref{ch:poisson}.\par
The lattice needs to be fine enough to resolve all the structure in the color fields.
In longitudinal direction, the size of fluctuations is governed by the correlation length parameter $\xi$.
The lattice spacing in longitudinal direction should thus be sufficiently smaller than this correlation length.
The overall size of the lattice in longitudinal direction is determined by the enveloping profile in the color charge correlator.
In transverse direction, the choice of UV cutoff $\Lambda_\mathrm{UV}$ gives an upper bound for the lattice spacing.
In fact, the lattice spacing introduces a hard UV cutoff $\Lambda_\mathrm{lat}\approx 8 \hbar c / a_T$ on its own, but the scale of this hard cutoff should be above the chosen $\Lambda_\mathrm{UV}$, such that modes affected by the hard cutoff are almost completely suppressed by $\Lambda_\mathrm{UV}$ anyway.
\begin{figure}
    \centering
    \includegraphics[scale=1.0]{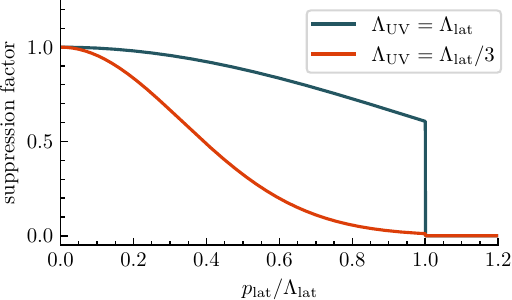}
    \caption{Suppression factor $\exp(p_\mathrm{lat}^2/(2\Lambda_\mathrm{UV}^2))$ in the transverse Poisson equation \eqref{eq:momentum_poisson_main} as a function of lattice momentum for different choices of UV cutoff $\Lambda_\mathrm{UV}$ (shown in units of the lattice cutoff $\Lambda_\mathrm{lat}$). For $\Lambda_\mathrm{UV} = \Lambda_\mathrm{lat}$ (blue), the UV cutoff only gives a small suppression, and the main cutoff comes from the lattice dimensions, which is undesirable. For $\Lambda_\mathrm{UV} = \Lambda_\mathrm{lat}/3$ (orange), the main contribution to the suppression comes from $\Lambda_\mathrm{UV}$, which is preferred.}
    \label{fig:uv}
\end{figure}
This is visualized in Figure~\ref{fig:uv}, where the suppression factor $\exp(p^2_\mathrm{lat}/(2\Lambda_\mathrm{UV}^2))$ in the solution to the Poisson equation \eqref{eq:momentum_poisson_main} is shown as a function of the lattice momentum. The main suppression should come from $\Lambda_\mathrm{UV}$, a scale that we can set freely, and not from the lattice. This is fulfilled if $\Lambda_\mathrm{UV}$ is sufficiently smaller than $\Lambda_\mathrm{lat}$.\par
In transverse direction, the extent of the color field is larger than the extent of the sources due to the properties of the Poisson equation.
The extent of the lattice needs to be large enough to accommodate all relevant contributions to the color fields.
Strictly speaking, the color fields have infinite extent in all directions.
In practice, we cut them off once they have fallen off significantly (less than 1\% of their maximum value). Labeling by $N_T$ the number of cells in transverse direction and by $N_L$ the number of cells in longitudinal direction, a typical dimension for a lattice on which the color fields of the nuclei are sampled in dilute Glasma computations is $N_T\times N_T\times N_L = 1024\times 1024\times 256$.\par
The formulae \eqref{eq:fpm}--\eqref{eq:fij} reference the color fields of the colliding particles at arbitrary three-dimensional coordinates.
As discussed above, we only have access to these quantities on discrete lattices.
In practice, we use linear interpolation to obtain values for the color fields between lattice sites.
This allows us to evaluate the field strength tensor at arbitrary spacetime points inside the forward lightcone.
We evaluate it on a regular lattice, which is completely independent of the lattice on which the color fields of the colliding particles reside.
The lattice on which the field strength tensor is evaluated is four-dimensional and parametrized in Milne coordinates.
This allows the computation of observables along hypersurfaces of constant proper time $\tau$ and studying their dependence on spacetime rapidity $\eta_s$ with one additional subtlety, which is discussed in the following section.\par
The lattices are represented on the computer as arrays, and all data are stored and manipulated in lattice units. A brief introduction to lattice units and the related concept of natural units in particle physics is given in Appendix~\ref{ch:units}.
\section{Shifted Milne coordinates}
\begin{figure}
    \centering
    \includegraphics[scale=1]{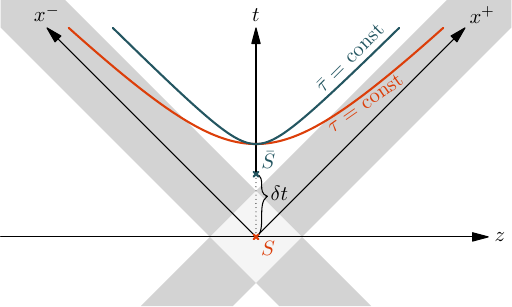}
    \caption{Different choices of the Milne frame origin and the corresponding hyperbolas of constant proper time. The origin of the frame $S$ (orange) is placed at the center of the collision region (light gray) in a symmetric collision of two relativistic nuclei. The corresponding $\tau = \mathrm{const}$ surface cuts into the tracks of the nuclei. The frame $\bar S$ is shifted with respect to $S$ by the distance $\delta t$ along $t$-direction, such that the origin of the frame $\bar S$ lies outside of the collision region. As a consequence, the $\bar \tau = \mathrm{const}$ surface does not cut into the tracks of the nuclei. Original figure from \cite{Ipp:2024ykh}.}
    \label{fig:milne_shift}
\end{figure}
\begin{figure}
    \centering
    \includegraphics[scale=1]{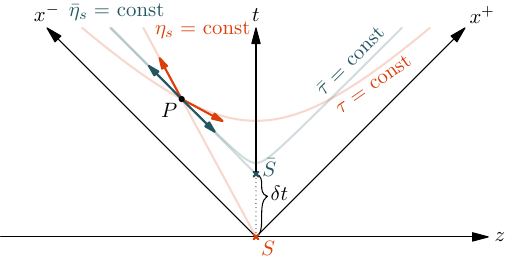}
    \caption{Contravariant basis vectors at the same spacetime point $P$ but with respect to Milne frames with different origins. The lines of constant $\tau$ and $\eta_s$ through $P$ are shown for the Milne frame with origin $S$ (orange) and analogously for the Milne frame with origin $\bar S$ (blue). The vectors are normalized and not to scale.}
    \label{fig:milne_basis_vectors}
\end{figure}
We present numerical results in Milne coordinates (see Section \ref{sec:milne}). In particular, we present observables at values of constant proper time $\tau$ and oftentimes as differential expressions in spacetime rapidity $\eta_s$. In the boost-invariant case, all the interactions between colliding particles take place at a single point in the $t$-$z$-plane. This collision point is an obvious choice for the origin of the Milne frame. In a collision of longitudinally extended particles, however, there is no obvious choice for the origin. As is illustrated in Figure \ref{fig:milne_shift} for a symmetric collision of two nuclei, different choices of the Milne frame origin have different consequences for the hypersurfaces of constant $\tau$. The frame $S$ (orange) is placed at the center of the collision region. Hyperbolas of constant $\tau$ in this frame cut into the tracks of the nuclei. This is problematic as our formulae for the field strength tensor of the Glasma are not valid in this region. Therefore, we shift the origin in $t$-direction by some offset $\delta t$ that is as large or slightly larger than the extent of the nuclei. This leads to a coordinate system anchored at $\bar{S}$ (blue). The hyperbolas of constant $\bar \tau$ in this coordinate system will never enter the tracks of the nuclei, and we can safely evaluate observables up to arbitrary values of $\bar \eta_s$. All results that are presented in Milne coordinates refer to such a shifted Milne frame. However, to reduce clutter, we omit the bars in $\bar \tau$ and $\bar \eta_s$.\par
The shifted Milne frame leads to some considerations when interpreting the results presented in Chapter~\ref{ch:results}. The $\tau=\mathrm{const}$ hypersurface in the shifted frame is a valid Cauchy surface. It can be used to evaluate observables and to initialize hydro simulations. However, since the Milne frame is a curvilinear coordinate system, the significance of tensor components changes between shifted Milne frames. For example, this means that $T^{\tau\tau}(x) \neq T^{\bar \tau \bar\tau}(x)$ even though the tensors are evaluated at the same physical point $x$. The basis vectors in the Milne frame at a given physical point depend on the origin of that frame. This is highlighted in Figure~\ref{fig:milne_basis_vectors}, where the contravariant basis vectors defined as $e_\mu = \partial_\mu P$ are shown for two Milne frames with different origins. In Chapter~\ref{ch:results}, we show $\tau$- and $\eta$-components of tensors to illustrate this issue. However, to present our data in an unambiguous way, we focus on transverse tensor components and Lorentz scalars, such as $\elrf$.
\section{Monte Carlo integration procedure}
\label{sec:mc_procedure}
As shown in Figures \ref{fig:backwards_milne_integral} and \ref{fig:3d_integrand}, and briefly discussed at the end of Section \ref{sec:dilute_glasma_f}, the integrals in the expressions \eqref{eq:fpm}--\eqref{eq:fij} for the dilute Glasma field strength tensor receive nonzero contributions only from a small subset of the integration volume. While this is strictly speaking only correct for colliding particles with compact support, this is certainly true for the numerical evaluation of these integrals. We, therefore, may, without changing the result of the numerical integration, restrict the integration bounds in \eqref{eq:fpm}--\eqref{eq:fij}. Employing polar coordinates $(|\vperp|, \theta_\vperp)$ for the $\vperp$-plane, the unrestricted bounds are
\begin{align}
    -\infty &< \eta' < \infty,\\
    0 &\leq |\vperp| < \infty,\\
    0 &\leq \theta_\vperp < 2\pi.
\end{align}
We start by constraining $\eta'$ and consider the longitudinal arguments of the functions
\begin{align}
\label{eq:beta}
    \beta^i_{A/B}(x^\pm -\frac{|\vperp|}{\sqrt{2}}\ee^{\pm \eta'}, \xperp - \vperp)
\end{align}
as they appear in \eqref{eq:V}--\eqref{eq:Vij}.
Nonzero contributions can only come from
\begin{align}
\label{eq:longitudinal_inequality}
    -\frac{L_L^{
    A/B}}{2} \leq x^\pm - \frac{|\vperp|}{\sqrt{2}}\ee^{\pm\eta'} \leq \frac{L_L^{A/B}}{2}.
\end{align}
Here, $L_L^{A/B}$ is the length of the lattice in $x^\pm$ direction for the color fields of $A$ and $B$.
This restricts $\eta' \in (\eta'_\mathrm{min},\eta'_\mathrm{max})$ with
\begin{align}
    \eta'_\mathrm{min} = \mathrm{max}\left(\ln\left[\frac{\sqrt{2}}{|\vperp|}\left(x^+-\frac{L_L^A}{2}\right)\right], -\ln\left[\frac{\sqrt{2}}{|\vperp|}\left(x^- + \frac{L_L^B}{2}\right)\right]\right)
\end{align}
and
\begin{align}
    \eta'_\mathrm{max} = \mathrm{min}\left(\ln\left[\frac{\sqrt{2}}{|\vperp|}\left(x^++\frac{L_L^A}{2}\right)\right], -\ln\left[\frac{\sqrt{2}}{|\vperp|}\left(x^- - \frac{L_L^B}{2}\right)\right]\right).
\end{align}
From \eqref{eq:longitudinal_inequality}, it follows that
\begin{align}
\label{eq:longitudinal_inequality_2}
    \sqrt{2}x^\pm+\frac{L_L^{
    A/B}}{\sqrt{2}} \geq  |\vperp|\ee^{\pm\eta'} \geq \sqrt{2}x^\pm-\frac{L_L^{A/B}}{\sqrt{2}}.
\end{align}
Note that the rightmost quantity is strictly positive, as we always evaluate the field strength tensor outside the tracks of the nuclei. This allows us to multiply the two inequalities for $x^+$ and for $x^-$ contained in \eqref{eq:longitudinal_inequality_2} term by term such that $\ee^{\pm \eta'}$ cancels, which yields
\begin{align}
    |\vperp|_\mathrm{min} \leq |\vperp| \leq |\vperp|_\mathrm{max}
\end{align}
with
\begin{align}
    |\vperp|^2_\mathrm{min} = \left(\sqrt{2}x^+-\frac{L_L^{A}}{\sqrt{2}}\right)\left(\sqrt{2}x^- -\frac{L_L^{B}}{\sqrt{2}}\right),\\
    |\vperp|^2_\mathrm{max}=\left(\sqrt{2}x^++\frac{L_L^{
    A}}{\sqrt{2}}\right)\left(\sqrt{2}x^- +\frac{L_L^{
    B}}{\sqrt{2}}\right).
\end{align}
To sample points uniformly over the integration volume, we have to take into account the length of the corresponding $\eta'$-interval at each value of $|\vperp|$ and the Jacobian in the volume element $\dd{V}= |\vperp| \dd{|\vperp|}\dd{\theta_\vperp}$. We therefore draw $|\vperp|$ from a distribution
\begin{align}
    f(|\vperp|) = \begin{cases}
        (\eta'_\mathrm{max} - \eta'_\mathrm{min})|\vperp| & \mathrm{if}\,\,\, |\vperp|_\mathrm{min} \leq |\vperp| \leq |\vperp|_\mathrm{max}\\
        0 & \mathrm{else}
    \end{cases}
\end{align}
(up to normalization) and select a value for $\eta'$ uniformly from the allowed interval.\par
\begin{figure}[t]
    \centering
    \includegraphics[scale=1]{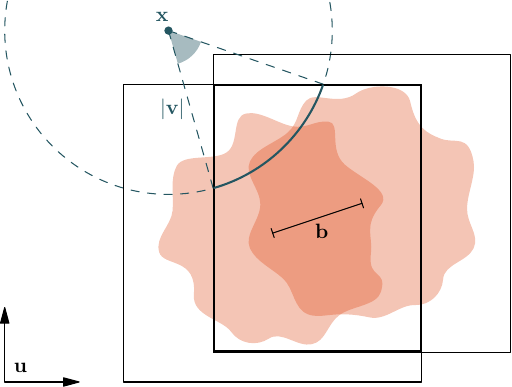}
    \caption{Example of a possible integration path for fixed $|\vperp|$ in the $\uperp$-plane. Only where the lattices of the two nuclei overlap can the integrand be nonzero. The circle with radius $|\vperp|$ centered at $\xperp$ only intersects this overlap for a small range of angles. Figure adapted from \cite{Ipp:2024ykh}.}
    \label{fig:u_plane_integration}
\end{figure}
The angle $\theta_\vperp$ can, in principle, be sampled uniformly. However, note that the $\theta_\vperp$-integration runs over a circle of radius $|\vperp|$ centered at $\xperp$ in the $\uperp$-plane. The coordinates $\uperp = \xperp - \vperp$ are local coordinates for the fields of the colliding particles in transverse direction.
This means that, in the case of vanishing impact parameter, the nuclei are centered at $\uperp = 0$.
The only difference due to a finite impact parameter is that nucleus $A/B$ is centered at $\uperp = \mp \bperp/2$.
This is illustrated in Figure~\ref{fig:u_plane_integration}.
The integrand can only be nonzero in the region where both nuclei are defined, i.e.,\ where the lattices of both nuclei overlap, indicated by the thick rectangular box.
The circle centered at $\xperp$ with radius $\vperp$ does not necessarily lie completely within this box.
The sampling of angles $\theta_\vperp$ is made more efficient by restricting the range of angles to those where the arc of the circle actually lies within the box.
\begin{figure}[t]
    \centering
    \includegraphics[width=\linewidth]{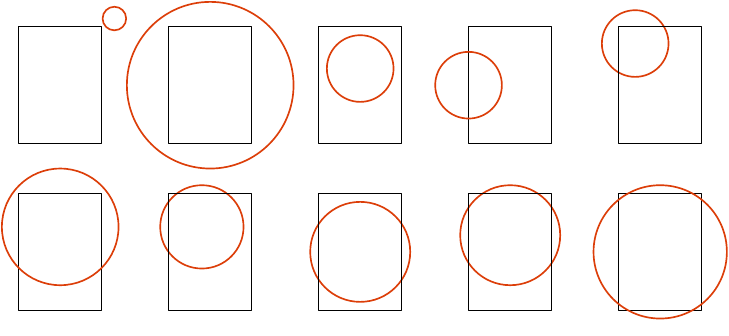}
    \caption{All possible ways in which a circle can (not) intersect a rectangle, ordered by increasing number of intersections. For the graphs featuring 2 intersections, the rectangle could also be a square, and the specific sides do not matter. Therefore, versions rotated by multiples of $\pi/2$ need to be considered as well. This also applies to the first graph with 4 intersections. The second graph with 4 intersections can only occur for a rectangle with unequal sides, and there are no other versions of this graph. The graph featuring 6 intersections also requires a rectangle with unequal sides, and a version rotated by $\pi$ has to be considered. The last graph features 8 intersections, and no other versions exist.}
    \label{fig:sections}
\end{figure}
There are, up to symmetries, 10 different scenarios that have to be taken into account, sketched in Figure~\ref{fig:sections}. These scenarios are all considered in the implementation of the Monte Carlo integral such that the angle $\theta_\vperp$ is always restricted to the region where the integrand can be nonzero.
\section{Jackknife method}
It is well known that the Monte Carlo integration error is proportional to $1/\sqrt{N}$, where $N$ is the number of samples (see Appendix~\ref{ch:mc} for a derivation).
The exact error carries a prefactor that depends on the specific function that is being integrated.
In the dilute Glasma, estimates of the field strength tensor are obtained through a Monte Carlo integration procedure.
However, the observables that we study are potentially complicated functions of two independent estimates for the field strength tensor.
The Monte Carlo integration error propagates through these computations in a highly nontrivial way. Therefore, we use the jackknife method to estimate the statistical error in our observables due to the Monte Carlo integration procedure.
This error is computed for a single collision event and is different from the event-by-event fluctuations of observables.\par
The Monte Carlo method of solving an integral consists of sampling values of the integrand and computing the average of these samples.
A single sample could be as simple as evaluating the integrand at a random point and multiplying with the integration volume.
In practice, we use importance sampling to draw random points from a non-uniform distribution and compensate for that bias by introducing a weight factor that is essentially the inverse probability of having drawn that random point.
To use the jackknife method, we divide the $N$ Monte Carlo samples into $N_\mathrm{bin}$ bins.\footnote{If the observable of interest is the energy-momentum tensor or a quantity derived from it, two independent estimates $f^{\mu\nu}_1$ and $f^{\mu\nu}_2$ of the field strength tensor need to be computed, as mentioned in Chapter~\ref{ch:implementation}. Then, each bin has to contain $N/N_\mathrm{bin}$ samples for $f^{\mu\nu}_1$ and the same number of independent samples for $f^{\mu\nu}_2$.}
We label the set of all samples in a bin as $x_i$ with $i=1,\dots, N_\mathrm{bin}$.
Next, we define the leave-1-out sets as
\begin{align}
    x^{\mathrm{l1o}}_i = \bigcup_{\substack{j=1\\j\neq i}}^{N_\mathrm{bin}}x_j.
\end{align}
These sets contain a fraction $(N_\mathrm{bin}-1)/N_\mathrm{bin}$ of all samples, so almost all of the information that was collected. The best estimate for an observable $\mathcal O$ is given by the jackknife estimator
\begin{align}
    \langle \mathcal O \rangle_\mathrm{jack} = \frac{1}{N_\mathrm{bin}}\sum_{i=1}^{N_\mathrm{bin}} \mathcal O(x^\mathrm{l1o}_i),
\end{align}
where by $\mathcal O(x^\mathrm{l1o}_i)$ we mean $\mathcal O$ evaluated using all of the samples in $x^\mathrm{l1o}_i$.
The jackknife variance is defined as
\begin{align}
    \sigma_\mathrm{jack}(\mathcal O) =\frac{N_\mathrm{bin}-1}{N_\mathrm{bin}} \sum_{i=1}^{N_\mathrm{bin}} \left(\mathcal O(x_i^\mathrm{l1o}) - \langle \mathcal O \rangle_\mathrm{jack}\right)^2
\end{align}
and gives a good estimate for the Monte Carlo error on the level of the observable $\mathcal O$.
We use $N_\mathrm{bin} = 10$ to compute the jackknife variance for all numerical results in this thesis.\par
\begin{figure}
    \centering
      \includegraphics[width=\textwidth]{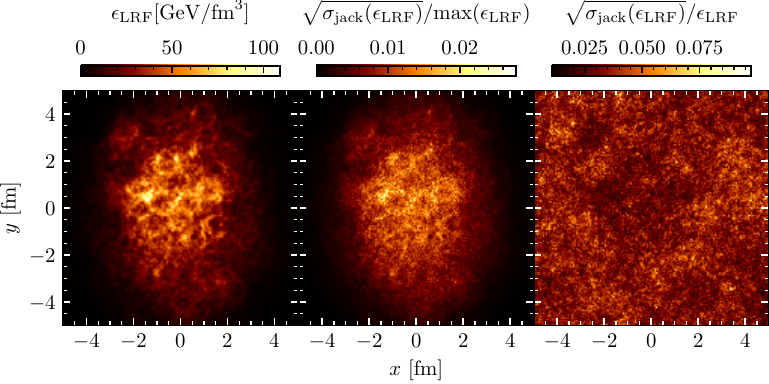}
    \caption{Local rest frame energy density $\elrf$ in the transverse plane at mid-rapidity (left). Simulation parameters are identical to the results shown in Figure~\ref{fig:elrf_slices}. In the center, the square root of the jackknife variance is shown as a fraction of the maximum value of $\elrf$ in the plane. On the right side, the square root of the jackknife variance is divided by the value of $\elrf$ at the respective point. The number of Monte Carlo samples was $2\times 10^5$.}
    \label{fig:jackknife_errors}
\end{figure}
We show the jackknife error for a given simulation result in Figure~\ref{fig:jackknife_errors}. The left plot depicts the local rest frame energy density $\elrf$ in the transverse plane at mid-rapidity $\eta_s = 0$. In the center and right plots, we show the square root of the jackknife variance as a measure of the propagated Monte Carlo integration error. This error has to be related to a typical value of $\elrf$ to obtain a relative error. In the center plot, we show the square root of the jackknife variance rescaled by the maximum value of $\elrf$ in the transverse plane. The maximum relative error is less than 3\% and lower for many points. This kind of relative error tends to be larger at points where the energy density is larger. The opposite is the case in the right plot, where we rescale the square root of the jackknife variance with the local value of $\elrf$. The relative error is lower in regions where $\elrf$ itself is large and is, per definition, larger overall. However, large relative errors on small values do not contribute much to integrated quantities and are not problematic. Most of the observables considered in the Glasma are integrated over the transverse plane. For these observables, the errors at the individual transverse points cancel to some degree, and the overall error on the observable is considerably smaller. Therefore, in practice, $10^5$ Monte Carlo samples are sufficient to obtain observables with good accuracy.
\chapter{Collisions of Woods-Saxon nuclei}
\label{ch:results}
In this section, we present numerical results based on the Woods-Saxon model of lead (Pb) and gold (Au) nuclei, determined by the correlator \eqref{eq:correlator_factorized} with \eqref{eq:U_mod} and \eqref{eq:long_normalized_WS}. The choice of nucleus enters only through the Woods-Saxon radius $R$ and skin depth $d$, which are taken from \cite{Schenke:2012hg}. We choose center-of-mass energies per nucleon-nucleon pair according to experimentally realized setups. Specifically, for Au-Au collisions we choose $\sqrt{s_{NN}} = \SI{200}{\giga\electronvolt}$ as realized at RHIC, whereas for Pb-Pb collisions we choose $\sqrt{s_{NN}} = \SI{5400}{\giga\electronvolt}$, which is in the ballpark of collision energies realized at the LHC. The center-of-mass energy enters our expressions through the Lorentz contraction factor $\gamma = \sqrt{s_{NN}}/(\SI{2}{\giga\electronvolt})$, where we have assumed a mass of $\SI{1}{\giga\electronvolt}$ per nucleon.
\begin{table}
  \centering
    \begin{tabular}{llll}
    \textbf{Param.} & \textbf{Name} & \textbf{Value(s)} & \textbf{Unit} \\
    \midrule
    $N_c$ & Number of colors & 3 & - \\
    $\gamma$ & Lorentz factor & 100 (RHIC), 2700 (LHC) & - \\
    $\sqrt{s_{NN}}$ & c.m.~energy & 200 (RHIC), 5400 (LHC) & GeV \\
    $R$   & WS radius & 6.38 (RHIC), 6.62 (LHC) & fm \\
    $d$   & WS skin depth & 0.535 (RHIC), 0.546 (LHC) & fm \\
    $\delta t$ & Milne origin shift & 0.069 (RHIC), 0.0027 (LHC) & fm \\
    $g$ & YM coupling & 1 & - \\
    $\mu$ & MV scale & 1 & GeV \\
    $m$   & IR cutoff & 0.2, 2.0 & GeV \\
    $\Lambda_\mathrm{UV}$ & UV cutoff & 10 & GeV \\
    $\xi$ & correlation length & 0.1, 0.5, 2.0 & $R_l$ \\
    $b$   & impact parameter & 0, 1  & $R$ \\
    $\tau$ & proper time & 0.2, 0.4, 0.6, 0.8, 1.0 & fm/c \\
    \end{tabular}
\caption{Parameters for the Woods-Saxon (WS) model calculations for RHIC and LHC setups. The correlation length $\xi$ and the impact parameter $b$ are given as multiples of the Woods-Saxon radius and are, therefore, different for RHIC and LHC setups. Note that the values for $\delta t$ here are rounded. The exact values are $(R+d)/\gamma$ for the respective collision energies. Table adapted from \cite{Ipp:2024ykh}.}
\label{tab:params}
\end{table}
Table \ref{tab:params} shows the physical parameters used for the Woods-Saxon model computations.
The correlation length $\xi$ and impact parameter $b$ are given in units of the Woods-Saxon radius, yielding slightly different values for the RHIC and LHC configurations.
Also, the impact parameter is given as a scalar quantity and is always aligned in $x$-direction.
For the correlation length parameter $\xi$, the value of $0.1\,R_l$ roughly corresponds to the size of a single nucleon. 
\section{Notions of energy density}
\label{sec:notions_energy_density}
As discussed before, there are different notions of energy density in the Glasma. The component $T^{\tau\tau}$ is a measure of energy density in the rest frame of an observer boosted with $\eta_s$. It is, however, not a meaningful measure of the energy density in the system's rest frame. One reason for this is transverse flow. Especially in peripheral regions, the Glasma expands outward in the transverse plane, exhibiting non-negligible flow velocity in $x$- and $y$-direction. Furthermore, boost invariance is broken in the (3+1)D Glasma, and therefore, the Milne coordinates do not fully capture the symmetries of the system. This is also clear from the fact that there is no single obvious choice to place the origin of the Milne frame. Instead, there is an extended collision region where gluons are produced. The energy density at any given point in the forward lightcone may come from different such points in the collision region.\par
These observations necessitate a coordinate-invariant notion of energy density, which was already introduced as the local rest frame energy density $\elrf$. We also consider a version of $\elrf$ that is rescaled with $\gamma$, the local Lorentz factor obtained from the fluid velocity (i.e.,\ the timelike eigenvector of the energy-momentum tensor).\par
Finally, we consider the transverse energy density, which is just the sum of transverse pressures $T^{xx}+T^{yy}$. It is a useful quantity, as it can be directly obtained from the energy-momentum tensor without solving the eigenvalue problem and does not suffer from the ambiguity of the coordinate system used for the longitudinal direction.
\begin{figure}
    \centering
        \centering
        \includegraphics{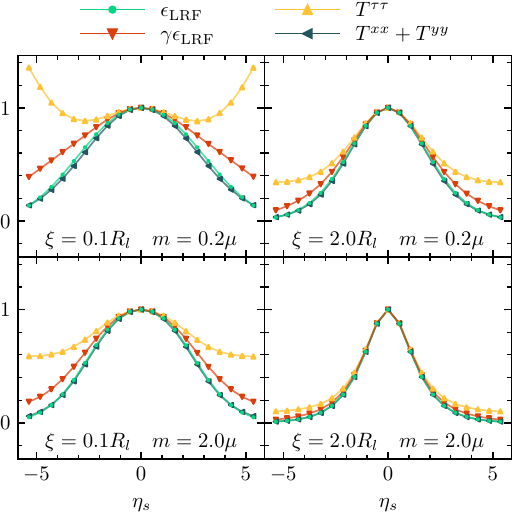}\\
        \vspace{0.4cm}
        \centering
        \includegraphics{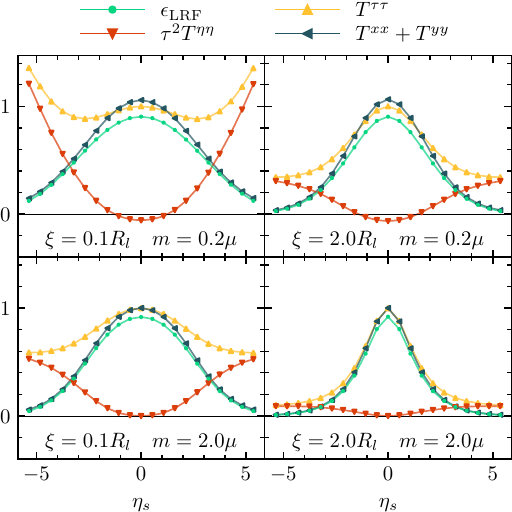}
    \caption{Spacetime rapidity dependence of (various notions of) energy density integrated over the transverse plane at proper time $\tau=0.4\,\mathrm{fm}/c$, averaged over 10 independent collision events at RHIC energy with zero impact parameter.
    The top profiles are rescaled to match at $\eta_s = 0$. For the bottom figure, the relative scale is unchanged, and the longitudinal pressure $\tau^2 T^{\eta\eta}$ is shown instead of $\gamma\elrf$. Original figures from \cite{Ipp:2024ykh}.}
    \label{fig:energy-comp}
\end{figure}
Figure~\ref{fig:energy-comp} shows a comparison between these different notions of energy density integrated over the transverse plane as a function of spacetime rapidity.
In the upper plot, profiles were rescaled to match at $\eta_s=0$.
Clearly, the component $T^{\tau\tau}$ becomes pathological for large rapidities and should not be used for studies of the (3+1)D Glasma apart from the mid-rapidity region.
The local rest frame energy density $\elrf$ is well-behaved and, regarding the shape of the profile, is very similar to the transverse energy. The latter can thus be used as a good proxy for $\elrf$ if the overall scale of the profile does not matter.
As can be seen from the lower plot, where profiles are not rescaled, $T^{xx} + T^{yy}$ exhibits slightly larger values than $\elrf$.
The lower plot also features the longitudinal pressure $\tau^2 T^{\eta\eta}$ instead of $\gamma\elrf$.
The longitudinal pressure shows the same pathologies as $T^{\tau\tau}$, which has to be the case to keep the energy-momentum tensor traceless.\par
\begin{figure}
    \centering
        \centering
        \includegraphics[width=245.52pt]{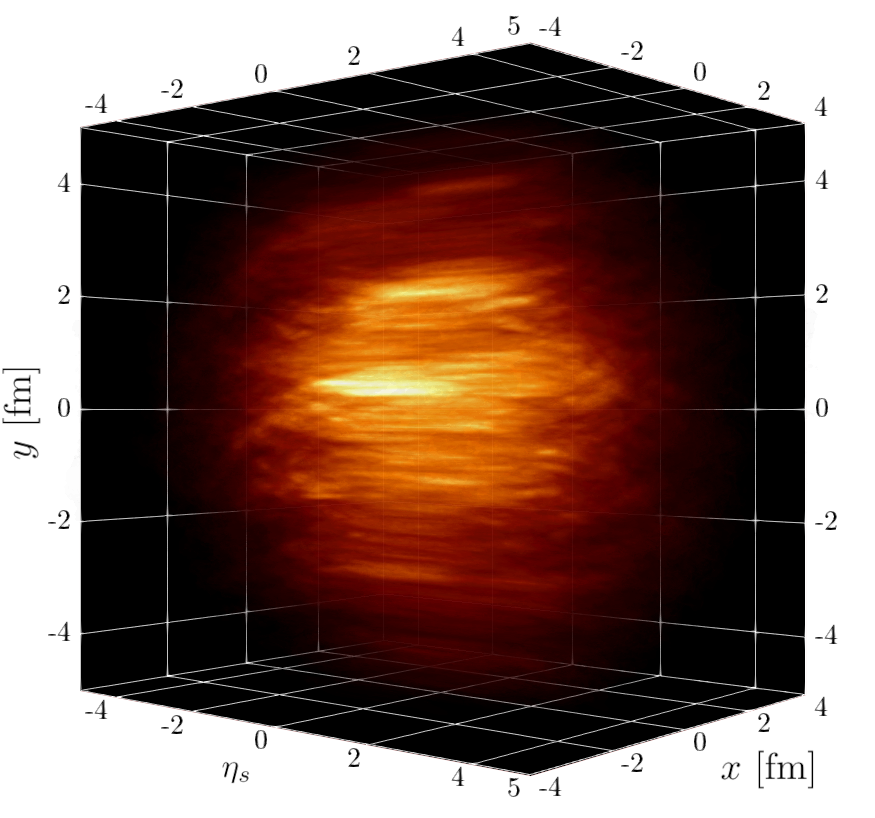}
    \caption{Three-dimensional rendering of the local rest frame energy density $\elrf$ at $\tau=0.4\,\mathrm{fm}/c$ in a collision at RHIC energy with $b=R$, $\xi=0.5R_l$, $m=0.2\,\mathrm{GeV}$. Original figure from \cite{Ipp:2024ykh}.}
    \label{fig:elrf_3d}
\end{figure}
In Figure~\ref{fig:elrf_3d}, we show a 3D rendering of the local rest frame energy density $\elrf$ for an event with finite impact parameter $b=R$. The transverse structure resembles the overlap region between two circular nuclei offset by the impact parameter. In longitudinal direction, long range structures akin to flux tubes can be seen.
\begin{figure}
        \centering
        \includegraphics[width=\textwidth]{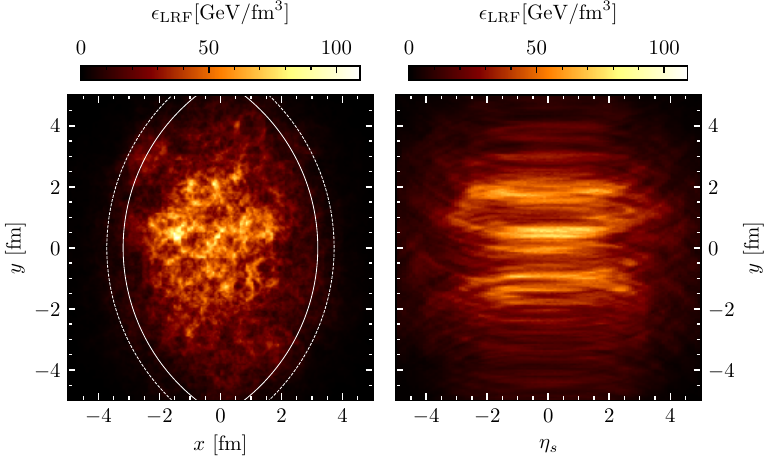}
    \caption{Local rest frame energy density $\elrf$ at $\tau = 0.4\,\mathrm{fm}/c$ in a collision at RHIC energy with $b=R$, $\xi=0.5R_l$, $m=0.2\,\mathrm{GeV}$. The left plot shows a longitudinal cut at mid-rapidity, and the right plot shows a transverse cut at $x=0$. The solid (dashed) white line represents the $R$ ($R+d$) boundary of the nuclei. Figure adapted from \cite{Ipp:2024ykh}.}
    \label{fig:elrf_slices}
\end{figure}
We show slices through the 3D plot of Figure~\ref{fig:elrf_3d} in Figure~\ref{fig:elrf_slices}. The edges of the nuclei in the transverse plane are drawn in white and bound the \say{almond-shaped} region where most energy density resides.
\section{Transverse flow}
\label{sec:transverse_flow}
\begin{figure}
    \centering
      \includegraphics[width=\textwidth]{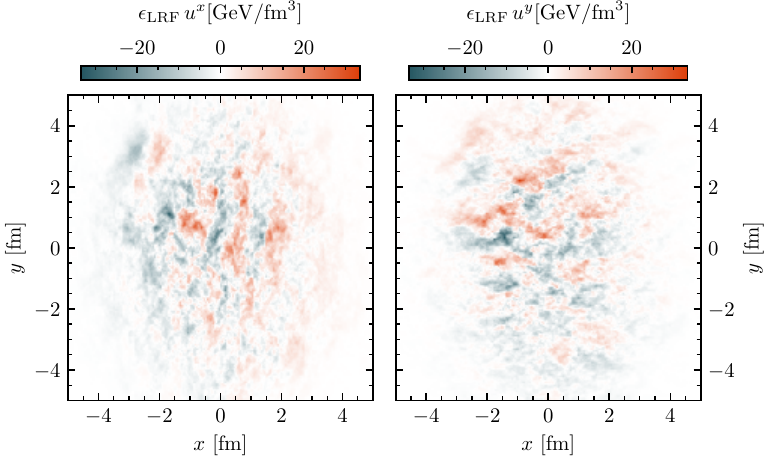}
    \caption{Transverse flow components $u^x$ and $u^y$ scaled with the local rest frame energy density $\elrf$ for the mid-rapidity slice shown in the left panel of Figure~\ref{fig:elrf_slices}.}
    \label{fig:transverse_flow}
\end{figure}
We expect a considerable amount of transverse flow in collisions of Woods-Saxon nuclei due to fluctuations in the initial conditions and the overall geometry of the nuclei.
In Figure~\ref{fig:transverse_flow}, we show the $\elrf$-weighted components $u^x$ and $u^y$ for the specific event that was studied in Section~\ref{sec:notions_energy_density} and depicted in Figure~\ref{fig:elrf_slices}.
A careful examination of the fluctuations reveals more flow in positive $x$-direction at positive $x$ coordinates and more flow in negative $x$-direction at negative $x$ coordinates.
The same holds for the $y$-direction.
This is to be expected for a fireball expanding outwards in all directions.
However, local fluctuations in energy density lead to a much noisier picture than what would be the case for an expanding homogeneous medium.
\section{Longitudinal flow}
\label{sec:longitudinal_flow}
\begin{figure}
    \centering
    \includegraphics[scale=1]{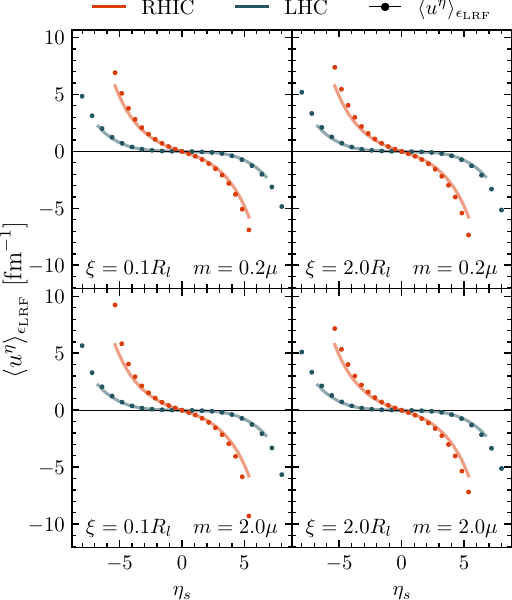}
    \caption{Longitudinal flow $u^\eta$ weighted with $\elrf$ and integrated over the transverse plane as a function of spacetime rapidity at $\tau = 0.4\,\mathrm{fm}/c$ (discrete points). The data are averaged over 10 independent central collision events. Lines are predictions from an analytic model taking the collision geometry into account. Original figure from \cite{Ipp:2024ykh}.}
    \label{fig:longitudinal_flow}
\end{figure}
In Figure~\ref{fig:longitudinal_flow}, we present the longitudinal flow in the dilute Glasma, weighted with the local rest frame energy density and integrated over the transverse plane, i.e.,
\begin{align}
    \langle u^\eta \rangle _{\elrf}= \intop_\xperp \dd[2]{\xperp} u^\eta(\tau, \eta_s, \xperp) \elrf(\tau, \eta_s, \xperp).
\end{align}
Note that this quantity is not dimensionless. We abstain from rescaling with $\tau$ here as we never compare values at different times. The longitudinal flow is shown for select values of $\eta_s$ as discrete data points for RHIC and LHC energy. The longitudinal flow is positive for negative spacetime rapidity and vice versa. This seems to indicate that the system expands more slowly than what is expected from Bjorken flow, for which $u^\eta = 0$. However, longitudinal flow is affected in two places by the ambiguities surrounding the Milne frame origin. First, the direction in which the rapidity component of a vector at a fixed spacetime location points depends on the choice of origin. This can artificially enhance or reduce longitudinal flow at extreme values of rapidity. Furthermore, even if Bjorken flow were produced at every point in the interaction region, it would overlay to something that does not look like Bjorken flow globally. This is a direct consequence of the extended collision region and is an issue that is completely absent in boost-invariant treatments. To estimate the size of this effect, the continuous lines in Figure~\ref{fig:longitudinal_flow} are obtained from simple model calculations assuming the production of Bjorken flow but taking into account the geometry of our setup.
\begin{figure}
    \centering
    \includegraphics[scale=1]{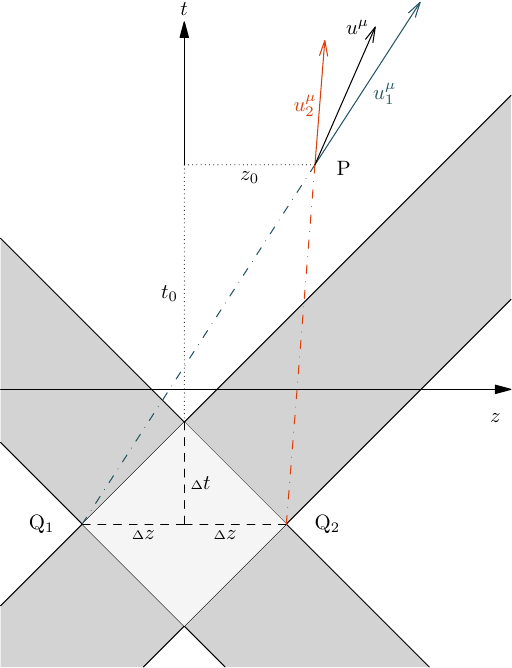}
    \caption{Model of Bjorken flow being produced at different locations in the collision region diamond and added up at the point $\mathrm{P}$. The coordinates $\Delta t$ and $\Delta z$ parametrize the collision region diamond (light gray) starting at the top corner. The point $\mathrm{P}$ is offset from this corner by $(t_0,z_0)$. The origin of the $t$-$z$-coordinate system is independent of the collision region diamond. Figure adapted from \cite{Ipp:2024ykh}.}
    \label{fig:sub_bjorken}
\end{figure}
The essence of this model is shown in Figure~\ref{fig:sub_bjorken}.
Everywhere within the collision region diamond, Bjorken flow is produced as if a boost-invariant collision (without transverse dynamics) had taken place at that location. The resulting flow is added up at the spacetime point $\mathrm{P}$.
We parametrize $\mathrm{P}$ using the coordinates $(t_0, z_0)$, which originate at the top corner of the collision region diamond. Results can then be translated from $(t_0, z_0)$ to $(t,z)$, where the latter are coordinates with respect to an arbitrary origin that does not have to coincide with the top corner of the collision region. Consider the flow produced at $\mathrm{Q}_1 = (-\Delta t, -\Delta z)$ and $\mathrm{Q}_2 = (-\Delta t, \Delta z)$. To add up their contributions to the total flow at $\mathrm{P}$, we need to weigh them with $\epsilon_{1/2}=\epsilon_0 / \tau_{1/2}$, where $\tau_{1/2}$ is the proper time elapsed between $\mathrm{Q}_{1/2}$ and $\mathrm{P}$. The value $\epsilon_0$ does not matter, but the difference in proper time is important. We compute
\begin{align}
\label{eq:eps_1_u_1}
    \epsilon_1 u_{1/2}^t &= \epsilon_0\frac{t_0 + \Delta t}{(t_0+\Delta t)^2 - (z_0\pm \Delta z)^2},\\
    \label{eq:eps_2_u_2}
    \epsilon_2 u_{1/2}^z &= \epsilon_0\frac{z_0 \pm \Delta z}{(t_0+\Delta t)^2 - (z_0\pm \Delta z)^2}.
\end{align}
Then, $\epsilon u^\mu = \epsilon_1 u_1^\mu + \epsilon_2 u_2^\mu$ is the weighted sum of these two contributions at $P$. We eliminate $\epsilon_0$ by considering the ratio $u^t/u^z$, which satisfies the inequality
\begin{align}
\label{eq:ut_uz_inequality}
    \frac{u^t}{u^z} = \frac{\frac{t_0 + \Delta t}{(t_0+\Delta t)^2 - (z_0+ \Delta z)^2}+\frac{t_0 + \Delta t}{(t_0+\Delta t)^2 - (z_0- \Delta z)^2}}{\frac{z_0 + \Delta z}{(t_0+\Delta t)^2 - (z_0+ \Delta z)^2}+\frac{z_0 - \Delta z}{(t_0+\Delta t)^2 - (z_0- \Delta z)^2}}> \frac{t_0}{z_0}
\end{align}
for positive $z$. We implicitly assumed $t_0 > z_0$ (i.e.,\ that $\mathrm{P}$ is not inside the tracks of the nuclei) and $\Delta t > \Delta z$ (which is true for all contributions from inside the collision region diamond) in evaluating the inequality \eqref{eq:ut_uz_inequality}.
The case $z<0$ is analogous with the \say{$>$} replaced by \say{$<$}. What we find is that the sum from the two ($z$-symmetric) Bjorken flow contributions, properly taking into account energy density, leads to a combined flow where $|u^t/u^z|$ is larger than it would be for Bjorken flow. This argument holds for any pair of points in the collision region, and we thus conclude that the total flow at P must have negative $u^\eta$ at positive $z_0$. If P were located at negative $z_0$ or, equivalently, at negative $\eta_s$, then $u^\eta$ would be positive. Overall, the longitudinal expansion of the system is slower than for Bjorken flow.\par
We can formalize this argument further and consider the result of contributions from all over the collision region diamond. That is, we consider the integral
\begin{align}
\label{eq:eps_u_integral}
    \epsilon_\mathrm{tot} u^\mu_\mathrm{tot} = \intop_{\scalebox{1.2}{$\diamond$}} \dd\Delta t\, \dd\Delta z\, \epsilon(\Delta t, \Delta z)\, u ^\mu(\Delta t, \Delta z),
\end{align}
where the $\epsilon(\Delta t, \Delta z)u^\mu(\Delta t, \Delta z)$ are the analogues of Eqs.~\eqref{eq:eps_1_u_1} and \eqref{eq:eps_2_u_2} changing continuously over the collision region \scalebox{1.5}{$\diamond$}. These velocities, and consequently the result of the integral, depend on $t_0$ and $z_0$.
After transforming to Milne coordinates and normalizing the total velocity, we obtain a result $u^\eta_\mathrm{tot}(\eta_s)$ for fixed $\tau$. We solve \eqref{eq:eps_u_integral} analytically and show the resulting curve as solid lines in Figure~\ref{fig:longitudinal_flow}.
The curves lie almost perfectly on top of the dilute Glasma simulation results. We conclude that the observed longitudinal flow is largely due to the extended collision region in the $t$-$z$-plane and the offset in the origin of the Milne frame as captured by our simple model. 
The size of the collision region in the model is set to twice the Lorentz contracted Woods-Saxon radius, and the offset in the origin of the Milne frame is the same as that in the numerical Glasma simulation.\par
\section{Time evolution}
\begin{figure}
        \centering
        \includegraphics{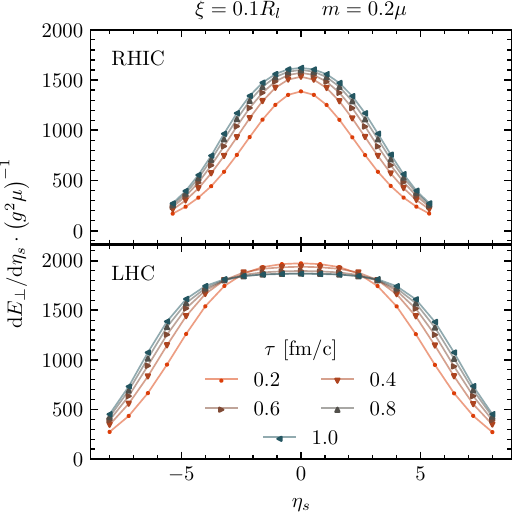}
    \caption{Differential transverse energy $\dd{E_\perp}/\dd{\eta_s}$ as a function of $\eta_s$ for different proper times $\tau$. Curves are averages over 10 independent collision events at RHIC and LHC energy. Initially, the profile changes as a function of $\tau$ but stabilizes for late times.
    Original figure from \cite{Ipp:2024ykh}.}
    \label{fig:time_evolution}
\end{figure}
Next, we study the time evolution of the energy distribution in the (3+1)D dilute Glasma. In Figure~\ref{fig:time_evolution} we analyze the differential transverse energy
\begin{align}
    \dv{E_\perp}{\eta_s}\coloneqq \tau \intop_\xperp \left(
    T^{xx}(\tau, \eta_s, \xperp) + T^{yy}(\tau, \eta_s, \xperp)
    \right).
\end{align}
We integrate over the transverse plane to capture all the energy per rapidity slice regardless of transverse expansion.
Note also that we have multiplied with a factor $\tau$, which accounts for the fact that the size of a unit $\eta_s$-interval scales with $\tau$. This factor compensates for the expected leading order time dependence due to the longitudinal expansion. In simulations based on the MV model, which is boost-invariant and features no transverse expansion, the energy density per unit rapidity scales with $1/\tau$ to good agreement at late times \cite{Lappi:2003bi, Lappi:2006hq}. If $u^\tau = 1$ everywhere in the Glasma, the energy density would exhibit perfect $1/\tau$ behavior. Deviations from this behavior can come from two sources. The first is transverse flow.
Even in the case of periodic boundary conditions, fluctuations in the initial conditions can lead to transverse flow, which decreases $u^\tau$. Our system has a finite size and consequently expands in transverse direction. This further increases transverse flow in the system (see Section~\ref{sec:transverse_flow}).
The second cause is the emergence of $u^\eta$ from the breaking of boost invariance. As was shown in Section~\ref{sec:longitudinal_flow}, this is in large part due to the extended collision region in the $t$-$z$-plane and the single choice of origin for the Milne frame.
The existence of longitudinal flow is a unique feature of the (3+1)D Glasma and might alter the time evolution of the energy density.\par
In Figure~\ref{fig:time_evolution} we show $\dd{E_\perp}/\dd{\eta_s}$ as a function of $\eta_s$ for different proper times $\tau$.
Notably, at early times, the profile changes as a function of $\tau$, which hints at a different time dependence than $1/\tau$.
Note, however, that for early times, this deviation from $1/\tau$ was also observed in the boost-invariant case \cite{Lappi:2003bi}.
At late times, the profiles appear to stabilize, again, similar to the boost-invariant case.
We conclude that neither the transverse nor the longitudinal flow in the (3+1)D Glasma significantly alters the $1/\tau$ behavior at late times.
The latter can be understood from the fact that as $\tau$ increases, the extent of the collision region becomes increasingly insignificant in comparison.\par
The effect of 3D initial conditions on the time evolution of the Glasma appears to be small.
We study the effect of the extended collision region on the time evolution more rigorously by employing a similar analysis to Section~\ref{sec:longitudinal_flow} and the model depicted in Figure~\ref{fig:sub_bjorken}. We consider bits of energy density produced everywhere in the collision region diamond and expanding with $u^\tau=1$. The resulting energy density at a point in the forward lightcone will be smaller by a factor of $1/\tau$, where $\tau$ is the proper time distance between the production and detection points. We only consider the resulting energy density at mid-rapidity, where $t=\tau$. Therefore, we need to integrate contributions like
\begin{align}
    \frac{\epsilon_0}{\sqrt{(t_0+\Delta t) - (\Delta z)^2}}.
\end{align}
We integrate these contributions from the entire collision region. The diameter of the collision region in $t$- and $z$-direction is twice the Lorentz contracted Woods-Saxon radius. The energy density at a point $(\tau, \eta_s=0)$ in the forward lightcone is then proportional to
\begin{align}
\label{eq:eps_diamond_int}
&\quad\intop_{\scalebox{1.2}{$\diamond$}} \dd\Delta t\, \dd\Delta z\, 
\frac{\epsilon_0}{\sqrt{(\tau+\delta t - R/\gamma+\Delta t)^2 - (\Delta z)^2}}\nn\\
&\propto
4\left(\tau + \delta t- \sqrt{(\tau + \delta t)^2 - R^2/\gamma^2}\right).
\end{align}
We have used $t_0 =  t + \delta t - R/\gamma = \tau + \delta t - R/\gamma$, which takes into account that the diameter of the collision region diamond is $R/\gamma$ but the origin of the coordinate system is shifted by $\delta t$ from the center of the collision region.
To see how the $1/\tau$ behavior hides in \eqref{eq:eps_diamond_int} we expand in powers of $1/\tau$ and find
\begin{align}
    \frac{2R^2}{\gamma^2}\left(\frac{1}{\tau} -\frac{\delta t}{\tau^2} +\frac{\delta t^2 + R^2/(4\gamma^2)}{\tau^3} + \mathcal O (\frac{1}{\tau^4})\right)\eqqcolon \frac{2R^2}{\gamma^2}\frac{\epsilon(\tau)}{\epsilon_0'},
\end{align}
where we have defined the $\tau$-dependent energy density $\epsilon(\tau)$, which is normalized with some (irrelevant) constant $\epsilon'_0$.
The prefactor $2R^2/\gamma^2$ is just the area of the collision region diamond, and the produced energy must be proportional to it. However, if we consider collisions with differently sized collision regions and the same produced energy, we need to normalize by dividing this prefactor off. Inside the parentheses, the expected $1/\tau$ behavior is clearly the dominant contribution if $\tau \gg R/\gamma \approx \delta t$. However, there are subleading corrections at finite values of $\tau$.\par
\begin{figure}
        \centering
        \includegraphics{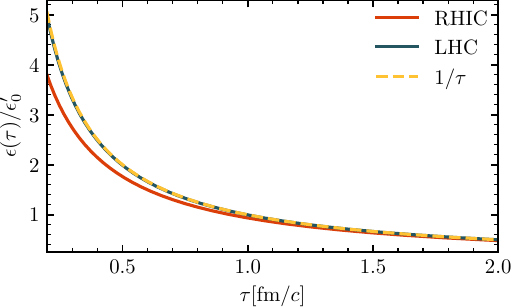}
    \caption{Time evolution of the energy density at mid-rapidity in the Glasma according to a simple model taking into account the extended collision region and disregarding transverse dynamics. Curves for RHIC and LHC energy are compared to the ideal $1/\tau$ behavior.}
    \label{fig:time_evolution_model}
\end{figure}
We study the falloff of the energy density with time at mid-rapidity according to this model in Figure~\ref{fig:time_evolution_model} and compare to the ideal $1/\tau$ behavior. The difference between RHIC and LHC energies is mainly in the Lorentz contraction factor, which leads to collision regions of different sizes.
As a consequence, the time evolution for RHIC energy deviates from the ideal $1/\tau$ behavior at early times. In contrast, the LHC curve falls on top of the $1/\tau$ curve, indicating that it is close to the boost-invariant case. At late times, the curves become indistinguishable as the leading $1/\tau$ factor takes over.\par
We conclude that the time evolution of the (3+1)D dilute Glasma roughly follows the boost-invariant Glasma. Small deviations due to the extended collision region in the $t$-$z$-plane occur at intermediate times and low collision energies. 
\section{Limiting fragmentation}
\begin{figure}
    \centering
    \includegraphics[scale=1]{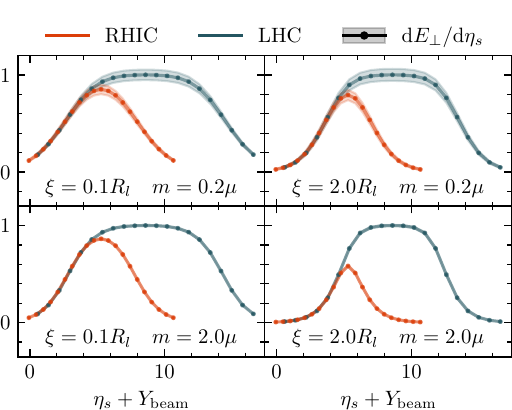}
    \caption{Differential transverse energy $\dd{E_\perp}/\dd{\eta_s}$ at fixed proper time $\tau=0.4\,\mathrm{fm}/c$ as a function of spacetime rapidity $\eta_s$. Curves are normalized to the LHC value at $\eta_s = 0$ for each combination of longitudinal correlation length $\xi$ and infrared regulator $m$. The rapidities are shifted by the beam rapidities $Y_\mathrm{beam}$ at RHIC and LHC energy, respectively. Results are averaged over 10 central collisions with error bands corresponding to one standard deviation in the event-by-event fluctuations. Figure adapted from \cite{Ipp:2024ykh}.}
    \label{fig:limiting_fragmentation}
\end{figure}
We study limiting fragmentation in our numerical results in Figure \ref{fig:limiting_fragmentation}. The flanks of the transverse energy profiles as a function of spacetime rapidity $\eta_s$ at different collision energies $\sqrt{s_{NN}}$ clearly overlap when the profiles are shifted by the respective beam rapidity $Y_\mathrm{beam}=\arcosh(\gamma)$. In regions of extreme rapidity (the fragmentation region), the rapidity profile does not depend on collision energy. While the exact shape of the profile varies, this claim holds for all studied combinations of longitudinal correlation length $\xi$ and infrared regulator $m$.\par
We expect this to be the case from theoretical considerations involving our analytic expressions for the field strength tensor (see Section~\ref{sec:limiting_fragmentation}). However, in these considerations, we implicitly assumed $\eta_s \gg 0$, and the regime of validity for this assumption is not entirely clear. From Figure~\ref{fig:limiting_fragmentation}, we see that the profiles for RHIC and LHC energies overlap close up to the region where the peak forms in the RHIC profiles and the plateau develops in the LHC curves. Notably, we see that the increase of collision energy does not lead to a uniform stretching of the profiles in rapidity but can, to good approximation, be understood as cutting the profile at mid-rapidity, leaving the flanks intact, and inserting a plateau, whose width depends on the collision energy.
\section{Eccentricity}
\begin{figure}
\centering
    \includegraphics[scale=1]{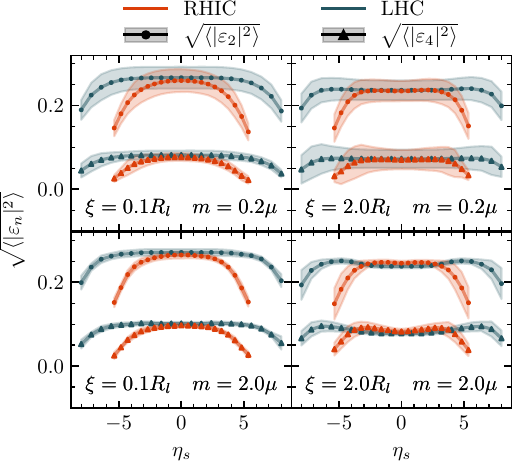}
    \caption{Eccentricities $\sqrt{\langle|\varepsilon_2|^2\rangle}$ and $\sqrt{\langle |\varepsilon_4|^2\rangle}$ as a function of spacetime rapidity $\eta_s$ at fixed proper time $\tau=0.4\,\mathrm{fm}/c$. The impact parameter was fixed in $x$-direction with a magnitude of $b=R$. Here, $\langle \cdot \rangle $ refers to an average over 10 events, and the error bands represent one standard deviation in the event-by-event fluctuations. Original figure from \cite{Ipp:2024ykh}.}
    \label{fig:epsilon_2_4}
\end{figure}
\begin{figure}
\centering
    \includegraphics[scale=1]{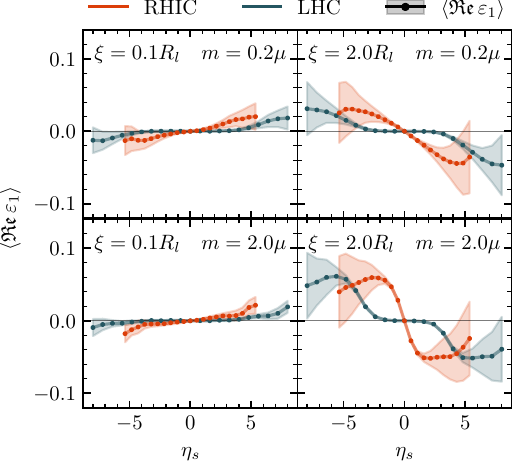}
    \caption{Real part of the eccentricity $\varepsilon_1$ as a function of spacetime rapidity $\eta_s$ at fixed proper time $\tau=0.4\,\mathrm{fm}/c$. The impact parameter was fixed in $x$-direction with a magnitude of $b=R$, and $\langle \cdot \rangle $ refers to an average over 10 events. The error bands represent one standard deviation in the event-by-event fluctuations. Original figure from \cite{Ipp:2024ykh}.}
    \label{fig:epsilon_1}
\end{figure}
We investigate the transverse structure of the local rest frame energy density $\elrf$ by studying eccentricity coefficients
\begin{align}
    \varepsilon_n(\tau,\eta_s)= 
    \frac{
        \int_\xperp \gamma(\tau, \eta_s, \xperp)\ \elrf(\tau, \eta_s, \xperp)\ \bar{r}^n \exp(\ii n\bar\phi)
        }{
        \int_\xperp \gamma(\tau, \eta_s, \xperp)\ \elrf(\tau, \eta_s, \xperp)\ \bar r^n,
    }  
\end{align}
where $\gamma(\tau, \eta_s, \xperp) = u^\tau(\tau, \eta_s, \xperp)$ is the local Lorentz factor, not to be confused with the Lorentz contraction factor related to the beam energy. The inclusion of this factor is a matter of convention and puts more emphasis on areas where the local flow velocity deviates from Bjorken flow. Also,
\begin{align}
    \bar r=\sqrt{(x-x_0)^2 + (y-y_0)^2}
\end{align}
and
\begin{align}
    \bar \phi=\arctan\frac{y-y_0}{x-x_0}
\end{align}
are polar coordinates in the center of mass frame. In general, the center of mass
\begin{align}
\xperp_0(\tau, \eta_s) &= \frac{\int_\xperp \gamma(\tau, \eta_s, \xperp)\ \elrf(\tau, \eta_s, \xperp)\ \xperp}{\int_\xperp \gamma(\tau, \eta_s, \xperp)\ \elrf(\tau, \eta_s, \xperp) }
\end{align}
at a given time depends on rapidity. However, we evaluate it at mid-rapidity, $\eta_s = 0$, and use the obtained value for all rapidity slices.\par
We show our results for the rapidity dependence of $\sqrt{\langle |\varepsilon_2|^2\rangle}$ and $\sqrt{\langle |\varepsilon_4|^2\rangle}$ in Figure \ref{fig:epsilon_2_4}.
The impact parameter was chosen as $b=R$ and aligned with the $x$-direction.
This leads to the typical \say{almond}-shape in the overlap region of the two nuclei and of the deposited energy density (see Figure~\ref{fig:elrf_slices}).
Consequently, we expect large contributions to $\varepsilon_2$ and a considerable $\varepsilon_4$, both of which can be seen in Figure \ref{fig:epsilon_2_4}. We show values at RHIC and LHC energies and for different combinations of the longitudinal correlation length $\xi$ and infrared regulator $m$. The results at LHC energy exhibit the typical plateau structure expected from large collision energies. At RHIC energy, there seems to be a plateau for $\xi = 2.0R_l$ as well, but it is certainly narrower than at LHC energy. At extreme rapidities, the eccentricities drop off, which could be an indication that fluctuations dominate the transverse energy distribution in that regime. The values of $\varepsilon_2$ and $\varepsilon_4$ at mid-rapidity seem to be mostly independent of $\xi$ and $m$.\par
In Figure \ref{fig:epsilon_1}, we study the real part of $\varepsilon_1$. Since we align the impact parameter with the $x$-axis, we do not expect contributions to the imaginary part of $\varepsilon_1$ (except from fluctuations, which cancel on average). Furthermore, owing to the symmetry of the collision, we expect no net contribution to $\varepsilon_1$ at all at mid-rapidity. This can be clearly seen in the data shown in Figure \ref{fig:epsilon_1}.
For a longitudinal correlation length $\xi = 0.1R_l$, which roughly corresponds to the typical size of nucleons within the nucleus, there is more energy density in regions of positive $x$ at positive rapidity $\eta_s$. For negative rapidity, there is more energy density in regions of negative $x$.
This skew of the energy density in the $x$-$\eta_s$-plane is what one might expect from the fact that the nucleus moving in positive $z$-direction is shifted to positive $x$ by the impact parameter if one imagines the nucleus to \say{drag} the Glasma with it. Curiously, the situation is reversed for $\xi = 2.0R_l$, i.e.,\ longitudinally coherent nuclei.
In the positive $\eta_s$ slices, there is more energy density at negative $x$. This establishes $\varepsilon_1$ as a parameter that is particularly sensitive to the longitudinal structure of nuclei, which, in the future, could shed light on the actual size of longitudinal correlations in real nuclei.
\chapter{Proton-proton collisions}
\label{ch:pp}
A real nucleus is not a homogeneous mass of nuclear matter but consists of individual nucleons. For the interactions taking place in heavy-ion collision experiments, the distinction between neutrons and protons is negligible on the surface level. A heavy nucleus can, therefore, be assumed to be made up of a number of protons corresponding to its mass number, e.g.,\ 208 for the lead isotope used at LHC \cite{Roland:2014jsa}. In the CGC framework, the total color charge of a nucleus can be modeled as a sum of color charges for the individual nucleons, i.e.,
\begin{align}
    \rho_{A/B} = \sum_i \rho_{A/B}^i,
\end{align}
where $\rho_{A/B}^i$ denotes the color charge density of the $i$-th nucleon in nucleus $A$ or $B$.
Since the Poisson equation \eqref{eq:poisson} is linear, this property carries over to the color fields of the individual nuclei before the collision. From the structure of the dilute Glasma field strength tensor \eqref{eq:fpm}--\eqref{eq:fij}, which is bilinear in the single-nucleus fields, it is then clear that the field strength tensor for a given nucleus-nucleus collision can be expressed as a sum over binary nucleon-nucleon collisions, i.e.,
\begin{align}
\label{eq:f_as_linear_sum}
    f^{\mu\nu}[\rho_A,\rho_B] = \sum_i\sum_j f^{\mu\nu}[\rho_A^i, \rho_B^j].
\end{align}
The double sum in \eqref{eq:f_as_linear_sum} effectively runs over all pairs of nucleons.
However, some of these nucleon pairs will be separated by a large impact parameter in the transverse plane.
In an isolated nucleon-nucleon collision experiment, such nucleon pairs would have a very low probability of interacting.
Thus, it makes sense to include this interaction probability in the model and reduce the sum in \eqref{eq:f_as_linear_sum} to those nucleon pairs that are determined to interact, given their transverse separation as will be explained in the following section.\par
A plethora of experimental data are available for proton-proton and proton-antiproton collisions.
These data guide the treatment of proton-proton collisions in the dilute Glasma.
The ultimate goal is then to reconstruct proton-nucleus and nucleus-nucleus collisions in the dilute Glasma as a superposition (on the level of the field strength tensor) of individual proton-proton collisions.
\section{Color charge correlator}
We assume that the color charges of two colliding nucleons $A$ and $B$ are distributed according to a Gaussian probability functional, just like the color charges of nuclei in the previous section. We demand the 1-point function to vanish,
\begin{align}
    \langle \rho_{A/B}^a(x^\pm, \xperp)\rangle = 0.
\end{align}
For the 2-point function, we employ \eqref{eq:correlator_general}, or, equivalently, \eqref{eq:correlator_factorized} with a Gaussian profile $T$, i.e.,
\begin{align}
\label{eq:correlator_single_nucleon}
    &\langle \rho^a_{A/B}(x^\pm,\xperp)\rho^b_{A/B}(y^\pm, \yperp)\rangle\nn\\
    &\hspace{2cm}= g^2\mu^2\delta^{ab}\frac{1}{\sqrt{2\pi}s_l}\ee^{-\frac{(x^\pm+y^\pm)^2}{8s_l^2}}\ee^{-
    \vphantom{\frac{(x^\pm+y^\pm)^2}{8s_l^2}}
    \frac{(\xperp+\yperp)^2}{8s^2}}\frac{1}{\sqrt{2\pi}\xi}\ee^{-
    \vphantom{\frac{(x^\pm+y^\pm)^2}{8s_l^2}}
    \frac{(x^\pm-y^\pm)^2}{2\xi^2}}\delta^{(2)}(\xperp-\yperp)
\end{align}
with the width parameter $s$ and its boosted version $s_l = s/(\sqrt{2}\gamma) $. A notation commonly used in the literature is $B_G = s^2$. This nucleon model still allows longitudinal fluctuations through the parameter $\xi$.
\section{Cross section}
In principle, the color charge densities $\rho_{A/B}$ of nucleons in our model have infinite spatial extent.
This means that a proton on Earth and a proton on the moon should, according to our model, have overlapping color fields and, therefore, interact and produce a very small but nonzero amount of energy density.
In practice, we put nucleons on a discrete lattice in boxes with finite extent, giving them finite support.
Consequently, nucleons can only interact if, when viewed head-on, their boxes overlap in the transverse plane.
This introduces an arbitrary cutoff on the allowed impact parameter between two nucleons.
While it is true that ultra-peripheral collisions will contribute very little to the overall energy density, it is desirable to have a precise notion of whether or not a collision takes place.
This allows us to do statistics over collision events and compare to experimental results.\par
We follow the procedure for determining nucleon-nucleon collision events outlined in \cite{dEnterria:2010xip}.
Our nucleons are essentially just fluctuations of color charge whose magnitude in the transverse direction is governed by the enveloping function
\begin{align}
\label{eq:nucleon_thickness}
    t(\xperp) = \frac{1}{2\pi s^2}\exp\left(-\frac{\xperp^2}{2s^2}\right),
\end{align}
called a thickness function,
as is evident from \eqref{eq:correlator_single_nucleon}.
We define the overlap function of two nucleons separated in the transverse direction by $\bperp$ as
\begin{align}
    \Theta(b) = \int \dd[2]{\xperp}\,t(\xperp+\bperp/2)t(\xperp-\bperp/2) = \frac{1}{4\pi s^2}\exp\left(-\frac{b^2}{4 s^2}\right),
\end{align}
with $b=|\bperp|$. Introducing a partonic cross section $\sigma_{gg}$ and a parton number $N_g$ per nucleon, we define the (average) number of parton-parton collisions
\begin{align}
    N_{gg} := N^2_g \sigma_{gg} \Theta(b).
\end{align}
Assuming that the number of parton-parton collisions follows a Poisson distribution with mean $N_{gg}$, the probability of at least one interaction is then
\begin{align}
\label{eq:pp_collision_probability}
    P(b) = 1-\exp\left(-N_g^2\sigma_{gg} \Theta(b)\right).
\end{align}
For a collision of solid disks, this function would be 1 in a circle the size of the cross section and 0 outside. The expression \eqref{eq:pp_collision_probability} can be seen as a generalization to that configuration, where the collision probability is smeared out and can take values between 0 and 1. Analogously to the example of solid disks, the cross section for a nucleon-nucleon collision $\sigma_{NN}$ can be found by integrating \eqref{eq:pp_collision_probability} over all impact parameters, i.e.,
\begin{align}
    \sigma_{NN} = \int\dd[2]{\bperp}P(b).
\end{align}
The cross section $\sigma_{NN}$ can also be viewed as the normalization factor, i.e.,\ the denominator, in the differential interaction probability
\begin{align}
\label{eq:differential_pp_ia_probability}
\dv[2]{P}{\bperp}(b) = \frac{1-\exp\left(-N_g^2\sigma_{gg} \Theta(b)\right)}{\int \dd[2]{\bperp} \left(1-\exp\left(-N_g^2\sigma_{gg} \Theta(b)\right)\right)}.
\end{align}
In practice, $\sigma_{NN}$ has to be equated to the experimental result at the corresponding collision energy $\sqrt{s}$. This fixes $N_g^2\sigma_{gg}$ for a given $\sqrt{s}$, and no further assumption is required about these parameters.\par
\begin{figure}
    \centering
    \includegraphics[scale=1]{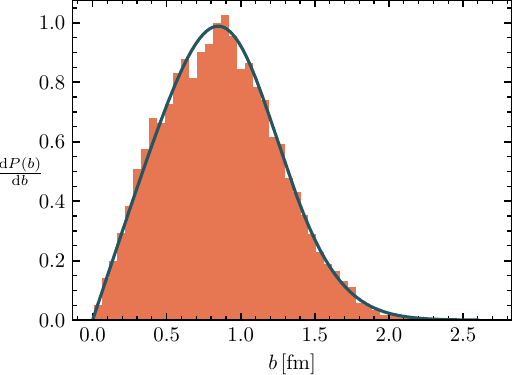}
    \caption{Distribution of impact parameters $b=|\bperp|$ in proton-proton collisions at $200\,\mathrm{GeV}$ with $\sigma_{NN}$ obtained from \eqref{eq:sigma_inel_interpolation}. The analytic distribution function $\mathrm{d}P(b)/\mathrm{d}b$ is shown in blue, and a histogram of $10^4$ randomly drawn values for $b$ is shown in orange.}
    \label{fig:dPdb}
\end{figure}
The process of randomly sampling a number $N_\mathrm{ev}$ of nucleon-nucleon collision events with arbitrary impact parameters is then straightforward. First, sample an impact parameter $\bperp$ from a two-dimensional uniform distribution. We simplify this process by only sampling the modulus $b$ from a linear distribution up to some maximum impact parameter $b_\mathrm{max}$, which has to be chosen large enough such that only a negligible amount of events are affected by the cutoff. A reasonable requirement is $N_\mathrm{ev}P(b_\mathrm{max})\ll 1$. For each impact parameter 
$b$ sampled this way, the corresponding event is accepted with probability $P(b)$. The resulting distribution in terms of the modulus $b=|\bperp|$ is
\begin{align}
\label{eq:dPdb}
    \dv{P}{b} = N\,b\left(1-\exp\left(-N_g^2\sigma_{gg} \Theta(b)\right)\right)
\end{align}
with a normalization constant $N$. Consequently, the distribution of $b$ features a linear rise for small $b$ but gets suppressed exponentially on the large end. The distribution \eqref{eq:dPdb} is visualized in Figure~\ref{fig:dPdb} together with a histogram of $10^4$ values sampled from the distribution. Finally, the azimuthal angle is drawn from a uniform distribution.
\section{Charged particle multiplicity}
We simulate a large number of nucleon-nucleon collisions with impact parameters obtained from the procedure outlined in the previous section. 
In particular, we simulate $10^4$ collision events each for different values of the IR cutoff at $\sqrt{s}=200\,\mathrm{GeV}$.
The nucleon size used in the thickness function \eqref{eq:nucleon_thickness} (and consequently the interaction probability and impact parameter distribution) is $\sqrt{B_G}=s=0.395\,\mathrm{fm}$, a value that was obtained through fits to HERA data \cite{Rezaeian:2012ji}. In the correlator \eqref{eq:correlator_single_nucleon}
we use the value $\sqrt{B_G}=s=0.358\,\mathrm{fm}$ also used in \cite{Mantysaari:2016jaz, Demirci:2022wuy} instead. For simplicity, we assume individual nucleons to be coherent, i.e., $\xi=2 s_l$. The inelastic proton-proton cross section at a given collision energy is obtained from the interpolation formula \cite{McLerran:2015qxa}
\begin{align}
\label{eq:sigma_inel_interpolation}
    \sigma_{NN} &= \big[2.52 + 0.005\,\ln(\sqrt{s}/\mathrm{GeV})+0.056\,\ln(\sqrt{s}/\mathrm{GeV})^2\nn\\
    &\qquad +4.52\,\big(\sqrt{s}/\mathrm{GeV}\big)^{-0.9}+3.38\,\big(\sqrt{s}/\mathrm{GeV}\big)^{-1.1}\big]\,\mathrm{fm}^2.
\end{align}
We take the event-by-event values for the local rest frame energy density $\elrf$ at mid-rapidity $\eta_s = 0$ and sum over the transverse plane. The resulting energy is used as a proxy for the charged particle multiplicity $N_\mathrm{ch}$ measured in experiments. 
\begin{figure}
    \centering
    \includegraphics[scale=1]{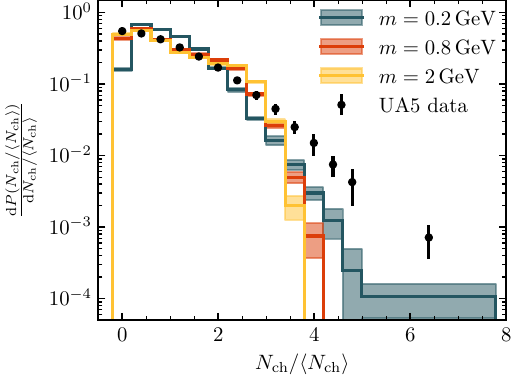}
    \caption{Distribution of charged particle multiplicity $N_\mathrm{ch}$ divided by the average charged particle number at mid-rapidity at $\sqrt{s}=200\,\mathrm{GeV}$. Discrete points are experimental results of charged particle multiplicities in the $-0.5< \eta_s < 0.5$ rapidity range from proton-antiproton collisions performed by the UA5 collaboration \cite{UA5:1988gup, hepdata.15457.v1/t3}. The histograms represent the distribution of $\elrf$ as a proxy for $N_\mathrm{ch}$ in dilute Glasma simulations of nucleon-nucleon collisions with $10^4$ events for each value of the IR cutoff. Error bands are estimated using the jackknife method.}
    \label{fig:200_hist_unscaled}
\end{figure}
Figure \ref{fig:200_hist_unscaled} shows the results of these simulations compared to actual charged particle multiplicity data obtained by the UA5 collaboration for proton-antiproton collisions \cite{UA5:1988gup, hepdata.15457.v1/t3}.
Both the simulated and measured multiplicity values are normalized with respect to their respective expectation values and shown as a histogram.
Although there is a small difference between $m=0.2\,\mathrm{GeV}$ and the other choices of the IR cutoff, all histograms show that the dilute Glasma simulations underrepresent large multiplicity events.
This was already discovered in the context of IP-Glasma simulations of proton-proton collisions \cite{Schenke:2013dpa, McLerran:2015qxa}.
Large multiplicity events are important for several reasons.
In the study of proton-proton collisions and small systems in general, these large multiplicity events exhibit some signatures of collective behavior and are a crucial ingredient in studying the existence of the quark-gluon plasma in small systems.
In the context of building proton-nucleus collisions or nucleus-nucleus collisions from a superposition of proton-proton collisions, the large multiplicity events will have the biggest impact on the resulting energy density and should be faithfully represented in the collision. In conclusion, the dilute Glasma, just like the IP-Glasma, fails to reproduce an adequate number of large multiplicity events and needs to be modified to accurately reproduce the experimental multiplicity histograms.
\subsection{Color charge fluctuations}
It was shown in IP-Glasma simulations that allowing fluctuations in the ratio between $Q_s$ and $g^2 \mu$ can repair the multiplicity histograms of proton-proton collisions and restore the large multiplicity events observed in experiment \cite{Schenke:2013dpa}. This was further refined into specific fluctuations of the saturation momentum $Q_s$ drawn from the distribution \cite{McLerran:2015qxa}
\begin{align}
\label{eq:Qs_fluctuation_dist}
    P\left(\ln(Q_s^2/\langle Q_s^2\rangle )\right)=\frac{1}{\sqrt{2\pi}\sigma}\exp\left( -\frac{\ln^2(Q_s^2/\langle Q_s^2\rangle)}{2\sigma^2}\right).
\end{align}
The saturation momentum of each individual proton is rescaled with a factor $\sqrt{Q_s^2/\langle Q_s^2\rangle}$ drawn from \eqref{eq:Qs_fluctuation_dist}. 
We follow this procedure, using $Q_s \propto g^2\mu$.
Since the MV parameter $\mu$ sets the overall scale of color charges in the model \eqref{eq:correlator_single_nucleon}, a rescaling of $Q_s$ for each nucleon corresponds to a rescaling of the color charges $\rho_{A/B}$, which are proportional to $\mu$.
As the dilute Glasma field strength tensor \eqref{eq:fpm}--\eqref{eq:fij} is linear in $\rho_{A/B}$ and the energy-momentum tensor is essentially the square of the field strength tensor, we rescale $T^{\mu\nu}$ for each event with two independent numbers $Q_s^2/\langle Q_s^2\rangle$ drawn from \eqref{eq:Qs_fluctuation_dist}.
Note that such a rescaling can be applied a posteriori to already obtained computation results and, since the rescaling commutes with the eigenvalue problem, can also be applied to $\elrf$.
The width $\sigma$ of the distribution \eqref{eq:Qs_fluctuation_dist} was previously restricted to $\sigma \approx 0.5$ for proton-proton collisions in the literature \cite{McLerran:2015qxa, Bzdak:2015eii}.
Furthermore, $\sigma$ exhibits a weak dependence on the collision energy. In the context of dilute Glasma computations, we do not constrain $\sigma$ and choose the value that best reproduces the experimental multiplicity distribution.
To quantify this, we use the Kullback-Leibler divergence
\begin{align}
    D_\mathrm{KL}=\sum_i P_i \ln \left(\frac{P_i}{Q_i}\right)
\end{align}
for two discrete distributions $P$ and $Q$.
In our case, $P_i$ is the experimentally measured probability for the $i$-th multiplicity bin as shown in Figure~\ref{fig:200_hist_unscaled}. Then, $Q_i$ is the corresponding probability for the $i$-th bin resulting from the histogram of simulated collision events.
The matching of multiplicity bins between the experimental and simulation data, however, is somewhat ambiguous.
From the experiment, there are probabilities for discrete numbers of charged particles. Most bins contain only one such number, starting at zero particles.\footnote{Because we consider only the most narrow $-0.5 < \eta_s < 0.5$ rapidity range, there is the possibility of a collision taking place but zero charged particles being produced in this rapidity range.} For the largest number of produced particles, several counts are lumped together in a single bin.
Either way, particle numbers are discrete integer numbers\footnote{In Figure~\ref{fig:200_hist_unscaled} $N_\mathrm{ch}$ is rescaled with $\langle N_\mathrm{ch}\rangle$ and therefore datapoints are not integers.} and matching them to bins of continuous variables is ambiguous.
A natural choice is to extend intervals symmetrically. The bin containing 2 particles (before rescaling with the average particle number) would be matched to the $1.5$--$2.5$ range.
This is how data are visualized in Figure~\ref{fig:200_hist_unscaled}. The first interval then goes from $-0.5$ to $0.5$.
Even if a proxy like energy density is used for particle numbers in the simulation, the observable cannot become negative.
Although the leftmost bin formally ranges from $-0.5$ to $0.5$, contributions from the simulation data can only come from the $0$--$0.5$ range.
Therefore, the leftmost bin is an outlier, and we do not consider it when computing the Kullback-Leibler divergence between distributions.
\begin{figure}
    \centering
    \includegraphics[scale=1]{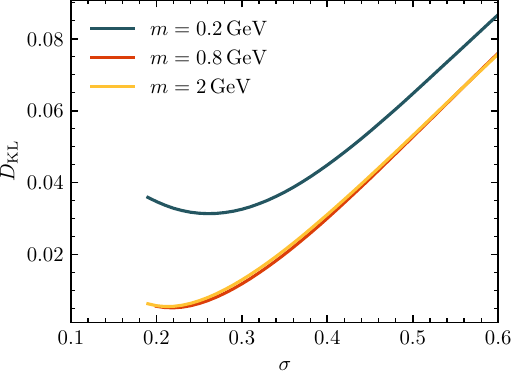}
    \caption{Kullback-Leibler divergence between experimental multiplicity distributions and dilute Glasma simulations with additional fluctuations for each nucleon following \eqref{eq:Qs_fluctuation_dist} with different values of $\sigma$. Experimental data are taken from the UA5 collaboration \cite{UA5:1988gup, hepdata.15457.v1/t3}. The dilute Glasma simulations were performed at $\sqrt{s}=200\,\mathrm{GeV}$ for different values of the IR cutoff. The first multiplicity bin containing zero is omitted in the computation of $D_\mathrm{KL}$. The curves do not extend to small values of $\sigma$ as fluctuations with these parameters can lead to multiplicity bins with zero counts, which cannot be accounted for using the Kullback-Leibler divergence.}
    \label{fig:KL}
\end{figure}
We then apply fluctuations to the energy-momentum tensor of the simulated events, drawing random numbers from \eqref{eq:Qs_fluctuation_dist} as explained before. We allow for different values of the $\sigma$ parameter and compute the Kullback-Leibler divergence for each.
The results are shown in Figure~\ref{fig:KL} and the minima of the $D_\mathrm{KL}(\sigma)$ curves mark the optimal value $\sigma_\mathrm{opt}$ of $\sigma$.
The values of $\sigma_\mathrm{opt}$ for the different IR cutoffs can be found in Table~\ref{tab:sigma_opt}.\par
\begin{table}[htbp]
  \centering
    \begin{tabular}{r|lll}
    $m$     & $0.2\,\mathrm{GeV}$ & $0.8\,\mathrm{GeV}$ & $2\,\mathrm{GeV}$ \\
    \midrule
    $\sigma_\mathrm{opt}$ & 0.26     & 0.22     & 0.21 \\
    \end{tabular}%
  \caption{Optimal values for the $\sigma$ parameter in the distribution \eqref{eq:Qs_fluctuation_dist} when applying fluctuations to $\rho$ for nucleon-nucleon collisions in the dilute Glasma at $\sqrt{s}=200\,\mathrm{GeV}$.}%
  \label{tab:sigma_opt}%
\end{table}
\begin{figure}
    \centering
    \includegraphics[scale=1]{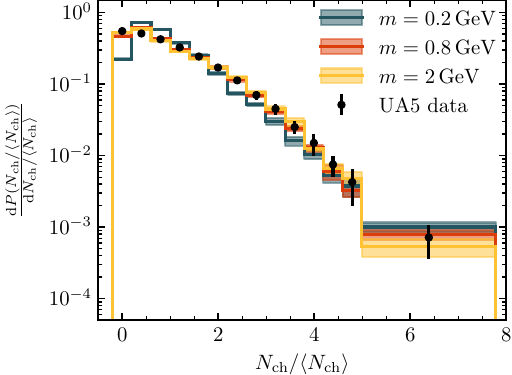}
    \caption{Distribution of charged particle multiplicity divided by the average charged particle number at mid-rapidity at $\sqrt{s}=200\,\mathrm{GeV}$. Discrete points are experimental results of charged particle multiplicities in the $-0.5< \eta_s < 0.5$ rapidity range from proton-antiproton collisions performed by the UA5 collaboration \cite{UA5:1988gup, hepdata.15457.v1/t3}. The histograms represent the distribution of $\elrf$ in dilute Glasma simulations of nucleon-nucleon collisions with $10^4$ events for each value of the IR cutoff. Fluctuations sampled from \eqref{eq:Qs_fluctuation_dist} were applied to the color charge of each nucleon. The optimal $\sigma$ values are taken from Table~\ref{tab:sigma_opt}. Error bands are estimated using the jackknife method.}
    \label{fig:200_hist_Qs_scaled}
\end{figure}
We perform rescalings of the simulation data shown in Figure~\ref{fig:200_hist_unscaled} with random numbers drawn from \eqref{eq:Qs_fluctuation_dist} and optimal $\sigma$ values from Table~\ref{tab:sigma_opt}. The resulting histograms are shown in Figure~\ref{fig:200_hist_Qs_scaled}. Clearly, the large multiplicity events are restored, and the histograms fit the experimental data much better than in Figure~\ref{fig:200_hist_unscaled}. Also note that in Figure~\ref{fig:200_hist_Qs_scaled}, the histogram for $m=0.2\,\mathrm{GeV}$ fits the experimental data slightly worse than larger values of the IR cutoff.
Note that the values $\sigma_\mathrm{opt}\approx 0.25$ are approximately half of the values $\sigma \approx 0.5$ used in the literature. In terms of the distribution \eqref{eq:Qs_fluctuation_dist}, sampling values of $Q_s^2/\langle Q_s^2\rangle$ with $\sigma = 0.25$ is the same as sampling values of $\sqrt{Q_s^2/\langle Q_s^2\rangle}$ with $\sigma = 0.5$. That means we could use a similar $\sigma$ as the literature in the dilute Glasma if we only rescale $\mu$ with the square root of the sampled rescaling factor. It is, however, not clear why this would be the correct prescription in the dilute Glasma.
\subsection{Protons consisting of $n_v = 3$ valence quarks}
Another source of fluctuations on the level of individual nucleons can be introduced by allowing for individual hot spots in a nucleon. Motivated by the idea of valence quarks, we choose the number of these hot spots to be $n_v=3$.
\begin{figure}
    \centering
        \includegraphics[scale=0.8]{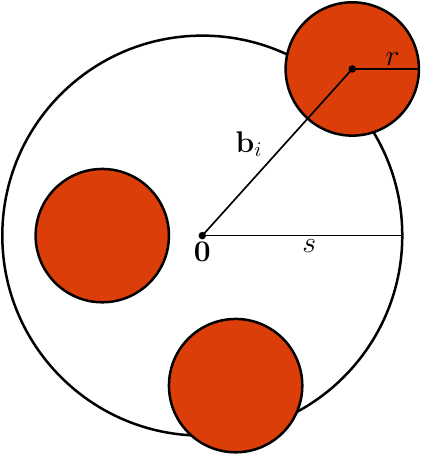}
    \caption{Two-dimensional projection of the hot spot model for a nucleon featuring $n_v=3$ hot spots. The point $\mathbf{0}$ marks the center of the nucleon that hot spot positions are sampled around. The parameter $s$ governs the size of the nucleon and determines the positioning of hot spot centers. The hot spot radius is determined by the parameter $r$. The position of each hot spot relative to $\mathbf 0$ is $\mathbf{b}_i$ with $i\in (1,2,3)$. }
    \label{fig:hot_spot_model}
\end{figure}
In Figure~\ref{fig:hot_spot_model}, we show schematically the transverse structure of a nucleon in the hot spot model. Three hot spots are placed around the center $\mathbf{0}$ according to some distribution governed by the nucleon size parameter $s$. The vector $\mathbf{b}_i$ describes the position of the $i$-th hot spot relative to the center. The parameter $r$ governs the size of individual hot spots. Note that the center does, in general, not correspond to the center of mass, i.e.,\ the average position of the hot spots. This issue will be addressed later. Also, note that there is equivalent structure in the longitudinal direction, taking into account the Lorentz contraction of particles moving at relativistic speed.\par
We formalize the hot spot model by specifying the 1- and 2-point functions of color charges.
As always, the 1-point function
\begin{align}
    \langle \rho^a(x^\pm, \xperp)\rangle = 0
\end{align}
of color charges vanishes. The 2-point function in the hot spot model is
\begin{align}
\label{eq:correlator_nucleon_hot_spot}
    &\langle \rho^a_{A/B}(x^\pm,\xperp)\rho^b_{A/B}(y^\pm, \yperp)\rangle\nn\\
    &= g^2\mu^2\delta^{ab}
    \frac{1}{\sqrt{2\pi}r_l}\frac{1}{\sqrt{2\pi}\xi}\ee^{-
    \vphantom{\frac{(x^\pm+y^\pm-2b^\pm_i)^2}{8r_l^2}}
    \frac{(x^\pm-y^\pm)^2}{2\xi^2}}\delta^{(2)}(\xperp-\yperp)
    \sum_{i=1}^{n_v}
    \ee^{-\frac{(x^\pm+y^\pm-2b^\pm_i)^2}{8r_l^2}}\ee^{-
    \vphantom{\frac{(x^\pm+y^\pm-2b^\pm_i)^2}{8r_l^2}}
    \frac{(\xperp+\yperp+2\bperp_i)^2}{8r^2}}
\end{align}
and the idea behind it is as follows: Each valence quark is modeled analogously to how a whole nucleon was modeled before, replacing $s\rightarrow r$ and $s_l\rightarrow r_l$. The parameters $r$ and $r_l$ are the quark (hot spot) radius in the transverse and longitudinal direction, respectively. The individual quarks are placed at the positions $(b^\pm_i, \bperp_i)$, which are drawn from a Gaussian distribution with width $s_l$ in longitudinal and $s$ in transverse direction. Thus, the parameters $s$ and $s_l$ still set the size of nucleons in the valence quark model. Note that the hot spot model still allows correlations governed by the parameters $\xi$ and $m$ for each individual hot spot.\par
We simulate $10^4$ collisions of nucleons sampled from the hot spot model at $\sqrt{s}=200\,\mathrm{GeV}$ and $2.5\times 10^4$ collisions at $\sqrt{s}=7000\,\mathrm{GeV}$. Just like before, $\sqrt{B_G}=s=0.395\,\mathrm{fm}$ is used to determine impact parameters.
There is now some ambiguity surrounding the impact parameter. The positions of hot spots are drawn relative to some virtual proton center. However, this proton center will not coincide with the center of mass of the $n_v$ hot spots after sampling. We consider the impact parameter to be the distance between the center of mass for each sampled proton.
If two nucleons are determined to interact, all hot spots in one nucleon interact with all hot spots in the other nucleon.
The distribution of hot spots in a nucleon is governed by $s = 0.416\,\mathrm{fm}$, and the radius of individual hot spots is $r=0.116\,\mathrm{fm}$. These values are taken from \cite{Mantysaari:2022ffw}, where a Bayesian analysis of HERA data under the assumption of a hot spot model with $n_v=3$ valence quarks per nucleon was performed.\par
\begin{figure}
    \centering
        \includegraphics{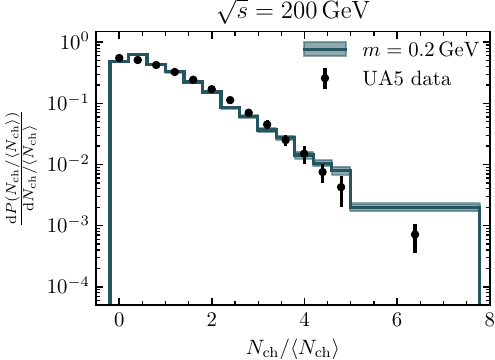}\\
        \vspace{0.5cm}
        \includegraphics{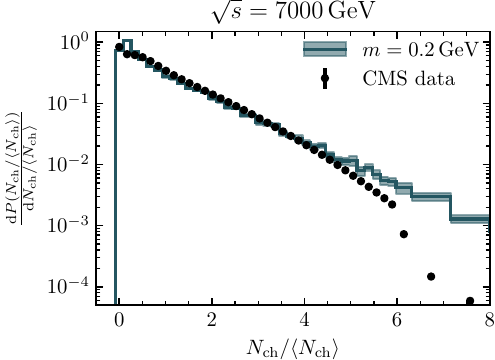}
    \caption{Distribution of charged particle multiplicity $N_\mathrm{ch}$ divided by the average charged particle number at mid-rapidity at $\sqrt{s}=200\,\mathrm{GeV}$ (top) and $\sqrt{s}=7000\,\mathrm{GeV}$ (bottom). Discrete points are experimental results of charged particle multiplicities in the $-0.5< \eta_s < 0.5$ rapidity range from proton-antiproton collisions performed by the UA5 collaboration \cite{UA5:1988gup, hepdata.15457.v1/t3} (top) and proton-proton collisions performed by the CMS collaboration \cite{CMS:2010qvf, hepdata.57909.v1/t12} (bottom). The histograms represent the distribution of $\elrf$ as a proxy for $N_\mathrm{ch}$ in dilute Glasma simulations of nucleon-nucleon collisions with $10^4$ (top) and $2.5\times 10^4$ (bottom) events. The IR cutoff is fixed to $m=0.2\,\mathrm{GeV}$, and the hot spot model was used for the color charges of nucleons. Error bands are estimated using the jackknife method.}
    \label{fig:hist_nv}
\end{figure}
The resulting histograms are shown in Figure~\ref{fig:hist_nv}. The simulation results at $\sqrt{s}=7000\,\mathrm{GeV}$ are compared to data measured by the CMS collaboration. For both collision energies, the simulation results fit the data better than in the simple Gaussian nucleon model without any rescalings shown in Figure~\ref{fig:200_hist_unscaled}. However, for the hot spot model nucleons, large multiplicity events are slightly overestimated.
\subsection{Nucleon center of mass}
As mentioned before, we draw hot spot positions within a nucleon from a Gaussian distribution, and we have to pick an arbitrary point as the nucleon center in this process. After sampling the $n_v$ hot spot positions, the average of these positions, i.e.,\ the nucleon center of mass, will, in general, not align with the previously selected center. For all practical purposes, it makes sense to consider the center of mass the actual center of the proton. However, the distribution of hot spot coordinates around the center of mass will be different than the distribution from which the positions were initially sampled. This problem was studied extensively in \cite{Mitchell:2016jio}. We illustrate some aspects in the following.\par
For each nucleon, we sample $n_v$ different hot spot positions from the three-dimensional distribution
\begin{align}
\label{eq:fv_gauss}
f_v(\vec{b})=
    f_v(b^\pm, \bperp) = \frac{1}{\sqrt{2\pi s_l^2}}\exp\left(-\frac{(b^\pm)^2}{2s_l^2}\right)\frac{1}{2\pi s^2}\exp\left(-\frac{\bperp^2}{2s^2}\right)
\end{align}
where the symbol $\vec{b}$ denotes the three-dimensional position vector $( b^\pm,\bperp)^T$ of the hot spot.
We consider the most relevant case $n_v=3$ with individual hot spot positions distributed with \eqref{eq:fv_gauss}. After sampling 3 hot spot positions, the coordinates of each hot spot are shifted by a vector $\vec{B}$, which corresponds to the center of mass of all 3 hot spots and, therefore, depends on all sampled positions. This means that the individual hot spots are not distributed independently. Instead, the collective distribution function is
\begin{align}
\label{eq:collective_hs_dist}
    f(\vec{b}_1, \vec{b}_2, \vec{b}_3) &=
    \left(\frac{2\pi}{3}\right)^{3/2}s^2s_l\,
    f_v(\vec{b}_1-\vec B)f_v(\vec{b}_2-\vec B)f_v(\vec{b}_3-\vec B)\nonumber\\
    &\qquad \times \delta^{(3)}(\frac{\vec{b}_1+\vec{b}_2+\vec{b}_3}{3}-\vec{B})\nonumber\\
    &=
    \frac{1}{\sqrt{3}^{3}(2\pi)^3s_l^2  s^4}
    \exp\!\left(-\frac{(b_1^\pm\!-\!B^\pm)^2 + (b_2^\pm\!-\!B^\pm)^2 + (b_3^\pm\!-\!B^\pm)^2}{2s_l^2}\right)\nonumber\\
    &\qquad \times \exp\left(-\frac{(\bperp_1 - \Bperp)^2+(\bperp_2 - \Bperp)^2 + ( \bperp_3-\Bperp)^2}{2s^2}\right)\nonumber\\
    &\qquad \times\delta^{(3)}(\frac{\vec{b}_1+\vec{b}_2+\vec{b}_3}{3}-\vec{B}),
\end{align}
where the prefactor ensures the correct normalization such that
\begin{align}
    \int \dd[3]{\vec b_1} \int \dd[3]{\vec b_2} \int \dd[3]{\vec b_3} f(\vec b_1, \vec b_2, \vec b_3) = 1.
\end{align}
The collective distribution function \eqref{eq:collective_hs_dist} formalizes the procedure outlined before. A random position is drawn for each hot spot from the distribution $f_v(\vec b)$. Then, all coordinates are shifted by the offset $\vec B$, which is determined as the average of all sampled positions.
We would now like to determine the resulting distribution function for a single hot spot position. We assume $\vec B$ fixed, and since \eqref{eq:collective_hs_dist} is completely symmetric in all hot spots, we can integrate out two arbitrary positions, say $\vec b_2$ and $\vec b_3$, and obtain a distribution for $\vec b_1$. The resulting effective distribution for the position $\vec b_1$ is
\begin{align}
    f_\mathrm{eff}(\vec b_1) &= \int \dd[3]{\vec b_2} \int \dd[3]{\vec b_3} f(\vec{b}_1, \vec{b}_2, \vec{b}_3)\nonumber\\
    &=\sqrt{\frac{3}{4\pi s_l^2}}\exp\left(-\frac{3(b^\pm_1-B^\pm)}{4s_l^2}\right)\frac{3}{4\pi s^2}\exp\left(-\frac{3(\bperp_1 - \Bperp)^2}{4s^2}\right).
\end{align}
It is now clear that the hot spot position $\vec b_1$ is distributed around the nucleon position $\vec{B}$ with a Gaussian, but the width of the corresponding Gaussian is $2s^2/3$ instead of $s^2$ in the transverse direction and analogously in the longitudinal direction.\par
This concludes the discussion of the hot spot model. It was shown that either color charge fluctuations or (valence quark) hot spots are required to more accurately reproduce the multiplicity histogram in proton-proton collisions. This paves the way for the study of proton-nucleus collision in the dilute Glasma, as will be highlighted in the outlook section.
\include{pA}
\chapter{Conclusion and Outlook}
\label{ch:conclusion}
\section{Summary}
We have discussed in great detail the derivation and some applications of the (3+1)D dilute Glasma framework. The (3+1)D dilute Glasma is the first analytic computation of rapidity-dependent observables in the Glasma incorporating 3D initial conditions, i.e.,\ considering longitudinally extended collision partners.
First, we introduced some theoretical background and discussed the boost-invariant Glasma.
Then, we showed how we linearize the Yang-Mills equations in the case of weak sources, which we call the dilute approximation.
In this limit, we found and simplified expressions for the Glasma field strength tensor.
The structure of the field strength tensor reveals the consequences of the dilute approximation: Gluons from both nuclei interact and produce gluons emitted on lightlike paths from everywhere within the collision region.
There are no additional interactions of emitted gluons, and the resulting Glasma becomes effectively Abelian.\par
We discussed observables that can be calculated from the field strength tensor and how the extended collision region in the $t$-$z$-plane in the case of longitudinally extended nuclei alters our understanding of the Milne frame. Since there is no obvious choice for the origin of the Milne frame, observables as a function of $\tau$ and $\eta_s$ are ambiguous. We shift the origin of the Milne frame into the future, out of the collision region. These considerations have important implications for future treatments of collisions where nuclei are modeled with finite longitudinal extent. We propose the local rest frame energy density $\elrf$ as an unambiguous measure of the energy density in treatments of the (3+1)D Glasma. Whenever observables are evaluated along curves of constant proper time, the origin of the respective Milne frame in relation to the collision region must be stated explicitly.
\par
Furthermore, we found that limiting fragmentation appears as a universal feature in the dilute Glasma. We showed in an analytic calculation that the shape of rapidity profiles for all observables constructed from the field strength tensor is universal in the fragmentation region, i.e.,\ for large values of spacetime rapidity. Applying a shift in rapidity corresponding to the respective beam rapidity, the profiles at two different collision energies overlap for $\eta_s \gg 0$. We also confirmed the occurrence of limiting fragmentation in numerical simulation results.\par
The dilute Glasma requires a model for the color charges pertaining to the hard degrees of freedom within the collision partners. We introduced a 3D Woods-Saxon model as a generalization to the McLerran-Venugopalan model. Our model includes tunable parameters that regulate the size of fluctuations in the transverse and longitudinal directions.\par
Since the dilute Glasma ultimately relies on numerics, we highlighted some details about the computation of dilute Glasma observables on computers.
At the heart of our implementation lies the computation of a 3D integral using Monte Carlo integration techniques. However, the peculiar shape of the integrand requires a careful selection of coordinate ranges and importance sampling distributions to ensure fast convergence.\par
We gave an extensive review of simulation results concerning the collision of two Woods-Saxon-shaped nuclei with tunable transverse and longitudinal correlations.
The study of rapidity profiles for different notions of energy density in the Glasma reveals that $T^{\tau\tau}$ is indeed not a suitable observable in collisions of nuclei with nontrivial longitudinal extent. Instead, the local rest frame energy density $\elrf$ is a good measure of energy density.
Regarding the transverse structure of the (3+1)D dilute Glasma, we qualitatively match results from (2+1)D simulations. A finite impact parameter leads to the typical \say{almond} shape of the energy density viewed in the transverse plane. The transverse flow components show the transverse expansion of the Glasma.\par
Furthermore, we discover significant longitudinal flow. Through a simple analytical model, we argue that the longitudinal flow can mostly be ascribed to the shifted Milne frame origin and the extended collision region in the $t$-$z$-plane. We use a similar model to estimate the effect of the shifted origin and the extended collision region on the time evolution of the Glasma. We find small deviations from the ideal $1/\tau$ behavior reported from simulations of the (2+1)D Glasma. These deviations grow larger as the collision energy goes down. However, the deviations from $1/\tau$ are small enough that we do not notice a clear deviation from the $1/\tau$ behavior in our numerical simulations after some initial time, regardless of the collision energy.\par
We studied the eccentricities $\sqrt{\langle |\varepsilon_2|^2\rangle}$ and $\sqrt{\langle |\varepsilon_4|^2\rangle}$ for collisions at finite impact parameter. As expected from (2+1)D simulations, we find significant contributions to these observables at mid-rapidity. However, in the (3+1)D dilute Glasma, we can study eccentricities over a large rapidity range and see a slight suppression at extreme rapidities. This is plausible from the picture that fluctuations in the energy density are more dominant in these regions. These random fluctuations can partially wash out the geometry that leads to large eccentricities. Finally, we study the real part of $\varepsilon_1$ as a measure of skew in the $x$-$\eta_s$-plane. Curiously, we find a strong dependence of $\mathfrak{Re}\, \varepsilon_1$ on the size of longitudinal fluctuations in the colliding nuclei. We therefore propose this observable as a means to constrain the parameters of nuclear models in future analyses.\par
Finally, we turned to the treatment of proton-proton collisions in the dilute Glasma framework. To reproduce experimental multiplicity distributions, some modifications must be made to the naive Ansatz of Gaussian protons. We showed how event-by-event fluctuations of the color charge, or a hot spot model taking into account individual (valence) quarks, can both tune the multiplicity distribution towards experimental results. This is an important step towards simulating proton-nucleus collisions in the dilute Glasma.
\section{Outlook}
The dilute Glasma framework offers a complete description of observables in the (3+1)D Glasma.
Its main allure is the existence of an analytic closed-form description of the Glasma field strength tensor. 
Indeed, there is potentially more to be learned about the structure of the dilute Glasma from analytic considerations. For example, a description of the field strength tensor in momentum space could yield further insights into the physical processes relevant in the dilute Glasma.\par
The dilute Glasma is also suited for numerical simulations of different collision systems involving protons and/or nuclei. There is still a lot to explore and improve with respect to these simulations.
We highlight a few possible directions for future research.
\subsection{Proton-nucleus collisions}
\begin{figure}
    \centering
    \includegraphics[scale=1]{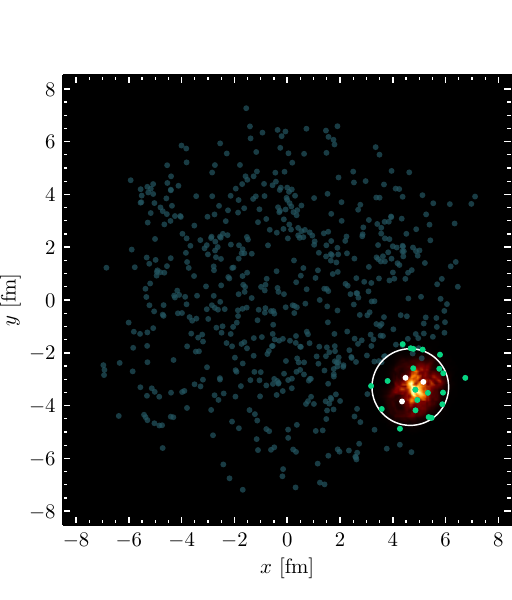}
    \caption{Sketch of the transverse geometry in a Pb-p collision event employing the hot spot model. Blue dots represent the transverse positions of all hot spots in the lead nucleus, which do not interact (spectators). Emerald dots correspond to interacting hot spots in the nucleus (participants). White dots mark the transverse locations of the three hot spots in the proton. The area within the white circle corresponds to the effective proton-proton cross section. In the background, a heatmap of the local rest frame energy density at mid-rapidity for the depicted collision event at $\tau=0.4\,\mathrm{fm}/c$ is shown.}
    \label{fig:pA}
\end{figure}
It is not yet clear whether the improvements to the proton model discussed in Chapter~\ref{ch:pp}, specifically the hot spot model, allow an accurate description of proton-nucleus collisions in the dilute Glasma. A first simulation of proton-nucleus collisions could give valuable hints as to whether the hot spot model needs further improvement.\par
First, the hot spot model determined by the correlator \eqref{eq:correlator_nucleon_hot_spot} for a single nucleon needs to be generalized to heavy nuclei consisting of nucleons with individual hot spots. Individual nucleons can be sampled and distributed according to a Woods-Saxon distribution, giving a more realistic model of heavy nuclei and their substructure than the simple Woods-Saxon model studied in Chapter~\ref{ch:results}.
Conveniently, the field strength tensor of the dilute Glasma is linear in the sources, and it is thus possible to treat each binary collision between pairs of hot spots separately and add the resulting field strength tensors.
In Figure~\ref{fig:pA}, we show the transverse geometry of a sample Pb-p collision. The lead nucleus features 208 nucleons with 3 hot spots each. Whether or not the proton interacts with a given nucleon is determined from the distance between the proton and nucleon center of mass employing the interaction probability \eqref{eq:differential_pp_ia_probability}. As a background in Figure~\ref{fig:pA}, we show the energy density produced in the collision event at mid-rapidity. Of course, in the dilute Glasma framework, the energy density can be computed at arbitrary values of spacetime rapidity. This allows the study of longitudinal event plane fluctuations, which were measured experimentally for Pb-p collisions at $\sqrt{s_{NN}}=5.02\,\mathrm{TeV}$ \cite{CMS:2015xmx}. Comparisons of simulation results with experimental data could help constrain the parameters in the hot spot model and provide additional insights into the structure of nucleons and nuclei.
\subsection{Refining the nuclear model using TMDs}
The proton or nuclear model serves as an input to the computation of observables in the dilute Glasma and is completely interchangeable with little effort.
Improved models for the color charges of protons or nuclei are a welcome and valuable addition to the dilute Glasma framework.
Ideally, these models allow the description of effects previously not describable in the dilute Glasma.\par
For example, an effect currently not captured by our proton models is related to event-by-event fluctuations in the shape of the rapidity profile \cite{Bzdak:2015eii, Bzdak:2016aii}. Specifically, there is a measurable asymmetry in the rapidity distribution of multiplicities that can be extracted from two-particle correlation functions. The magnitude of this effect has been constrained experimentally. Ideally, a refined approach to proton-proton simulations in the dilute Glasma would accurately reproduce these effects. However, it is not clear what the required modifications in the proton model are.\par
In general, an improved model for nuclei and protons could be constructed by including more information about the structure of nucleons or nuclei in the model.
This approach is taken in \cite{Schlichting:2020wrv}, which incorporates transverse momentum distribution functions (TMDs) into the correlator of color charges.
Specifically, the authors of \cite{Schlichting:2020wrv} equate the gluon TMD in the Golec-Biernat W\"usthoff (GBW) model \cite{Golec-Biernat:1998zce}
\begin{align}
\label{eq:GBW}
    \left.x_2 G^{(2)}(x_2,\kperp)\right|_{x_2=x_0}=\frac{N_c S_\perp}{2\pi^3\alpha_s}\frac{\kperp^2}{Q_s^2(x_2)}\exp\left(-\frac{\kperp^2}{Q_s^2(x_2)}\right)
\end{align}
with the Glasma TMD, which contains the correlator of color charges to first order in the dilute limit.
The saturation momentum in \eqref{eq:GBW} can be parametrized as
\begin{align}
    Q_s(x)=Q_0 x^{-\lambda}(1-x)
\end{align}
for small $x$, that is, the saturation scale (which determines the size of transverse fluctuations) depends in some way on the longitudinal momentum fraction $x$ (which is linked to the size of longitudinal fluctuations).
Incorporating the GBW model into the color charge correlator in this way would yield a model that is more closely tied to experimental results of the nucleon structure.
The resulting model could describe effects like the rapidity asymmetries in proton-proton collisions.
\subsection{Coupling to hydrodynamics}
A complete description of heavy-ion collision events requires modeling all distinct phenomenological stages of such collisions. Most importantly, the Glasma energy-momentum tensor has to be understood as initial conditions for a subsequent simulation of the hydrodynamic phase. This is easily achievable by performing Landau matching and extracting a hydrodynamic energy-momentum tensor from the dilute Glasma. Although the $\tau=\mathrm{const}$ hypersurfaces, along which we evaluate the energy-momentum tensor, relate to the arbitrary origin of the Milne frame, they are valid Cauchy surfaces on which the initial value problem solved in hydrodynamics is properly defined.\par
Finally, including particlization and the hadronic transport stage yields observables that are directly comparable to experimental results.
This could shed more light on the connection between the initial state (specifically the longitudinal and transverse structure of protons and nuclei) and final state observables, which is currently an avid area of research. 

\appendix
\chapter{Classical Yang-Mills theory}
\label{ch:YM}
\section{Field content}
\label{sec:YM_field_content}
The Lagrangian of Yang-Mills theory with gauge group $\text{SU}(N_c)$ is given as
\begin{align}
    \label{eq:YM_Lagrangian}
    \mathcal{L}_\text{YM}(x) \coloneqq -\frac{1}{2}\Tr \left[F_{\mu\nu}(x)F^{\mu\nu}(x)\right],
\end{align}
where $F_{\mu\nu}(x)$ is the $\mathfrak{su}(N_c)$-valued field strength tensor on Minkowski space $M$. It is defined in terms of the $\mathfrak{su}(N_c)$-valued gauge field $A_\mu(x)$ as
\begin{align}
\label{eq:Fmunu_abstract}
    F_{\mu\nu}(x) := \partial_\mu A_\nu(x) - \partial_\nu A_\mu(x) - \ii g\, \comm{A_\mu(x)}{A_\nu(x)}.
\end{align}
The constant $g$ is called the coupling constant. The sign in front of the last term is up to convention.\footnote{This sign ambiguity extends to the gauge transformation behavior of the gauge field and the gauge covariant derivative. However, all of these signs are linked together, so fixing the sign in the field strength tensor fixes all subsequent signs. Another (unrelated) choice of convention is the placement of $g$. Sometimes, the gauge field and field strength tensor are defined with an additional factor of $g$ relative to the definition above. This removes $g$ from \eqref{eq:Fmunu_abstract} and \eqref{eq:gauge_field_transformation}, but introduces a factor $1/g^2$ in the Lagrangian \eqref{eq:YM_Lagrangian}.}
The Lagrangian \eqref{eq:YM_Lagrangian} is invariant under a local $\text{SU}(N_c)$ gauge transformation $U(x)$, which acts on the gauge field as
\begin{align}
\label{eq:gauge_field_transformation}
A_\mu(x) \rightarrow \tilde A_\mu(x) = U(x)\left(A_\mu(x) - \frac{1}{\ii g} \partial_\mu\right)U^\dagger(x).
\end{align}
It follows that the field strength tensor transforms as
\begin{align}
    F_{\mu\nu}(x)\rightarrow \tilde F_{\mu\nu}(x) = U(x)F_{\mu\nu}(x)U^\dagger(x),
\end{align}
from which it is clear that $\mathcal{L}_\text{YM}(x)$ is gauge invariant. One can express the group element as  $U(x) = \exp(-\ii g \omega(x))$, where $ \omega: M \rightarrow \mathfrak{su}(N_c)$ is the generator of $U(x)$. 
For infinitesimal $\omega(x)$ to leading order the transformation becomes
\begin{align}
\label{eq:infinitesimal_gauge_trafo_fundamental}
    \delta_\omega A_\mu(x) = \tilde A_\mu(x) - A_\mu (x) =  -\partial_\mu \omega(x) -  \ii g\comm{\omega(x)}{A_\mu(x)}.
\end{align}
From here on, we occasionally omit the spacetime arguments of fields to simplify the notation.
The gauge covariant derivative
\begin{align}
    D_\mu \coloneqq \p_\mu - \ii gA_\mu
\end{align}
acts on $\mathfrak{su}(N_c)$-valued expressions via the commutator. For example, one may write
\begin{align}
    \delta_\omega A_\mu =-\partial_\mu \omega +  \ii g\comm{A_\mu}{\omega}= -D_\mu \omega.
\end{align}
A useful relation containing the gauge covariant derivative is
\begin{align}
    \label{eq:Fmunu_DD}
    F_{\mu\nu} = \frac{\ii}{g}\comm{D_\mu}{D_\nu}.
\end{align}
Note that the commutator of two gauge covariant derivatives is an algebraic derivative. Equation \eqref{eq:Fmunu_DD} could also be taken as the definition of $F_{\mu\nu}$ with \eqref{eq:Fmunu_abstract} a consequence of that definition. The Bianchi identity
\begin{align}
\label{eq:bianchi}
    \comm{D_\mu}{F_{\nu\rho}} + \comm{D_\nu}{F_{\rho\mu}} + \comm{D_\rho}{F_{\mu\nu}} = 0
\end{align}
is then simply a consequence of the Jacobi identity.\par
One may choose a basis of $N_c^2-1$ generators $t^a$ for $\mathfrak{su}(N_c)$ and expand with respect to that basis
\begin{align}
    A_\mu(x) = A_\mu^a(x) t^a,\qquad\qquad F_{\mu\nu}(x) = F_{\mu\nu}^a(x) t^a.
\end{align}
The commutator of two generators is another generator
\begin{align}
\label{eq:fundamental_generator_commutator}
    \comm{t^a}{t^b}=\ii\,f^{abc}t^c,
\end{align}
where $f^{abc}$ are the structure constants of $\mathfrak{su}(N_c)$. Note that nothing so far hinges on a specific representation. To be compatible with the fermionic matter fields, which will be introduced shortly, we take our generators to be in the fundamental representation of $\mathfrak{su}(N_c)$. Furthermore, we normalize them such that
\begin{equation}
    \Tr[t^at^b] = \frac{1}{2}\delta^{ab}.
\end{equation}
From the relation $\Tr\left[\comm{t^a}{t^b}t^c\right]=\frac{\ii}{2} f^{abc}$ it then follows that the $f^{abc}$ are completely antisymmetric. Note that for $N_c=2$ this implies $f^{abc}\propto \epsilon^{abc}$. Furthermore, \eqref{eq:Fmunu_abstract} implies
\begin{align}
    F_{\mu\nu}^a = \partial_\mu A_\nu^a - \partial_\nu A_\mu^a + g f^{abc}A_\mu^bA_\nu^c.
\end{align}
One can then rewrite the Yang-Mills Lagrangian \eqref{eq:YM_Lagrangian} as
\begin{align}
    \label{eq:YM_Lagrangian_explicit}
    \mathcal{L}_\text{YM} = -\frac{1}{4}F^a_{\mu\nu}F^{\mu\nu,a}.
\end{align}
The covariant derivative acting on an $\mathfrak{su}(N_c)$-valued field $\omega(x) = \omega^a(x)t^a$ can now be written as
\begin{align}
    D_\mu(\omega^at^a) = \left(\partial_\mu \omega^a - gf^{abc}\omega^bA^c_\mu\right)t^a = (D_\mu)^{ab}\omega^b t^a,
\end{align}
where
\begin{align}
    \label{eq:cov_dev_components}
    (D_\mu)^{ab} \coloneqq \delta^{ab}\partial_\mu - gf^{abc}A^c
\end{align}
is a version of the covariant derivative acting on the $N_c^2-1$-component vector field $\omega^b(x)$. Note that \eqref{eq:cov_dev_components} acts without a commutator. Instead, it is often said that the gauge field acts in its adjoint representation $f^{abc}A^c(x)$ on $\omega^b(x)$.
The infinitesimal gauge transformation \eqref{eq:infinitesimal_gauge_trafo_fundamental} can now be expressed as
\begin{align}
    \delta_\omega A_\mu^a &= -\partial_\mu \omega^a + g f^{abc}\omega^bA_\mu^c\nn\\
    &= -(D_\mu)^{ab}\omega^b
\end{align}
and for the field strength tensor is given as
\begin{align}
    \delta_\omega F^a_{\mu\nu} = g f^{abc}\omega^b F^c_{\mu\nu}.
\end{align}
The conceptual shift of considering $(N_c^2-1)$-dimensional vector fields like $A^a_\mu(x)$ instead of $(N_c\times N_c)$-dimensional matrix fields like $A_\mu(x) = A^a_\mu(x) t^a$ can be extended to finite gauge transformations as well. For example, the field strength tensor transforms as
\begin{align}
    \tilde F^b_{\mu\nu}t^b = F^b_{\mu\nu}Ut^bU^\dagger.
\end{align}
With $U=\exp(-\ii g \omega^a t^a)$ we employ the formula
\begin{align}
    \ee^X Y e^{-X} = \sum_{n=0}^\infty \frac{1}{n!}\comm{X}{Y}_n,
\end{align}
where $\comm{X}{Y}_n = \comm{X}{\comm{X}{Y}_{n-1}}$ and $\comm{X}{Y}_0 = Y$. Since $X = -\ii g \omega^a t^a$ and $Y=t^b$ it follows that
\begin{align}
\label{eq:BCH_application}
    \comm{X}{Y}_{n} &= X^a \comm{X}{Y}_{n-1}^c\comm{t^a}{t^c}\nn\\
    &= \ii X^a \comm{X}{Y}_{n-1}^c f^{acd}t^d\nn\\
    &= \left((-\ii X\cdot f)^n\right)^{dc} Y^c t^d\nn\\
    &= \left((-\ii X\cdot f)^n\right)^{ab} t^a
\end{align}
Here, $X\cdot f$ means the matrix obtained by contracting the first index of $f^{abc}$ with $X^a$ and interpreting the remaining two indices as matrix indices. Note that the free index $b$ appears from $Y^c = \delta^{cb}$. Although the intermediate steps in \eqref{eq:BCH_application} are only valid for $n >0 $, the final result also applies to $n=0$ and hence
\begin{align}
    \tilde F^a_{\mu\nu}t^a &= F^b_{\mu\nu} \sum_{n=0}^\infty \frac{1}{n!}\left((-\ii X\cdot f)^n\right)^{ab} t^a\nn\\
    &= \exp(-\ii X\cdot f)^{ab} F^b_{\mu\nu} t^a\nn\\
    &= \exp( - g\omega \cdot f)^{ab}F^b_{\mu\nu} t^a\nn\\
    &= U^{ab}F^b_{\mu\nu} t^a,
\end{align}
where in the last step we introduced the $(N_c^2-1)\times (N_c^2-1)$ adjoint representation gauge transformation matrix field $U^{ab}\coloneqq \exp(-g\omega\cdot f)$, which can be used to concisely write $\tilde{F}^{a}_{\mu\nu} = U^{ab}F^b_{\mu\nu}$. Note that $U^{ab}$ can be obtained from the fundamental representation gauge transformation by replacing $(t^a)^{bc} \rightarrow -\ii f^{abc}$, which is exactly the change from fundamental to adjoint representation. The generators $-\ii f^{abc}$ also obey a commutation relation analogous to \eqref{eq:fundamental_generator_commutator}.
\section{Equations of motion and currents}
To treat the interactions of the gauge field with charged matter, one requires an additional matter contribution to the Lagrangian, for example
\begin{align}
    \label{eq:dirac_lagrangian}
    \mathcal{L}_\mathrm{matter}= \psi_f^\dagger(\ii \gamma^\mu D_\mu - m_f )\psi_f,
\end{align}
where $\psi_f=\psi_f(x)$ is a set of Dirac spinor fields with $N_c$-dimensional color vectors\footnote{To be precise, the color vectors are (acted on by) the fundamental representation of $\mathrm{SU}(N_c)$.} attached at each point. The index $f$ (which, in the case of QCD, labels the different quark flavors) is summed over. Note that the covariant derivative acts on the color vector via matrix multiplication in the fundamental representation of $SU(N_c)$. Integrating the Yang-Mills and matter Lagrangian over all of Minkowski space $M$ then gives the action
\begin{align}
    \label{eq:YM_matter_action}
    S[A_\mu,\psi] = \int_M \dd[4]{x}\left( -\frac{1}{4}F^a_{\mu\nu} F^{\mu\nu,a} + \psi_f^\dagger(\ii \gamma^\mu D_\mu - m_f )\psi_f\right).
\end{align}
Since under an $\text{SU}(N_c)$ gauge transformation the spinor $\psi_f$ transforms as
\begin{align}
\label{eq:spinor_gauge_trafo}
    \psi_f(x) \rightarrow U(x)\psi_f(x),
\end{align}
the action \eqref{eq:YM_matter_action} is gauge invariant thanks to the covariant derivative in the matter action.
Varying with respect to $A_\mu^a$ and dropping boundary terms yields the equations of motion (EOM) for the gauge field
\begin{align}
    \label{eq:YM_EOM}
    \partial_\mu F^{\mu\nu,a} - gf^{abc}F^{\mu\nu,b}A^c_\mu =(D_\mu)^{ab}F^{\mu\nu,b}=- j_\mathrm{matter}^{\nu,a},
\end{align}
where the definition
\begin{align}
    \label{eq:j_matter}
   j_\mathrm{matter}^{\nu,a} \coloneqq  \frac{ \delta \mathcal{L}_\mathrm{matter}}{\delta A^a_\nu}= g\psi_f^\dagger\gamma^\nu t^a\psi_f
\end{align}
was made. The right-hand side of \eqref{eq:j_matter} for specific values of $\nu$ and $a$ is a scalar field as there are two matrix multiplications in the spinor and color space that each collapse to a scalar. Multiplying both sides of \eqref{eq:YM_EOM} with $t^a$ the EOM can be written as
\begin{align}
    \label{eq:EOM_commutator}
    \comm{D_\mu}{F^{\mu\nu}} = -j_\mathrm{matter}^\nu.
\end{align}
Note that
\begin{align}
    \comm{D_\nu}{\comm{D_\mu}{F^{\mu\nu}}}&=\frac{1}{2}\left( \comm{D_\nu}{\comm{D_\mu}{F^{\mu\nu}}} - \comm{D_\mu}{\comm{D_\nu}{F^{\mu\nu}}}\right)\nn\\
    &=\frac{1}{2}\left(\comm{D_\nu}{\comm{D_\mu}{F^{\mu\nu}}} + \comm{D_\mu}{\comm{F^{\mu\nu}}{D_\nu}}\right)\nn\\
    &=-\frac{1}{2} \comm{F^{\mu\nu}}{\comm{D_\nu}{D_\mu}} = -\frac{\ii g}{2}\comm{F^{\mu\nu}}{F_{\mu\nu}}=0,
\end{align}
where the Jacobi identity was used for the first expression in the last line. Thus,
\begin{align}
    \comm{D_\nu}{j^\nu_\mathrm{matter}}=0,
\end{align}
that is, the matter current is covariantly conserved. The same equation in components is
\begin{equation}
    (D_\nu)^{ab}j^{\nu,b}_\mathrm{matter} = \partial_\nu j^{\nu,a}_\mathrm{matter} -gf^{abc}j^{\nu,b}_\mathrm{matter}A^c_\nu = 0.
\end{equation}
Starting from \eqref{eq:spinor_gauge_trafo}, the behavior of $j^\nu_\mathrm{matter}$ under a gauge transformation is given by
\begin{align}
    \tilde j^{\nu,a}_\mathrm{matter}= g\psi_f^\dagger U^\dagger \gamma^\nu t^aU\psi_f.
\end{align}
Using \eqref{eq:BCH_application} yields
\begin{align}
\label{eq:matter_current_gauge_trafo}
    \tilde j^{\nu,a}_\mathrm{matter} = U^{ab} j^{\nu, b}_\mathrm{matter},
\end{align}
completely analogous to the field strength tensor.\par
A consequence of \eqref{eq:YM_EOM} is that
\begin{align}
    -\p_\nu \p_\mu F^{\mu\nu,a}=\p_\nu\left( j^{\nu,a}_\mathrm{matter} - g f^{abc} F^{\mu\nu,b} A^c_\mu \right)=0.
\end{align}
Apparently, there exists an on-shell conserved current given by the quantity in parentheses. This is nothing but Noether's theorem, and the current corresponds (up to a constant) to the Noether current for the symmetry of rigid (i.e.,\ global) gauge transformations
\begin{align}
    j^\nu_\text{Noether} \propto j^\nu_\mathrm{matter} +\ii g\comm{F^{\mu\nu}}{A_\mu}.
\end{align}
This Noether current is not gauge invariant, and thus, there is no physical charge associated with it.\par
While on the fundamental level, the matter current has to be sourced by charged matter, it can be convenient in some applications to be agnostic about the origin of the matter current, and instead of $\mathcal L_\mathrm{matter}$ add $-A^a_\mu J^{\mu,a}$ to $\mathcal L_\text{YM}$. This yields the EOM
\begin{align}
    \comm{D_\mu}{F^{\mu\nu}} = J^\nu.
\end{align}
It should, however, be noted that even with the transformation behavior \eqref{eq:matter_current_gauge_trafo} for the current, the combination $A^a_\mu J^{\mu, a}$ is not gauge invariant.
\section{Curved background and energy-momentum tensor}
So far, Yang-Mills theory has been considered on a flat Minkowski spacetime. However, this can be generalized, and one could just as well study the theory on a curved spacetime manifold. While this goes far beyond the scope of this work, there is another relevant application of this generalization. Sometimes, it is convenient to parametrize Minkowski space with a curvilinear coordinate system. It is not at all obvious how the Yang-Mills EOM change as a consequence. A clean way of rederiving the EOM in a curvilinear coordinate system is given by the general formalism of Yang-Mills theory on a curved background. The Yang-Mills action with matter on a curved background is given by
\begin{align}
    \label{eq:S_curved}
    S[A_\mu, \psi] = \int_M \dd[4]{x}\sqrt{-g}\left(-\frac{1}{4}F_{\mu\nu}^aF_{\rho\sigma}^ag^{\mu\rho}g^{\nu \sigma} + \mathcal{L}'_\mathrm{matter}[A_\mu,\psi]\right),
\end{align}
where $\mathcal{L}'_\mathrm{matter}[A_\mu,\psi]$ can be taken as the Dirac Lagrangian on a curved background, which is more complicated than \eqref{eq:dirac_lagrangian}. The gauge covariant derivative can now be generalized to a spacetime and gauge covariant derivative
\begin{align}
    \mathcal D_\mu := \nabla_\mu -\ii gA_\mu
\end{align}
with the Levi-Civita covariant derivative $\nabla_\mu$. The spacetime manifold $M$ is no longer restricted to Minkowski space. Note, however, that in the action \eqref{eq:S_curved}, the metric is not a dynamical field.\footnote{One can add the Einstein-Hilbert action to \eqref{eq:S_curved}, which in total gives an action that contains a dynamical metric field. The corresponding equations of motion are the Einstein equations with the energy-momentum tensor of the gauge and matter fields as a source term.} The field strength tensor on the curved background
\begin{align}
    F_{\mu\nu} &:= \nabla_\mu A_\nu - \nabla_\nu A_\mu - \ii g\comm{A_\mu}{A_\nu} \nn\\
    &= \p_\mu A_\nu +\Gamma_{\mu\nu}^\rho A_\rho - \p_\nu A_\mu - \Gamma_{\nu\mu}^\rho A_\rho - \ii g\comm{A_\mu}{A_\nu}\nn\\
    &= \p_\mu A_\nu - \p_\nu A_\mu - \ii g\comm{A_\mu}{A_\nu}
\end{align}
is the same as on the flat background for a Levi-Civita connection with $\Gamma^\rho_{\mu\nu} = \Gamma^\rho_{\nu\mu}$. The EOM for the gauge field are once again obtained by varying with respect to $A^a_\mu$ and partially integrating, such that
\begin{align}
    \label{eq:curved_eom_partial}
    \p_\mu (\sqrt{-g}F^{\mu\nu,a})-\sqrt{-g}gf^{abc}F^{\mu\nu,b}A_\mu^c + \sqrt{-g} j^{\nu,a}_\mathrm{matter}=0.
\end{align}
Note that the $g$ under the square root is the determinant of the metric, while the $g$ without a square root is the coupling constant. Since
\begin{align}
\nabla_\mu F^{\mu\nu} &=\partial_\mu F^{\mu\nu} + \Gamma^{\mu}_{\mu\rho}F^{\rho\nu} + \Gamma^\nu_{\mu\rho} F^{\mu\rho} = \partial_\mu F^{\mu\nu} + \frac{1}{2}g^{\mu\lambda}\partial_\rho g_{\mu\lambda} F^{\rho\nu}\nn\\
&=\frac{\partial_\mu(\sqrt{-g}F^{\mu\nu})}{\sqrt{-g}}
\end{align}
for any antisymmetric tensor $F^{\mu\nu}$, we can replace the derivative in the first term of \eqref{eq:curved_eom_partial} by a spacetime covariant derivative in front of $F^{\mu\nu,a}$. Also multiplying with $t^a$ while dividing by $\sqrt{-g}$ then gives
\begin{align}
    \nabla_\mu F^{\mu\nu} - \ii  g \comm{A_\mu}{F^{\mu\nu}} = -  j^\nu_\mathrm{matter},
\end{align}
which can be simplified to
\begin{align}
    \comm{\mathcal D_\mu}{ F^{\mu\nu}} = - j^\nu_\mathrm{matter}.
\end{align}
Putting the theory on a curved background also makes it easy to find the symmetric energy-momentum tensor
\begin{equation}
    \label{eq:EMT_formula}
    T^{\mu\nu} = -\frac{2}{\sqrt{-g}}\frac{\delta \mathcal L}{\delta g_{\mu\nu}},
\end{equation}
where $\mathcal L$ stands for everything under the action integral, including the factor of $\sqrt{-g}$. Ignoring any matter contribution, \eqref{eq:EMT_formula} can be used to compute the energy-momentum tensor of the gauge field
\begin{equation}
    T^{\mu\nu} = 2\Tr\left[-F^{\mu}{}_\rho F^{\nu\rho}+ \frac{1}{4}g^{\mu\nu}F_{\rho\sigma}F^{\rho\sigma}\right].
\end{equation}
Using the Bianchi identity \eqref{eq:bianchi} (with $\mathcal D$ instead of $D$) one can show that on-shell
\begin{align}
\label{eq:cov_conservation_EMT}
    \nabla_\mu T^{\mu\nu} = 0
\end{align}
in the absence of sources.\footnote{Strictly speaking, \eqref{eq:cov_conservation_EMT} is still valid in the presence of sources, but in that case, the energy-momentum tensor of the sources needs to be added to that of the gauge field.
This is intuitively clear from energy conservation.
If the energy of the sources is not accounted for, the energy of the system can increase as the sources pump energy into the system.
However, the sources lose energy in this process, and accounting also for the energy of the sources recovers energy conservation.}
The fact that
\begin{equation}
    T^\mu{}_\mu = 0
\end{equation}
shows the scale invariance of classical Yang-Mills theory.
However, once the interaction with quarks is taken into account, the corresponding quantum theory is not scale invariant.
\section{Wilson lines and Wilson loops}
The gauge covariant derivative $D_\mu$ introduces a notion of parallel transport. A color vector $\psi(x)$ at $x$ is parallel transported along a path $\gamma: [0,T] \rightarrow M$ in Minkowski space starting at $x$ if
\begin{align}
    (\p_t\gamma^\mu(t)) D_\mu \psi(\gamma(t)) = 0,
\end{align}
that is, the covariant derivative of the matter field in the direction of the tangent vector to the curve $\gamma$ must vanish anywhere on $\gamma$. In terms of the gauge field, this means that
\begin{align}
    (\partial_t\gamma^\mu(t))\partial_\mu \psi(\gamma(t))-\ii g (\p_t\gamma^\mu(t)) A_\mu(\gamma(t)) \psi(\gamma(t))=0,
\end{align}
which is equivalent to
\begin{align}
\label{eq:wilson_line_equation}
    \dv{}{t} \psi(\gamma(t))=\ii g A_\gamma(t)\psi(\gamma(t)),
\end{align}
where the abbreviation $A_\gamma(t):=(\p_t \gamma^\mu(t))A_\mu(\gamma(t))$ was introduced. 
Integrating this expression yields
\begin{align}
\label{eq:psi_gamma_t_implicit}
    \psi(\gamma(t)) = \psi(\gamma(0)) +\ii g \intop_0^t\dd{s}A_\gamma(s) \psi(\gamma(s)).
\end{align}
Of course, this is not an explicit solution for $\psi(\gamma(t))$, but it can be applied iteratively to obtain (after one additional iteration)
\begin{align}
\label{eq:psi_gamma_t_3_terms}
    \psi(\gamma(t)) &= \psi(\gamma(0))+\ii g\intop_0^t \dd{s} A_\gamma(s)\psi(\gamma(0))\nn\\
    &\quad+ (\ii g)^2\intop_0^t\dd{s_1}\intop_0^{s_1}\dd{s_2} A_\gamma(s_1)A_\gamma(s_2)\psi(\gamma(s_2)).
\end{align}
This suggests that just inserting \eqref{eq:psi_gamma_t_implicit} into itself an infinite number of times would yield
\begin{align}
\label{eq:psi_gamma_t_solution_1}
    \psi(\gamma(t)) = \sum_{n=0}^\infty (\ii g)^n\intop_0^t\dd{s_1}\intop_0^{s_1}\dd{s_2}\dots \intop_0^{s_{n-1}}\dd{s_n} A_\gamma(s_1)A_\gamma(s_2)\dots A_\gamma(s_n) \psi(\gamma(0)).
\end{align}
This sum converges, and it solves \eqref{eq:psi_gamma_t_implicit}, but it is not in the form the solution is usually written in. It is, however, equivalent to
\begin{align}
\label{eq:psi_gamma_t_solution_2}
    \psi(\gamma(t))= \sum_{n=0}^\infty \frac{(\ii g)^n}{n!} \intop_0^t\dd{s_1}\intop_0^{t}\dd{s_2}\dots \intop_0^{t}\dd{s_n} \mathcal P\,A_\gamma(s_1)A_\gamma(s_2)\dots A_\gamma(s_n) \psi(\gamma(0)).
\end{align}
In \eqref{eq:psi_gamma_t_solution_1}, the integration variables were constrained by $t>s_1>s_2>\dots >s_{n-1}> s_n$, which means that the $A_\gamma$ are ordered such that the arguments decrease from left to right. In \eqref{eq:psi_gamma_t_solution_2}, the integration variables are only constrained by $0< s_i < t$. The ordering of the $A_\gamma$ such that their arguments decrease from left to right is now taken care of by the path-ordering operator $\mathcal P$, which does exactly that. It takes all the elements that come after and reorders them such that those evaluated earlier on the path are placed to the right of those evaluated later on the path. The integrand of \eqref{eq:psi_gamma_t_solution_2} is then the same as that of \eqref{eq:psi_gamma_t_solution_1}, except for an overcounting by $n!$, which is taken care of by an additional prefactor in \eqref{eq:psi_gamma_t_solution_2}. The path-ordering operator can just as well be drawn in front of the sum, which can then be succinctly written as an exponential
\begin{align}
  \psi(\gamma(t)) = \mathcal P \exp\left(\ii g \intop_0^t\dd s\, (\p_s \gamma^\mu(s))A_\mu(\gamma(s))\right)\psi(\gamma(0)),
\end{align}
which is defined through its power series and should be viewed as a shorthand for \eqref{eq:psi_gamma_t_solution_2}. Labeling the endpoints of $\gamma$ as $\gamma(0)=x$ and $\gamma(T)=y$, there is now a prescription for parallel transporting the matter field $\psi$ from $x$ to $y$
\begin{align}
    \psi(y) = W(\gamma,A) \psi(x)
\end{align}
simply by applying the gauge group element
\begin{align}
    W(\gamma, A) = \mathcal P \exp\left(\ii g \intop_0^T\dd s\, (\p_s \gamma^\mu(s))A_\mu(\gamma(s))\right),
\end{align}
which is called the Wilson line along the path $\gamma$. Since the notation is misleading here, it should be stressed that $\psi(y)$ and $\psi(x)$ denote color vectors at two specific points, and $\psi(y)$ is the result of parallel transporting $\psi(x)$ along the path $\gamma$. A property of the Wilson line is that
\begin{align}
    W(\gamma,A) = W(\beta\alpha,A)=W(\beta,A)W(\alpha,A),
\end{align}
with $\alpha(T)=\beta(0)$, so the paths $\alpha$ and $\beta$ are composable to the single path $\gamma$. One can check that under a gauge transformation
\begin{align}
    W(\gamma,A') = U(\gamma(T))W(\gamma,A)U(\gamma(0))^{\dagger}
\end{align}
the Wilson line transforms nonlocally unless one considers a closed path $\gamma(0)=\gamma(T)=x$, in which case
\begin{align}
    W(\gamma, A') = U(x)W(\gamma,A)U(x)^{\dagger}.
\end{align}
Obviously, the quantity
\begin{align}
    \Tr W(\gamma,A)
\end{align}
is then gauge invariant for any closed path $\gamma$. It is called a Wilson loop and is considered an important physical observable.
\chapter{The 2D Poisson equation}
\label{ch:poisson}
\section{Continuum Poisson equation}
For a single nucleus or hadron moving at the speed of light, the only nontrivial component of the gauge field in covariant gauge is related to the color charge density of the nucleus by (we drop indices $A/B$ and the longitudinal coordinate in the notation)
\begin{align}
\label{eq:poisson_eq}
    -\nabla_\perp^2 \phi(\xperp) = \rho(\xperp),
\end{align}
where $\nabla_\perp$ denotes the gradient with respect to $\xperp$.
Ignoring boundary conditions for now, a general solution to this equation is given by
\begin{align}
    \phi(\xperp) = \int \dd[2]{\yperp} G(\xperp-\yperp)\rho(\yperp) + \phi_\text{hom}(\xperp)
\end{align}
with $\phi_\text{hom}(\xperp)$ the solution to the 2D Laplace equation $\nabla_\perp^2 \phi_\text{hom}(\xperp)=0$ (i.e.,\ a homogeneous solution) and $G(\xperp-\yperp)$ the Green's function with the property
\begin{align}
\label{eq:greens_fct_defining}
    -\nabla^2_\perp G(\xperp-\yperp) = \delta^{(2)}(\xperp - \yperp).
\end{align}
Thus, $G(\xperp - \yperp)$ is the contribution to $\phi(\xperp)$ sourced by a normalized point charge located at $\yperp$. Writing $G$ and $\delta$ in terms of their Fourier transform,
\begin{align}
G(\xperp - \yperp) &= \int \frac{\dd[2]{\kperp}}{(2\pi)^2}\tilde G(\kperp)\ee^{-\ii \kperp\cdot(\xperp-\yperp)},\\
\delta^{(2)}(\xperp - \yperp) &= \int \frac{\dd[2]{\kperp}}{(2\pi)^2}\ee^{-\ii \kperp\cdot(\xperp-\yperp)}
\end{align}
puts \eqref{eq:greens_fct_defining} in the form
\begin{align}
    \tilde G(\kperp) = \frac{1}{\kperp\cdot\kperp}.
\end{align}
Transforming back yields the expression
\begin{align}
    G(\xperp-\yperp) = \int \frac{\dd[2]{\kperp}}{(2\pi)^2}\frac{1}{\kperp\cdot\kperp}\ee^{-\ii \kperp\cdot(\xperp-\yperp)}
\end{align}
for the Green's function, which in polar coordinates $(k=|\kperp|, \theta)$ reads
\begin{align}
    G(\xperp-\yperp) = \int_0^\infty \dd{k}\int_0^{2\pi}\dd{\theta} \frac{k}{(2\pi)^2}\frac{1}{k^2}\ee^{-\ii  k\left|\xperp-\yperp\right|\cos{\theta}}.
\end{align}
The $\theta$-integral can be carried out right away, yielding
\begin{align}
    \label{eq:greens_fctn_k_int}
    G(\xperp - \yperp) = \frac{1}{2\pi}\int_\epsilon^\infty \dd{k}\frac{1}{k}J_0(k\left|\xperp-\yperp\right|)
\end{align}
with $J_0$ the zeroth Bessel function of the first kind. Note also that the lower bound for the $k$-integral was replaced by a small positive constant $\epsilon$ as the integral would diverge otherwise. The integral can now be performed analytically and gives
\begin{align}
   &G(\xperp - \yperp) \nn\\
   &\hspace{0.3cm}= \frac{1}{2\pi}\left[-\gamma +\frac{\epsilon^2}{8}\left|\xperp - \yperp\right|^2{}_2F_3\left(1,1;2,2,2;-\frac{\epsilon^2}{4}\left|\xperp-\yperp\right|^2\right) + \ln{\left(\frac{2}{\epsilon \left|\xperp - \yperp\right|}\right)} \right] 
\end{align}
with ${}_2F_3$ a generalized hypergeometric function and $\gamma$ the Euler-Mascheroni constant. What matters more, however, is this expression's expansion in $\epsilon$, which is
\begin{align}
\label{eq:greens_fctn_eps_expansion}
    G(\xperp - \yperp) = \frac{1}{2\pi}\left(-\gamma -\ln \epsilon + \ln 2 - \ln \left|\xperp-\yperp\right| \right)+\mathcal{O}(\epsilon^2).
\end{align}
Just like the overall solution, $G(\xperp - \yperp)$ is only determined up to a solution of the Laplace equation, which includes constants. Thus, we can subtract all the constant terms (including $\ln\epsilon$) and then perform the $\epsilon \rightarrow 0$ limit, which finally gives
\begin{align}
    \label{eq:greens_ln}
    G(\xperp - \yperp) = \frac{-\ln |\xperp - \yperp|}{2\pi}
\end{align}
as the Green's function for the Poisson equation in two dimensions. However, one might want to leave $\epsilon$ explicit and keep the lowest order $\epsilon$-term, which gives
\begin{align}
    G_{\epsilon}(\xperp - \yperp) = \frac{-\ln\left(\epsilon |\xperp - \yperp|\right)}{2\pi}.
\end{align}
Since $\xperp$ and $\yperp$ have a dimension of length, the $\epsilon$ (which has inverse length) is required to make the argument of the logarithm dimensionless.
Any solution to the Poisson equation can now be found by calculating the convolution of the Green's function with the source and adding an arbitrary homogeneous solution such that the boundary conditions of the problem are satisfied.\footnote{In practice, it might be convenient to add a homogeneous solution to the Green's function instead of adding it to the overall result.} Note that \eqref{eq:greens_ln} diverges as $\left|\xperp- \yperp\right|\rightarrow \infty$.

There is another way of regularizing the divergent integral for the Green's function by making use of a mass-like parameter $m$ that modifies the Poisson operator $-\nabla_\perp^2 \rightarrow -\nabla_\perp^2 + m^2$. This replaces the integral \eqref{eq:greens_fctn_k_int} with
\begin{align}
    G(\xperp - \yperp) = \frac{1}{2\pi}\int_0^\infty \dd{k}\frac{k}{k^2+m^2}J_0(k\left|\xperp-\yperp\right|)
\end{align}
and there is now no need for a cutoff in $k$. This integral gives
\begin{align}
    \label{eq:greens_fctn_k0}
    G(\xperp - \yperp) = \frac{1}{2\pi} K_0\left(m\left|\xperp-\yperp\right|\right)
\end{align}
with $K_0$ the zeroth modified Bessel function of the second kind. For finite $m$ and $\left|\xperp- \yperp\right| \rightarrow \infty$ this Green's function scales like
\begin{align}
    G(\xperp-\yperp) \sim \frac{\ee^{-m\left|\xperp-\yperp\right|}}{2\sqrt{2\pi} \sqrt{m\left|\xperp-\yperp\right|}},
\end{align}
so it goes to zero asymptotically. Finally, a good consistency check is to see what happens in the limit $m \rightarrow 0$ for fixed finite $\left|\xperp-\yperp\right|$. Then
\begin{align}
    G(\xperp-\yperp) = \frac{1}{2\pi}\left(-\gamma -\ln m + \ln 2 - \ln \left|\xperp-\yperp\right|\right) + \mathcal{O}(m^2).
\end{align}
Note that this is the same as in \eqref{eq:greens_fctn_eps_expansion} with $m$ instead of $\epsilon$. After subtracting an ($m$-dependent) constant, one can again perform the limit $m \rightarrow 0$ and recover the form \eqref{eq:greens_ln} for the Green's function. The letter $m$ for the regulator was chosen to highlight its role as a screening mass. It suppresses long-ranged interactions and confines them to a distance $\sim 1/m$. Therefore, the presence of $m$ should not just be understood as a mathematical trick but as a fundamental physical quantity that is part of the model.\par
There are two possible interpretations for the introduction of the screening mass. Either one considers a new modified Poisson equation $(-\nabla_\perp^2 + m^2)\phi(\xperp) = \rho(\xperp)$, for which the Green's function is \eqref{eq:greens_fctn_k0}. Alternatively, one could still consider the unmodified Poisson equation but with a new source term obtained by a modification that is most easily expressed in momentum space as
\begin{align}
    \tilde \rho (\kperp) \rightarrow \frac{\kperp^2}{\kperp^2 + m^2}\tilde \rho(\kperp).
\end{align}
The appropriate Green's function to use with this new source term is given by \eqref{eq:greens_ln}. The two different Green's functions have different (asymptotic) behavior, as discussed before, but this is exactly compensated by the correction factor for the (Fourier transformed) source.
\section{Discrete Poisson equation}
The method of Green's functions is good for building intuition about the solution, but for finding $\phi$ on a discretized lattice, there is a more convenient way. First, the derivative in the Poisson equation needs to be replaced with one of its discrete analogs. We choose to write it as
\begin{align}
\label{eq:Poisson_discretized}
    -\frac{\phi_{x+1,y} -2\phi_{x,y} + \phi_{x-1,y}}{(a^1)^2} - \frac{\phi_{x, y+1} -2\phi_{x,y} + \phi_{x, y-1}}{(a^2)^2} = \rho_{x,y}.
\end{align}
This equation can now be solved using the discrete Fourier transformation (DFT). For a two-dimensional grid with $n_i$ points in the $i$-th direction, the DFT can be defined as
\begin{align}
\tilde \phi_{j,k} \coloneqq \sum_{x=0}^{n_1-1} \sum_{y=0}^{n_2-1} \phi_{x,y} \exp\left[-2\pi\ii \left(\frac{xj}{n_1} + \frac{yk}{n_2}\right)\right],
\end{align}
where $j$ runs from $0$ to $n_1 -1$ and $k$ runs from $0$ to $n_2 -1$.\footnote{Along with the distribution of the normalization between forward and backward transformation, the range of $j$ and $k$ is purely conventional. Note that if $j=0$ is replaced by $j=n_1$, the result for $\tilde \phi_{j,k}$ does not change.} The backtransformation (inverse DFT) is given by
\begin{align}
\label{eq:inverse_DFT}
    \phi_{x,y} =\frac{1}{n_1n_2} \sum_{j=0}^{n_1-1}\sum_{k = 0}^{n_2-1} \tilde \phi_{j,k}\exp\left[2\pi \ii \left(\frac{xj}{n_1} + \frac{yk}{n_2}\right)\right].
\end{align}
Inserting it into the discrete Poisson equation gives
\begin{align}
   & -\frac{1}{n_1n_2}\sum_{j=0}^{n_1-1}\sum_{k = 0}^{n_2-1}\tilde \phi_{j,k}\exp\left[2\pi \ii \left(\frac{xj}{n_1} + \frac{yk}{n_2}\right)\right]\nn\\
    &\hspace{3cm}\times\left[\frac{\exp(2\pi \ii j /n_1) - 2 + \exp(-2\pi \ii j/n_1)}{(a^1)^2} \right.\nn\\
    &\hspace{5cm}\left.+ \frac{\exp(2\pi \ii k /n_2) - 2 + \exp(-2\pi \ii k/n_2)}{(a^2)^2} \right]\nn\\
    &=\frac{1}{n_1n_2} \sum_{j=0}^{n_1-1}\sum_{k = 0}^{n_2-1} \tilde \rho_{j,k}\exp\left[2\pi \ii \left(\frac{xj}{n_1} + \frac{yk}{n_2}\right)\right].
\end{align}
Thus,
\begin{align}
\label{eq:mom_space_discrete_poisson}
    \tilde \rho_{j,k} &= -2\left(\frac{\cos(2\pi j/n_1) -1}{(a^1)^2}+\frac{\cos(2\pi k/n_2) -1}{(a^2)^2}\right) \tilde \phi_{j,k}\nn\\
    &=4\left(\frac{\sin^2(\pi j/n_1) }{(a^1)^2}+\frac{\sin^2(\pi k/n_2)}{(a^2)^2}\right) \tilde \phi_{j,k}\nn\\
    &=(p^2_{j,k})_\text{lat}\tilde \phi_{j,k} ,
\end{align}
where we introduced the squared lattice momentum
\begin{align}
(p^2_{j,k})_{\text{lat}} \coloneqq  4\left(\frac{\sin^2(\pi j/n_1) }{(a^1)^2}+\frac{\sin^2(\pi k/n_2)}{(a^2)^2}\right).
\end{align}
In comparison, the continuum momentum associated with the $(j,k)$-mode of the Fourier decomposition can be determined by comparing to \eqref{eq:inverse_DFT} as
\begin{align}
\label{eq:squared_mode_momentum}
    (p^2_{j,k})_{\text{con}} = 4\pi^2\left[\left( \frac{j}{a^1n_1}\right)^2 +  \left(\frac{k}{a^2n_2}\right)^2\right].
\end{align}
However, this expression is only valid for $j < n_1/2$ and $k < n_2/2$. Values $j > n_1/2$ and $k > n_2/2$ are associated with negative momenta while $j = n_1/2$ (for even $n_1$) and $k = n_2/2$ (for even $n_2$) can be viewed as either positive or negative frequency contributions (they correspond to the Nyquist frequency). Equation \eqref{eq:squared_mode_momentum} is valid for all possible values of $j$ and $k$ if one replaces
\begin{align}
    j \rightarrow j - n_1\qquad\qquad \text{for}\;j > n_1/2,\\
    k \rightarrow k - n_2\qquad\qquad \text{for}\;k > n_2/2.
\end{align}
It is straightforward to see that in the case of small momenta/large wavelengths, the approximation $\sin x \sim x$ leads to the same expressions for the mode momentum and the lattice momentum. This should not be surprising as long wavelength modes can be represented well enough on the discrete lattice. The difference increases for larger momenta, where the squared continuum momentum will generally be larger, but it will never be larger than the squared lattice momentum by more than a factor of $\pi^2/4$.
For a square lattice in lattice units, i.e.,\ $n_1 = n_2 = n$ and $a^1=a^2 = 1$, the squared lattice momentum becomes
\begin{align}
    (p^2_{j,k})_\text{lat} = 4\left[\sin^2\left(\frac{\pi j}{n}\right)+\sin^2\left(\frac{\pi k}{n}\right)\right].
\end{align}
The discretized Poisson equation can now be solved in three simple steps. First, compute the DFT of the charge density $\rho$. Then, divide by the squared lattice momentum $(p^2_{j,k})_\text{lat}$. Finally, compute the inverse DFT and obtain the solution $\phi$. However, this approach should be modified since it exhibits two problems. First of all, there will, in general, be a zero mode with $(p^2_{0,0})_\text{lat}= 0$ in the charge density that leads to an infrared (IR) divergence in $\tilde \phi$. Note that any sensible model of a nucleus will be globally color-neutral. If we enforce this condition on each transverse slice separately, we can simply set the zero mode $\tilde \rho_{0,0}$ to zero. However, it will still be beneficial to introduce an IR cutoff parametrized by the screening mass $m$. It is also reasonable to worry about the ultraviolet (UV) case.
High momentum modes are not represented well on the lattice, and the lattice spacing introduces a hard cutoff, after which momenta cannot be represented on the grid.
The maximum lattice momentum that can be resolved on the lattice is
\begin{align}
    (p^2)_\mathrm{lat}^\mathrm{max} = 4\left[\sin^2\left(\frac{\pi (n-1)}{n}\right)+\sin^2\left(\frac{\pi (n-1)}{n}\right)\right] \approx 8.
\end{align}
This provides a crude mechanism of UV regularization, but the physical value of this cutoff will depend on the lattice spacing. To eliminate this dependence on the arbitrary lattice spacing, an additional exponential UV cutoff with some scale $\Lambda_\text{UV}$ may be introduced. The solution to the discrete momentum space Poisson equation \eqref{eq:mom_space_discrete_poisson} with IR and UV regulators is then
\begin{align}
\label{eq:DFT_Poisson}
    \tilde \phi_{j,k} = \frac{1}{(p^2_{j,k})_\text{lat} + m^2}\tilde \rho_{j,k} \exp\left(-\frac{(p^2_{j,k})_\text{lat}}{2\Lambda_\text{UV}^2}\right).
\end{align}
Finally, transforming back using the inverse DFT gives the field $\phi$ on the lattice in position space.
As argued before, there are two possible viewpoints on the IR and UV regulators. One is to interpret the Poisson equation as being modified. For example, the IR regulator corresponds to replacing $-\nabla_\perp^2 \rightarrow -\nabla_\perp^2 + m^2$, as discussed for the continuum case. The other viewpoint is to consider it a modification of the charge density before the equation is even solved. The modification is conveniently expressed for the Fourier transformed charge density as
\begin{align}
    \tilde \rho_{j,k}\rightarrow \frac{(p^2_{j,k})_\text{lat}}{(p^2_{j,k})_\text{lat} + m^2}\tilde \rho_{j,k} \exp\left(-\frac{(p^2_{j,k})_\text{lat}}{2\Lambda_\text{UV}^2}\right),
\end{align}
that is, one dampens the very high and very low modes (and removes the zero mode). The unmodified discretized Poisson equation is then solved for this new charge density.\par
It is also worth discussing boundary conditions. In the previous treatment of the Poisson equation in the continuum, the choice of a Green's function dictated the asymptotic behavior. By solving the Poisson equation with the DFT on the lattice, one never gets to make a choice. Instead, the solution consists of complex exponentials, which are periodic on the spacetime region covered by the lattice, as can be seen from \eqref{eq:inverse_DFT}. The solution will, therefore, always exhibit periodic boundary conditions and, due to the removal of the zero mode, will be zero if integrated over the entire plane.
\chapter{Gaussian probability functionals}
\label{ch:gaussian}
The McLerran-Venugopalan model \eqref{eq:MV_2d_simple_1}--\eqref{eq:MV_2d_simple_2} and its generalizations presented in this thesis are based on the concept of Gaussian probability functionals for color charge configurations. In general, a probability functional ascribes a probability to each conceivable color charge configuration. In this appendix, we give a brief introduction to Gaussian probability functionals and how to extract the 1- and 2-point functions from them.\par
Suppose we have a stochastic variable $\rho(x)$, which is a function of the (potentially multidimensional) coordinate $x$. For simplicity, we do not consider any color structure in this section. If this variable is distributed via a Gaussian probability distribution, it has a partition function
\begin{align}
\label{eq:gauss_partition_function}
    Z[J(\cdot)] = \int D\rho\, \exp( -\frac{1}{2}\int \dd{x} \int \dd{y} \rho(x) M(x,y) \rho(y) + \int\dd{x}J(x)\rho(x))
\end{align}
where we have included a source term $J(\cdot)$, which is a function of spacetime, but, as the notation suggests, we do not pick a specific spacetime coordinate on the left-hand-side of \eqref{eq:gauss_partition_function}.
Note that only the part of $M(x,y)$ that is symmetric in its arguments contributes, so we assume that $M(x,y)$ is symmetric. To obtain explicit expressions for the 1-point function
\begin{align}
    \langle \rho(u) \rangle &= \frac{1}{Z[0]}\left.\frac{\delta Z}{\delta J(u)}\right|_{J=0}\nn\\
    &= \frac{1}{Z[0]}\int D\rho\,\rho(u) \exp( -\frac{1}{2}\int \dd{x} \int \dd{y} \rho(x) M(x,y) \rho(y))
\end{align}
and 2-point function
\begin{align}
    \langle \rho(u) \rho(v)\rangle &= \frac{1}{Z[0]}\left.\frac{\delta^2 Z}{\delta J(u)\delta J(v)}\right|_{J=0}\nn\\
    &= \frac{1}{Z[0]}\int D\rho\,\rho(u)\rho(v) \exp( -\frac{1}{2}\int \dd{x} \int \dd{y} \rho(x) M(x,y) \rho(y))
\end{align}
we have to solve the functional integral in the partition function. We first perform a change of variables $\eta(x) = \rho(x) - \int \dd{y}M^{-1}(x,y)J(y)$, where $M^{-1}(x,y)$ is the symmetric function defined by $\int \dd{y} M^{-1}(x,y) M(y,z) = \delta(x-z)$. This leads to
\begin{align}
    Z[J(\cdot)] &= \exp( \frac{1}{2} \int \dd{x}\int\dd{y}J(x)M^{-1}(x,y)J(y))\nn\\
    &\quad\times\int D\eta \exp(-\frac{1}{2} \int \dd{x}\int\dd{y}\eta(x)M(x,y)\eta(y)),
\end{align}
where the functional integral is now just a Gaussian integral, which can be carried out, giving a single prefactor. We can assume this prefactor to cancel with the measure $D\eta$ in such a way that the functional integral gives 1. This also conveniently normalizes $Z[0]=1$, as we now have
\begin{align}
    Z[J(\cdot)] = \exp( \frac{1}{2} \int \dd{x}\int\dd{y}J(x)M^{-1}(x,y)J(y)).
\end{align}
From this explicit form for the partition function, we obtain
\begin{align}
    \langle \rho(u) \rangle =Z[J(\cdot)] \left.\int \dd{x} J(x) M^{-1}(x,u)\right|_{J=0} = 0
\end{align}
and
\begin{align}
    &\langle \rho(u)\rho(v) \rangle\nn \\
    &\hspace{0.5cm}= \left.Z[J(\cdot)]\left(  M^{-1}(u,v)+\int \dd{x} J(x) M^{-1}(x,u)\int \dd{y} J(y) M^{-1}(y,v)\right)\right|_{J=0}\nn\\
    &\hspace{0.5cm}= M^{-1}(u,v).
\end{align}
Note that for a valid correlator
\begin{align}
    \langle \rho(x) \rho(y)\rangle = M^{-1}(x,y)
\end{align}
the matrix $M(x,y)$ must be invertible for a corresponding Gaussian probability functional to exist.

\chapter{Boost-invariant limit of the dilute Glasma}
\label{ch:boost-invariant_limit}
The expressions \eqref{eq:fpm}--\eqref{eq:fij} for the field strength tensor of the dilute Glasma recover the correct literature result \cite{Guerrero-Rodriguez:2021ask} in the boost-invariant limit. Consider the prototypical integral
\begin{align}
\label{eq:I_prototypical}
    I(x)=\int_{-\infty}^\infty\! \dd{\eta'}\intop_\vperp \comm{\beta^i_A(x^+\!-\frac{|\vperp|}{\sqrt{2}}\ee^{+\eta'}, \xperp - \vperp)}{\beta^j_B(x^-\!-\frac{|\vperp|}{\sqrt{2}}\ee^{-\eta'},\xperp - \vperp)} w^k\ee^{\pm \eta'},
\end{align}
from which the components of the field strength tensor can be constructed (performing index contractions, multiplying with prefactors, and omitting $w^k$ or $\ee^{\pm \eta'}$ as appropriate).
The boost-invariant limit corresponds to the replacement
\begin{align}
\label{eq:boost-invariant_limit_replacement}
    \beta_{A/B}^i(x^\pm - \frac{|\vperp|}{\sqrt{2}}\ee^{\pm \eta'})\rightarrow \delta(x^\pm - \frac{|\vperp|}{\sqrt{2}}\ee^{\pm \eta'})\alpha^i_{A/B}(\xperp - \vperp).
\end{align}
Note that
\begin{align}
    \delta(x^+-\frac{|\vperp|}{\sqrt{2}}\ee^{+\eta'})\delta(x^- - \frac{|\vperp|}{\sqrt{2}}\ee^{-\eta'})=\frac{\delta(\tau - |\vperp|)}{\tau}\delta(\eta_s-\eta').
\end{align}
We can now make the replacement \eqref{eq:boost-invariant_limit_replacement} in \eqref{eq:I_prototypical} and carry out the $\eta'$-integral, which yields
\begin{align}
    I(x) = \ee^{\pm \eta_s}\intop_\vperp \frac{\delta(\tau - |\vperp|)}{\tau} \comm{\alpha^i_A( \xperp - \vperp)}{\alpha^j_B(\xperp - \vperp)} w^k.
\end{align}
Writing the two-dimensional integral in polar coordinates $(|\vperp|, \theta_\vperp)$, we may carry out the radial part and are left with
\begin{align}
    I(x)=\ee^{\pm \eta_s}\int \dd{\theta_{\mathbf v}}\comm{\alpha^i_A( \xperp - \tau \eperp(\theta_{\mathbf{v}}) )}{\alpha^j_B(\xperp - \tau \eperp(\theta_{\mathbf{v}}))} \eperp(\theta_{\vperp})^k\,
\end{align}
where $\eperp(\theta_\vperp)$ is the unit vector in $\theta_\vperp$-direction. To compare with \cite{Guerrero-Rodriguez:2021ask} we would like to express
\begin{align}
    T^{\tau\tau} = \Tr\left[f^{+-}f^{+-} + \ee^{-2\eta_s}f^{+i}f^{+i} + \ee^{+2\eta_s}f^{-i}f^{-i}+\frac{1}{2}f^{ij}f^{ij} \right]
\end{align} in the boost-invariant limit. We collect the individual terms,
\begin{align}
    f^{+-}f^{+-} &= -\frac{g^2}{4\pi^2}\delta^{ij}\delta^{kl}\int \dd{\theta_\uperp} \int \dd{\theta_\vperp}
    \comm{\alpha^i_A( \xperp - \tau \eperp(\theta_{\mathbf{u}}) )}{\alpha^j_B(\xperp - \tau \eperp(\theta_{\mathbf{u}}))}\nn\\
    &\hspace{4.5cm}\times
    \comm{\vphantom{\alpha^j_A}\alpha^k_A( \xperp - \tau \eperp(\theta_{\mathbf{v}}) )}{\alpha^l_B(\xperp - \tau \eperp(\theta_{\mathbf{v}}))},\\
    \ee^{-2\eta_s}f^{+i}f^{+i}&=-\frac{g^2}{8\pi^2}
    (\epsilon^{mn}\epsilon^{ij}-\delta^{mn}\delta^{ij}) (\epsilon^{mp}\epsilon^{kl}-\delta^{mp}\delta^{kl})\nn\\
    &\hspace{0.5cm}\times\int \dd{\theta_\uperp} \int \dd{\theta_\vperp}
    \comm{\alpha^i_A( \xperp - \tau \eperp(\theta_{\mathbf{u}}) )}{\alpha^j_B(\xperp - \tau \eperp(\theta_{\mathbf{u}}))}\nn\\
    &\hspace{2.3cm}\times
    \comm{\vphantom{\alpha^j_A}\alpha^k_A( \xperp - \tau \eperp(\theta_{\mathbf{v}}) )}{\alpha^l_B(\xperp - \tau \eperp(\theta_{\mathbf{v}}))}
    \mathbf{e}(\theta_\uperp)^n\mathbf{e}(\theta_\vperp)^p,\\
    \ee^{+2\eta_s}f^{-i}f^{-i}&=-\frac{g^2}{8\pi^2}
    (\epsilon^{mn}\epsilon^{ij}+\delta^{mn}\delta^{ij}) (\epsilon^{mp}\epsilon^{kl}+\delta^{mp}\delta^{kl})\nn\\
    &\hspace{0.5cm}\times\int \dd{\theta_\uperp} \int \dd{\theta_\vperp}
    \comm{\alpha^i_A( \xperp - \tau \eperp(\theta_{\mathbf{u}}) )}{\alpha^j_B(\xperp - \tau \eperp(\theta_{\mathbf{u}}))}\nn\\
    &\hspace{2.3cm}\times
    \comm{\vphantom{\alpha^j_A}\alpha^k_A( \xperp - \tau \eperp(\theta_{\mathbf{v}}) )}{\alpha^l_B(\xperp - \tau \eperp(\theta_{\mathbf{v}}))}
    \mathbf{e}(\theta_\uperp)^n\mathbf{e}(\theta_\vperp)^p,\\
    \frac{1}{2}f^{ij}f^{ij}&=-\frac{g^2}{4\pi^2}
    \epsilon^{ij}
    \epsilon^{kl}
    \int \dd{\theta_\uperp} \int \dd{\theta_\vperp}
    \comm{\alpha^i_A( \xperp - \tau \eperp(\theta_{\mathbf{u}}) )}{\alpha^j_B(\xperp - \tau \eperp(\theta_{\mathbf{u}}))}\nn\\
    &\hspace{4.5cm}\times
    \comm{\vphantom{\alpha^j_A}\alpha^k_A( \xperp - \tau \eperp(\theta_{\mathbf{v}}) )}{\alpha^l_B(\xperp - \tau \eperp(\theta_{\mathbf{v}}))}.
\end{align}
Then,
\begin{align}
    T^{\tau\tau} &= \Tr\left[f^{+-}f^{+-} + \ee^{-2\eta_s}f^{+i}f^{+i} + \ee^{+2\eta_s}f^{-i}f^{-i}+\frac{1}{2}f^{ij}f^{ij} \right]\\
    &=-\frac{g^2}{4\pi^2}(\delta^{ij}\delta^{kl}+\epsilon^{ij}\epsilon^{kl})\nn\\
    &\hspace{0.5cm}\times
    \int \dd{\theta_\uperp} \int \dd{\theta_\vperp}
    \Tr\left\{
    \comm{\alpha^i_A( \xperp - \tau \eperp(\theta_{\mathbf{u}}) )}{\alpha^j_B(\xperp - \tau \eperp(\theta_{\mathbf{u}}))}\right.\nn\\
    &\hspace{2.5cm}\times\left.
    \comm{\vphantom{\alpha^j_A}\alpha^k_A( \xperp - \tau \eperp(\theta_{\mathbf{v}}) )}{\alpha^l_B(\xperp - \tau \eperp(\theta_{\mathbf{v}}))}\right\}
    \left(1+\cos (\theta_\uperp - \theta_\vperp)\right),
\end{align}
where it was used that $\mathbf{e}(\theta_\uperp)^i\mathbf{e}(\theta_\vperp)^i = \cos (\theta_\uperp - \theta_\vperp)$. This result matches the expression (64) in \cite{Guerrero-Rodriguez:2021ask}. Note that the boost-invariant limit reduces the number of integrals in the field-strength tensor from three to one. This also simplifies the corresponding numerical integration procedure tremendously. The main allure of the dilute Glasma, however, is the semi-analytical treatment of the general case, away from the boost-invariant limit and the study of rapidity-dependent observables.

\chapter{Monte Carlo integration}
\label{ch:mc}
Monte Carlo integration is a method of numerically evaluating definite integrals that is particularly useful for high-dimensional integrals. This section contains a brief introduction to the basic Monte Carlo method and the concept of importance sampling.
\section{Standard Monte Carlo integration}
Take the definite integral
\begin{align}
\label{eq:mc_integral}
    F=\int_D \dd{x} f(x) 
\end{align}
over the domain $D$. The Monte Carlo method of evaluating this integral is to sample a number $N$ of random values $x_n$ for $x$ and evaluate $f(x)$ for each of them. The integral can then be approximated by
\begin{align}
    F\approx E \coloneqq \frac{V}{N}\sum_{n=1}^N f(x_n),
\end{align}
where $V= \int_D \dd{x}$ is the volume of $D$. We can repeat this process of drawing $N$ random numbers and using them to obtain $E$ any number of times and compute the average of $E$ over all these processes, which is given by
\begin{align}
    \langle E\rangle = \frac{1}{V^N}\int_D \dd{x_1} \dotsm \int_D \dd{x_N}\,\frac{V}{N}\sum_{n=1}^N f(x_n) = \frac{1}{V^{N-1}N}V^{N-1}\sum_{n=1}^N F = F.
\end{align}
Most importantly, $E$ is not biased and will, on average, be equal to the true value $F$ of the integral. However, for any finite $N$, the estimate $E$ will have some error, so we compute the variance of $E$. First note that
\begin{align}
    \sigma_E^2 = \langle E^2 \rangle - \langle E \rangle^2 = \langle E^2\rangle - F^2 = \langle (E-F)^2\rangle,
\end{align}
which we evaluate as
\begin{align}
    \sigma_E^2 &= \frac{1}{V^N} \int_D \dd{x_1} \dotsm \int_D \dd{x_N}\left(\frac{V}{N} \sum_{n=1}^N f(x_n) -F\right)^2\nn\\&= \frac{1}{V^{N-2}N^2} \int_D \dd{x_1} \dotsm \int_D \dd{x_N}\left[\sum_{n=1}^N \left(f(x_n) -\frac{F}{V}\right)\right]^2\nn\\
    &=\frac{1}{V^{N-2}N^2}\sum_{n=1}^N\sum_{m=1}^N \int_D \dd{x_1} \dotsm \int_D \dd{x_N} \left(f(x_n) -\frac{F}{V}\right)\left(f(x_m) -\frac{F}{V}\right).
\end{align}
Clearly, this is zero for $n \neq m$, so
\begin{align}
    \sigma^2_E &= \frac{1}{V^{N-2}N^2}\sum_{n=1}^N\int_D \dd{x_1} \dotsm \int_D \dd{x_N} \left(f(x_n) -\frac{F}{V}\right)^2=\frac{V^2}{N} \sigma_f^2,
\end{align}
where
\begin{align}
    \sigma^2_f := \frac{1}{V} \int_D \dd{x} \left(f(x) - \frac{F}{V}\right)^2=\frac{1}{V}\int_D\dd{x}f(x)^2-\frac{F^2}{V^2}
\end{align}
is called the variance of $f$. It only depends on the specifics of the function $f$. However, this quantity is only accessible if the true value $F$ of the integral is known already. Instead, one can compute the sample variance
\begin{align}
    s_f^2 &:= \frac{1}{N-1}\sum_{n=1}^N \left(f(x_n) - \frac{E}{V}\right)^2\nn \\
    &=\frac{1}{N-1}\sum_{n=1}^N \left(
    f(x_n)^2- 2 f(x_n)\frac{E}{V} + \frac{E^2}{V^2}\right)\nn\\
    &=\frac{1}{N-1}\sum_{n=1}^N f(x_n)^2 - \frac{2N}{N-1}\frac{E^2}{V^2}+\frac{N}{N-1}\frac{E^2}{V^2}\nn\\
    &= \frac{1}{N-1}\left(\sum_{n=1}^Nf(x_n)^2- \frac{NE^2}{V^2}\right)
\end{align}
and use it instead of $\sigma^2_f$ whenever an estimate of the Monte Carlo error is required. This Monte Carlo error is then given by
\begin{align}
    \sigma_E = \frac{V}{\sqrt{N}} \sigma_f \approx \frac{V}{\sqrt{N}}s_f.
\end{align}
Note in particular the $\sqrt{N}$-dependence, which is the same no matter the number of dimensions in the integral. This makes the Monte Carlo method particularly well suited for high-dimensional integrals, where other numerical integration schemes scale poorly. Still, decreasing the Monte Carlo error by a factor of two requires four times the number of samples. There are also several techniques for reducing the Monte Carlo error without increasing computation time, which require some knowledge about the integrand. One of these techniques is known as importance sampling.
\section{Importance sampling}
The basic idea behind importance sampling is simple. There might be a region where the integrand is large and another region where it contributes next to nothing to the overall integral. It is then beneficial to choose random points not evenly distributed over the domain $D$ but predominantly in the region where the integrand is large. In other words, we suppose our integrand $f(x)$ factorizes in some arbitrary function $g(x)$ and some positive distribution function $p(x)$, such that
\begin{align}
 F = \int_D \dd{x} g(x) p(x),\qquad\qquad \int_D \dd{x} p(x) = 1.
\end{align}
We approximate this integral by
\begin{align}
    \label{eq:estimate_importance_sampling}
    E'  =\frac{1}{N} \sum_{n=1}^N g(x_n),
\end{align}
where the random numbers $x_n$ are drawn from the distribution described by $p(x)$. Of course
\begin{align}
    \langle E'\rangle = \int_D \dd{x_1}p(x_1) \dotsm \int_D \dd{x_N}p(x_N)\,\frac{1}{N}\sum_{n=1}^N  \frac{f(x_n)}{p(x_n)} = F,
\end{align}
but we can also compute the variance of $E'$ as
\begin{align}
    \sigma^2_{E'} 
    &= \int_D \dd{x_1}p(x_1) \dotsm \int_D \dd{x_N}p(x_N) \left(\frac{1}{N} \sum_{n=1}^N \frac{f(x_n)}{p(x_n)} -F\right)^2\nn\\
    &=\frac{1}{N^2}\int_D \dd{x_1}p(x_1) \dotsm \int_D \dd{x_N}p(x_N)\left[  \sum_{n=1}^N  \left(\frac{f(x_n)}{p(x_n)} -F\right)\right]^2\nn\\
    &=\frac{1}{N}\left(\int_D \dd{x} \frac{f(x)^2}{p(x)} -F^2\right).
\end{align}
We can now take a look at
\begin{align}
\label{eq:sigmaE_minus_sigmaE'}
    \sigma_E^2 - \sigma_{E'}^2 = \frac{1}{N}\int_D\dd{x}\left(V\,f(x)^2-\frac{f(x)^2}{p(x)}\right).
\end{align}
If this quantity is positive, importance sampling with the distribution $p(x)$ reduces the integration error. The effectiveness of importance sampling depends on the specifics of $f(x)$ and $p(x)$. It is clear from \eqref{eq:sigmaE_minus_sigmaE'} that $p(x)$ should be large wherever $\left| f(x) \right|$ is large and $p(x)$ can be small wherever $\left| f(x) \right|$ is small, subject, of course, to the overall normalization condition for $p(x)$. Note also that a bad choice of $p(x)$ can increase the Monte Carlo integration error, and some knowledge about the shape of $f(x)$ is required to choose a suitable distribution $p(x)$.
\section{Square of an integral}
Suppose we are interested in the square $F^2$ of some integral
\begin{align}
    F = \int_D\dd{x}f(x).
\end{align}
We might want to make use of importance sampling, and in that case, a reasonable first approach is to compute the estimate \eqref{eq:estimate_importance_sampling} and take the square of it, i.e.,\ compute
\begin{align}
    (E')^2 = \frac{1}{N^2}\left( \sum_{n=1}^Ng(x_n)\right)^2,
\end{align}
where $N$ random numbers $x_n$ are drawn. However, the expectation value of this estimate
\begin{align}
    \langle (E')^2\rangle &= \frac{1}{N^2}\int_D \dd{x_1}p(x_1) \dotsm \int_D \dd{x_N}p(x_N)\,\sum_{n=1}^N\sum_{m=1}^N \frac{f(x_n)}{p(x_n)}\frac{f(x_m)}{p(x_m)}\nn\\
    &=\frac{1}{N^2}\int_D \dd{x_1}p(x_1) \dotsm \int_D \dd{x_N}p(x_N)\left[\sum_{n=1}^N\sum_{\substack{m=1\\m\neq n}}^N \frac{f(x_n)}{p(x_n)}\frac{f(x_m)}{p(x_m)}
    \right.\nn\\&\hspace{7cm}\left.\vphantom{\sum_{\substack{m=1\\m\neq n}}^N}
    +\sum_{n=1}^N\left(\frac{f(x_n)}{p(x_n)}\right)^2\right]\nn\\
    &=\frac{N(N-1)}{N^2}F^2 + \frac{1}{N}\int_D\dd{x}\frac{f(x)^2}{p(x)}\nn\\
    &=F^2 + \sigma_{E'}^2
\end{align}
is not equal to the true value $F^2$. Using this technique, we will always overestimate the value of $F^2$ by some amount $\sigma^2_{E'}$. Therefore, $(E')^2$ is a biased estimator, and we should not be using it to compute the value of $F^2$. Instead, we should compute $E'$ twice with completely independent sets of random numbers $x_n$. In that case, the expectation value factorizes, and we get
\begin{align}
    \langle E'_1 E'_2 \rangle &= \frac{1}{N}\int_D \dd{x_1}p(x_1) \dotsm \int_D \dd{x_N}p(x_N)\,\sum_{n=1}^N\frac{f(x_n)}{p(x_n)}\nn\\
    &\qquad \times \frac{1}{N}\int_D \dd{x'_1}p(x'_1) \dotsm \int_D \dd{x'_N}p(x'_N)\,\sum_{n=1}^N\frac{f(x'_n)}{p(x'_n)}\nn\\
    &=F^2.
\end{align}
The standard deviation of this estimator is given by
\begin{align}
    \sigma_{E_1'E_2'}^2 &= 
    \langle (E_1'E_2')^2\rangle- \langle E_1' E_2'\rangle^2\nn\\
    &=\langle (E_1')^2\rangle\langle (E_2')^2\rangle- \langle E_1'\rangle^2 \langle E_2'\rangle^2\nn\\
    &= (F^2 + \sigma^2_{E'})^2 - F^4\nn\\
    &=\sigma^2_{E'}(2F^2 + \sigma^2_{E'}).
\end{align}
The $1\sigma$ interval for the estimator $E_1'E_2'$ is thus given by $F^2 \pm \sigma^2_{E'}\sqrt{1+2F^2/\sigma_{E'}^2}$.
\chapter{Natural and lattice units}
\label{ch:units}
\section{Natural units}
It is customary in particle physics to express quantities in natural units where the reduced Planck constant $\hbar$ and the vacuum speed of light $c$ are set to $1$. This means that the three SI basis units $\unit{\meter}$, $\unit{\second}$, and $\unit{\kilogram}$ are no longer independent. Instead, all physical quantities that are usually expressed in terms of $\unit{\meter}$, $\unit{\second}$, and $\unit{\kilogram}$ can now be expressed in terms of various powers of a single base unit. Picking, for example, meters as the single base unit, one finds that
\begin{align}
    \SI{1}{\meter} &= \SI{1}{\meter},\\
    \SI{1}{\second}&=\frac{c}{c}\,\unit{\second} = \frac{\SI{299792458}{\meter\per\second}}{c}\,\unit{\second} = \SI{299792458}{\meter},\\
    \SI{1}{\kg} &= \frac{\hbar c}{\hbar c}\,\unit{\kg} = \frac{\hbar}{c}\,\SI{2.84278849e42}{\per\meter\per\kg\kg} = \SI{2.84278849e42}{\per\meter}.
\end{align}
The trick is to always multiply by 1, expressed as some suitable combination of $\hbar$ and $c$. One can then replace some of these factors with their actual numerical value and units while setting the other factors to 1 such that the units cancel in the desired way.\par
In the world of particle physics, one meter is an astronomically large distance and very impractical to use as a base unit. Instead, one can express all lengths, times, masses, and derived quantities in terms of $\unit{\femto\meter}$ (femtometers or fermi), which are much better adapted to (sub-)atomic length scales. Once all physical quantities are expressed as some power of meters, it is a trivial task to convert them to powers of femtometers. Another popular choice in the particle physics community is to express everything in terms of energy. A suitable base unit in that case is $\unit{\GeV}$ (Gigaelectronvolts). Quantities can be converted from $\unit{\fermi}$ to $\unit{\GeV}$ and back by using
\begin{align}
    \hbar c = \SI{0.1973269804}{\GeV\fermi} = 1,
\end{align}
from which the relations
\begin{align}
    \SI{1}{\GeV}&= \SI{5.0677307177}{\per\fermi},\\
    \SI{1}{\fermi}&= \SI{5.0677307177}{\per\GeV}
\end{align}
can be derived.\par
Working in natural units comes at a price. One loses the power of dimensional analysis to verify results, and thus, special care is required. In order to perform any calculations, all variables need to be expressed in powers of the same base unit. In principle, this could be any SI unit or combination thereof, but in practice, either $\unit{\fermi}$ or $\unit{\GeV}$ is used. The numerical result has to be interpreted with respect to the same basis unit. Also, special care is required when some quantities are given in $\unit{\fermi}$ while others are given in $\unit{\GeV}$. However, if one can find an additional natural constant to set to $1$, all quantities will be dimensionless.\footnote{In (quantum) gravity, this is accomplished by setting Newton's constant $G$ to $1$. This makes the Planck length the fundamental length scale of the system. While this is the most \say{natural} choice (all scales come from natural constants), it is somewhat arbitrary in a framework where $G$ never enters any of the equations. In that case, a length scale that is more natural to the system is a better choice.} This leads to the notion of lattice units, which are helpful whenever a lattice of some fixed length scale is present.
\section{Lattice units}
Discretizing physical fields on a lattice introduces another natural choice of length and energy scales. Expressing physical quantities in terms of these scales solves the aforementioned issue with $\unit{\fermi}$ and $\unit{\GeV}$ not being equivalent and can also simplify some numerical manipulations, like taking derivatives. The lattice constant $a$ introduces a notion of a fundamental distance on the lattice. All lengths can be expressed as multiples of the lattice constant. The numerical value $x_{\text{l.u.}}$ of a length $x$ in lattice units is then
\begin{align}
\label{eq:x_l_u}
    x_{\text{l.u.}} \coloneqq \frac{x}{a},
\end{align}
where $x$ can be given in any unit as long as $a$ is given in the same unit.
Note that $x_{\text{l.u.}}$ is a dimensionless number that holds no information about the absolute length scale of the system.
This information is encoded in $a$, and only the product $x_{\text{l.u.}}a$ is a physical length. As was shown in the previous section, a length can be related to an energy using the factor $\hbar c=1$.
Therefore, the lattice also defines a fundamental energy scale
\begin{align}
    e_0 \coloneqq \frac{\hbar c}{a}.
\end{align}
There are two ways of interpreting this definition. One can set $\hbar c = 1$ and take $a$ to be given in units of inverse energy, e.g.,\ $\unit{\per \GeV}$. This gives $e_0$ in that energy unit. Otherwise, one can take $a$ to be given as a length and insert $\hbar c$ as a numerical value with units. This will also yield a dimension of energy for $e_0$ once the length units of $\hbar c$ and $a$ are appropriately canceled. Either way, one arrives at a fundamental lattice energy $e_0$ and the numerical value $E_\text{l.u.}$ of an energy $E$ in lattice units is then
\begin{align}
    E_\text{l.u.} = \frac{E}{e_0}.
\end{align}
Note that $e_0$ has the dimension of an energy, so it consists of a numerical value and a unit, while $E_\text{l.u.}$ is dimensionless and completely agnostic of the unit in which $E$ was originally described. Alternatively, one can use the fundamental lattice energy to relate
\begin{align}
    x_\text{l.u.} = x\, e_0 = x\, \frac{\hbar c}{a}
\end{align}
if $x$ is given in units of inverse energy (first equality). The second equality (where on the right-hand side $a$ should be viewed as a length and $\hbar c$ as an energy times length) shows that this is consistent with first rewriting $x$ from an inverse energy to a length (by multiplying with $\hbar c$) and then expressing it in lattice units via \eqref{eq:x_l_u}.\par
The upshot is that converting quantities to lattice units is always straightforward. One takes the physical quantity given as a power of length or energy and makes it dimensionless by multiplying with appropriate powers of $a$ or $e_0$, respectively. The result will always be a dimensionless quantity in lattice units.

\hyphenation{pseudo-rapidity}
\printbibliography

\end{document}